\documentclass[
aps,pra,
twocolumn
]{revtex4}
\usepackage{graphicx}
\usepackage{subfigure}

\usepackage{amsmath}
\usepackage{amssymb}
\usepackage{amsthm}
\usepackage{braket}
\usepackage{color}
\usepackage{microtype}
\usepackage{tikz}
\usepackage{makecell}
\usepackage{array}
\usepackage{tabularx}
\usepackage{makecell}

\usepackage{hyperref}
\hypersetup{colorlinks,linkcolor={blue},citecolor={red},urlcolor={blue}}

\theoremstyle{plain}

\theoremstyle{plain}

\begin{document}

\title{ Photonic Chips for Universal Quantum Computing: Retrospective and Prospective}

\author{Wei Wang$^{1, 2}$}
\author{Menglong Fang$^{1, 2}$}
\author{Hao Yu$^{1, 2}$}
\author{Sijin Li$^{1, 2}$}
\author{Zan Tang$^{1, 2}$}
\author{Hong Cai$^{1, 2}$}
\author{Lip-Ket Chin$^{1, 2}$}
\author{Peter van Loock$^{3}$}
\author{Daiqin Su$^{1, 2}$}
\email{daiqin.su@polyu.edu.hk}
\author{Ai-Qun Liu$^{1, 2}$}
\email{aiqun.liu@polyu.edu.hk}

\affiliation{$^{1}$Research Institute For Quantum Technology (RIQT)}
\affiliation{$^{2}$Department of Electrical and Electronic Engineering, The Hong Kong Polytechnic University, Hong Kong}
\affiliation{$^{3}$Institute of Physics, Johannes-Gutenberg University of Mainz, Staudingerweg 7, 55128 Mainz, Germany}

\date{\today}

\begin{abstract}
Integrated photonic quantum chips have emerged as a promising platform for realizing fault-tolerant photonic quantum computers. Compared with bulk-optics systems, integrated photonics offer compact device footprints, enhanced programmability, and exceptional stability, attributes that are essential for executing large-scale and computationally demanding quantum tasks. In this review, we provide a comprehensive overview of recent advances in both discrete-variable and continuous-variable universal photonic quantum computation. We survey progress in fault-tolerant architectures and state-of-the-art integrated platforms and highlight the key challenges that must be addressed to scale photonic systems toward full fault tolerance. Finally, we outline future opportunities and research directions aimed at bridging the gap between current proof-of-principle demonstrations and fully scalable quantum microprocessors.
\end{abstract}

\maketitle

\section{Introduction}

By harnessing intrinsically quantum phenomena, most notably superposition and entanglement, quantum computers offer capabilities beyond those of classical processors for certain problem classes, with anticipated impacts in areas including cryptography, optimization, drug discovery and artificial intelligence. Among the physical platforms under active development, including superconducting circuits~\cite{Devoret2013SuperconductingCircuits, Krantz2019Quantum, Arute2019QuantumSupremacy, Acharya2025QuantumErrorCorrection}, trapped ion~\cite{,Cirac1995Quantum, Monroe2013ScalingIonTrap, Bruzewicz2019TrappedIonQC}, neutral atom~\cite{Evered2023HighFidelity, Radnaev2025Universal, Bluvstein2025FaultTolerant}, silicon spin~\cite{Weinstein2023Universal, Edlbauer2025An11Qubit} and topological material~\cite{Kitaev2003Fault, Nayak2008NonAbelian}, quantum photonics has emerged as a compelling route toward scalable and practical quantum computers. Photons exhibit exceptionally weak coupling to their environment, enabling long-lived coherence; they can be generated, routed and manipulated at room temperature with ultrahigh bandwidth; and, as natural carriers of quantum information, they are inherently compatible with networking and modular architectures.

The first scalable model for photonic quantum computing was proposed in 2001~\cite{Knill2001SchemeEfficient}. Thereafter, measurement-based quantum computation (MBQC)~\cite{Raussendorf2001OneWay, Raussendorf2003Measurement}, in which quantum algorithms are implemented through adaptive local measurements on an entangled cluster state, emerged as the leading paradigm for realizing photonic quantum computers~\cite{Nielsen2004Optical, Browne2005Resource, Kieling2007Percolation}. Foundational proof-of-principle experiments, including two-qubit entangling gates~\cite{OBrien2003ControlledNOT}, photonic cluster states~\cite{Walther2005Experimental, Lu2007ExperimentalEntanglement, Yokoyama2013Ultra, Larsen2019Deterministic, Asavanant2019TimeDomain}, and elementary measurement-based gates~\cite{Walther2005Experimental, Asavanant2021TimeDomain, Larsen2021Deterministic}, have been demonstrated using bulk-optics platforms. However, these table-top implementations remain many orders of magnitude smaller than what is required for practical applications; for instance, breaking the 2048-bit Rivest-Shamir-Adleman (RSA) encryption would require the number of qubits at least at the order of $10^4$~\cite{Gidney2025Factor2048LessMillion, Webster2026PinnacleArchitecture, Cain2026Shor10kQuits}. Furthermore, bulk-optics systems inherently suffer from limited programmability, low stability, and high complexity, which severely constrain their usefulness for executing large-scale quantum algorithms.

Photonic quantum chips offer a promising route to overcome the intrinsic limitations of bulk-optics systems in scalability, programmability, and operational stability. Photon sources, either based on nonlinear optical processes or strong light-matter interactions, and high-performance photon detectors are compatible with mature silicon photonics technologies. This compatibility enables the monolithic integration of photon sources, linear optical networks, photon detectors, and even on-chip lasers, thereby reducing coupling and propagation losses, mitigating mechanical and thermal drift, and substantially improving stability and repeatability~\cite{zhou2023prospects, dutt2024nonlinear}. The miniaturization of key components allows large numbers of devices to be fabricated and integrated onto a single chip, greatly enhancing scalability. Close co-integration with electronics provides high-speed modulation, precise phase control, and low-latency feedforward, while advances in wafer-scale processing and packaging technologies address critical considerations of cost, footprint, and manufacturability, factors central to the deployment of photonic quantum computing~\cite{dutt2024nonlinear, wang2020integrated}.

\begin{figure*}[t]
\includegraphics[width=2 \columnwidth]{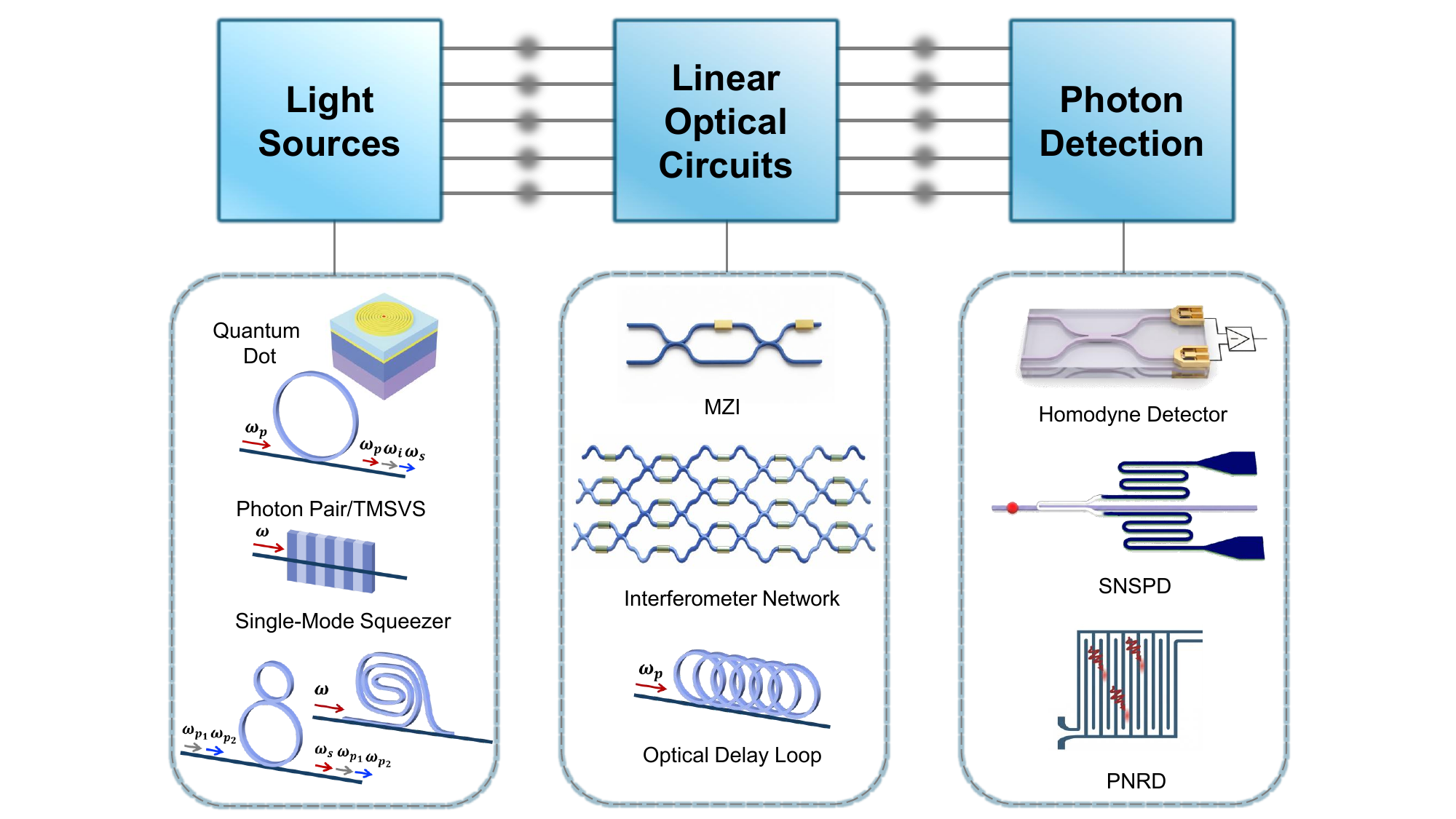}
\caption{ Schematic of the high-level architecture for photonic quantum computation. MZI: Mach-Zehnder Interferometer; SMSVS: two-mode squeezed vacuum state; SNSPD: superconducting nanowire single-photon detector; PNRD: photon-number-resolving detector. }
\label{fig:architecture}
\end{figure*}

In recent years, substantial efforts have been devoted to demonstrating the fundamental building blocks for universal quantum computing on photonic quantum platforms. These advances include on-chip generation of single photons and squeezed vacuum states, the creation of entangled photonic cluster states, the realization of large-scale reconfigurable linear optical networks, and the integration of high-performance photon detectors such as homodyne, single-photon and photon-number-resolving detectors. Several previous reviews have examined the foundational principles underlying universal photonic quantum computation~\cite{Kok2007Linear, OBrien2007Optical, Weedbrook2012Gaussian} or focused on technological progress in photonic quantum technologies~\cite{OBrien2009PhotonicQuantum, wang2025scalable, Zhu2026RecentProgress}. In contrast, this review surveys recent advances in fault-tolerant architectural design and the on-chip implementation of universal photonic quantum computing and quantum microprocessors. We further outline the key theoretical and technological challenges that must be addressed and highlight promising research directions toward the ultimate goal of realizing large-scale, fault-tolerant photonic quantum computers.

In this review, we begin with summarizing a high-level architecture for realizing photonic quantum computing. We then review discrete-variable (DV) photonic quantum computation, examining measurement-based and fusion-based architectures, the generation of single photons and small entangled resource states, and the integration of photon detectors. We next turn to continuous-variable (CV) measurement-based photonic quantum computation, focusing on the computational architectures and the on-chip generation of CV resource states, including CV cluster states and non-Gaussian states such as Gottesman-Kitaev-Preskill (GKP) states. Finally, we highlight key directions for future research that could accelerate progress toward large-scale, fault-tolerant photonic quantum computers.

\section{High-level Architecture}

Several distinct paradigms have been developed for realizing photonic quantum computing, including DV and CV approaches. Although these models differ conceptually, they can be reconciled within a unified high-level hardware architecture, as illustrated in Fig.~\ref{fig:architecture}. This architecture comprises three foundational modules: light sources, linear optical circuits, and photon detection.

The light-source module generates the quantum states of light that encode quantum information and serve as the computational resource. These states include single photons, entangled photon pairs, single-mode squeezed vacuum states, and two-mode entangled states. In practice, such resources are produced using several physical platforms. Nonlinear optical processes, most notably second-order $\chi^{(2)}$ processes such as spontaneous parametric down-conversion (SPDC) and third-order $\chi^{(3)}$ processes such as spontaneous four-wave mixing (SFWM), provide high-quality squeezed states and entangled photon pairs with well-controlled spectral properties. Alternatively, solid-state quantum emitters, such as semiconductor quantum dots (QDs), offer deterministic, on-demand generation of high-purity and highly indistinguishable single photons. 

The linear-optical-circuit module is responsible for coherently manipulating the photons emitted by the light-source module. Its key functions include implementing quantum interference, routing photons across spatial or temporal modes, and providing short- or long-lived storage. These operations rely on a suite of optical components, including phase shifters, Mach-Zehnder interferometers (MZIs), programmable multiport interferometer networks, high-speed optical switches, and tunable delay lines. Collectively, these elements enable the realization of linear unitary transformations. 

The photon-detection module performs measurements required either for implementing measurement-based quantum gates or for reading out the final outcomes of a computation. The required detection technology depends on the chosen information-encoding scheme. Homodyne detectors (HDs) enable high-efficiency quadrature measurements needed for CV protocols; single-photon detectors (SPDs) provide the single-photon sensitivity required for measurements of DV photonic qubits; and photon-number-resolving detectors (PNRDs) enable access to full photon-number statistics, which is essential for non-Gaussian state preparation.

\section{DV photonic quantum computation}

The basic unit of a quantum computer is the qubit, which is physically realized as a two-level quantum system. In DV photonic quantum microprocessors, single photons serve as natural carriers of quantum information, with qubits encoded in their various degrees of freedom. Coherent preparation, manipulation and measurement of multiple single photons enable scalable quantum computation.

DV measurement-based~\cite{Browne2005Resource} and fusion-based quantum computation (FBQC)~\cite{Bartolucci2023fusion} are particularly promising paradigms for realizing large-scale, fault-tolerant quantum computing and are particularly well-suited for photonic platforms. Their central principle is to employ entanglement as a computational resource, and implement quantum gates and error correction through adaptive single-photon measurements or projective entangling measurements. Realizing fault-tolerant DV photonic quantum computation relies on three key pillars:
(a) the generation of essential resource states, including small entangled resource states and large-scale DV cluster states;
(b) the implementation of universal quantum gates through sequences of single-photon and projective entangling measurements on these cluster states; and
(c) the suppression of errors via quantum error-correcting codes with high performance. 

Among the many degrees of freedom available for encoding a qubit in a single photon, including path, polarization, time-bin, orbital angular momentum and others, path encoding and polarization encoding are the two most commonly employed approaches. A path-encoded photonic qubit is physically realized using a single photon propagating through a tunable two-port MZI. To implement arbitrary single-qubit gates, phase shifters are typically required both before and after the central beam splitter, as well as inside the interferometer arms, to fully control the amplitudes and relative phase between the two spatial modes~\cite{Bartlett2020Universal}. A polarization-encoded photonic qubit is implemented using a single photon and wave plates. Arbitrary single-qubit states can be prepared using a single half-wave plate (HWP) and a quarter-wave plate (QWP) with appropriate orientations~\cite{Simon1990Minimal, Knill2001SchemeEfficient}. To realize arbitrary single-qubit gates, which correspond to any SU(2) transformations, the commonly used configuration is a QWP-HWP-QWP sandwich structure, where polarization rotations enable the desired unitary operation. Within the framework of linear optical quantum computation, two-qubit entangling gates cannot be implemented deterministically for either path- or polarization-encoded qubits. This limitation arises because purely linear optical elements cannot induce direct nonlinear interactions between photons. Consequently, any two-photon entangling operation must rely on post-selection, which succeeds only with a low probability. This fundamental constraint once led to serious doubts about whether photonic platforms could support truly scalable quantum computers. 

\begin{figure*}[t]
\includegraphics[width=2.0 \columnwidth]{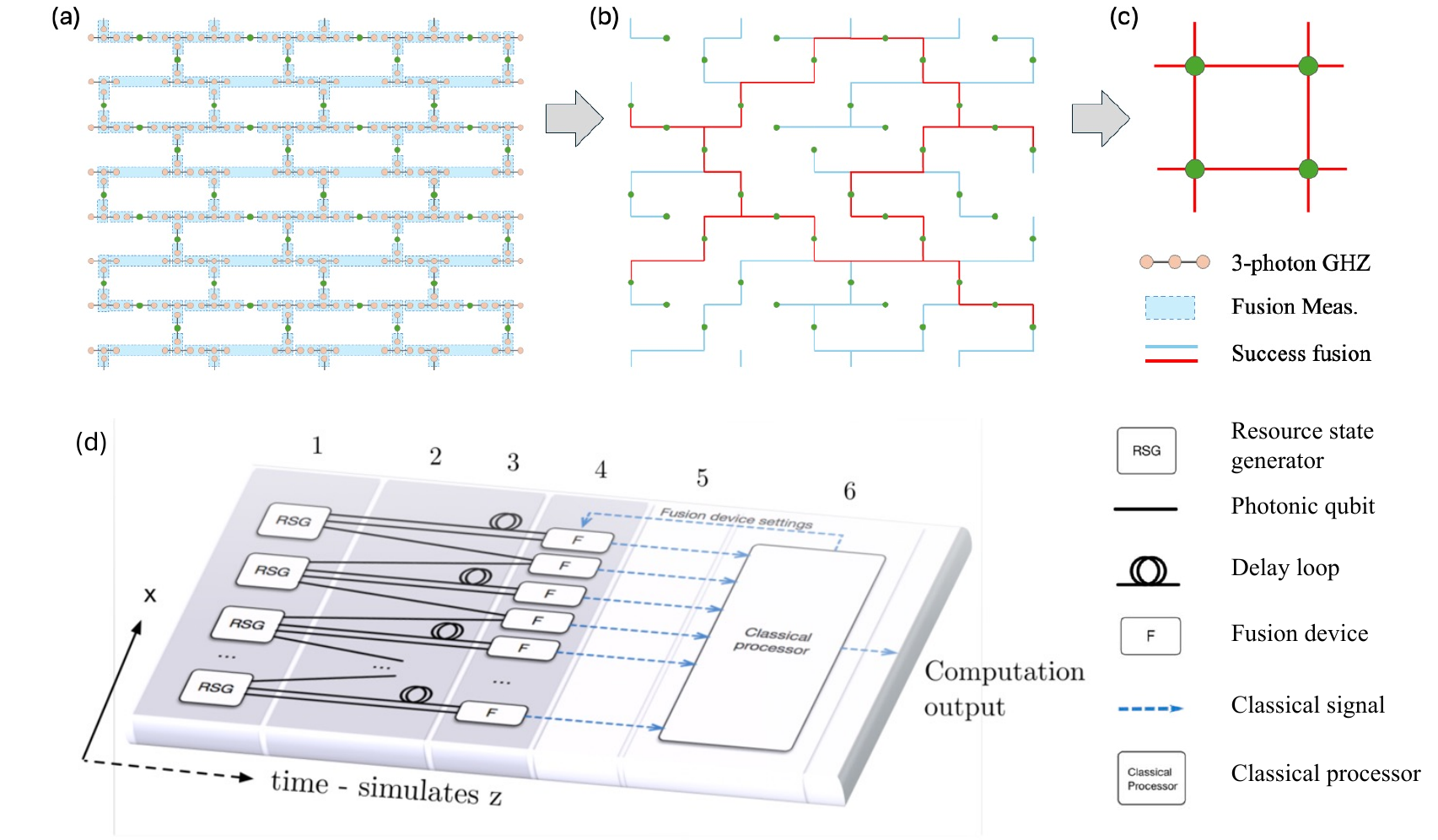}
\caption{ Schematic of ballistic generation of photonic DV cluster state~\cite{Pant2019percolation} and FBQC architecture~\cite{Bartolucci2023fusion}: (a) The connected three circles represent three photonic qubits prepared in the 3-photon GHZ state, each blue-shaded box represents a fusion measurement on the two qubits inside the box, the orange circles represent photonic qubits that are subject to projective measurement, and the green circles denote qubits that are not measured. (b) If the fusion measurement succeeds then a link is established at the corresponding location, while if the fusion measurement fails then no link is added. A selected configuration that is useful for measurement-based quantum computation is highlighted in red. (c) Renormalized photonic DV cluster state. (d) Layout for FBQC architecture~\cite{Bartolucci2023fusion}.  }
\label{fig:cluster_percolation}
\end{figure*}

In their seminal work~\cite{Knill2001SchemeEfficient}, Knill, Laflamme and Milburn (KLM) demonstrated the viability of overcoming this challenge. They showed that, even though two-qubit gates are inherently non-deterministic, linear optical quantum computation remains scalable in principle by employing ancillary photons and techniques such as quantum teleportation. However, the resource overhead required to boost the success probability of two-qubit gates to a practical level is extremely large. In the original KLM scheme, high-fidelity entangling gates typically demand a substantial number of ancillary photons and optical modes, which was widely regarded as a major bottleneck for realizing large-scale quantum computation in the early stages of the field. Shifting from the gate-based approach of KLM to MBQC can significantly reduce the resource requirements. By avoiding the cumulative resource overhead associated with complex inline gate operations, MBQC offers a highly promising route toward the construction of large-scale photonic quantum computers~\cite{Nielsen2004Optical, Browne2005Resource, Kieling2007Percolation, Gimeno2015From}

The fundamental substrate for DV MBQC is a large-scale cluster state, which is a multi-qubit entangled state~\cite{Raussendorf2003Measurement}. The preparation of cluster states requires a special type of two-qubit entangling gates, which is known as the controlled-phase (CZ) gate. Due to the non-deterministic nature of the two-qubit entangling gates, it is not efficient to prepare cluster states by directly applying CZ gates to a set of qubits. Instead, a set of small entangled resource states, e.g., 3-photon Greenberg-Horne-Zeilinger (GHZ) state, are first prepared and then are fused to form a cluster state using fusion measurements~\cite{Gimeno2015From}. The fusion measurement projects two qubits onto entangled basis states, e.g., the Bell states, and is implemented by linear optical components and single-photon detection. The implementation of the fusion measurement is also non-deterministic, e.g., the success probability for the type-II fusion measurement is only 50\%~\cite{Kieling2007Percolation}. However, the success probability can be boosted by introducing ancillary single photons or entangled photon pairs, e.g., the success probability can be boosted to 25/32 ($\approx 78.1\%$) by adding eight ancillary single photons~\cite{Ewert2014Efficient}. 

A natural question arises as to whether the probabilistic fusion measurements can convert a set of small entangled resource states into a cluster state that is universal for MBQC. The  answer is affirmative. This can be nicely interpreted using the percolation theory, which acts as a preliminary theory for phase transition in graphs and has been extensively studied in statistical mechanics. When preparing cluster states with small entangled resource states and fusion measurements, the unmeasured qubits are considered as the vertices of a graph and the application of fusion measurements establishes links (CZ gates) between qubits. According to the percolation theory, there exists a threshold, known as the percolation threshold, for the success probability of the fusion measurement, above which a connected graph is formed. This connected graph can be renormalized to represent a one-dimensional cluster state, which enables propagation of quantum information through sequential quantum gate teleportation. An example of generating a two-dimensional DV cluster state through 3-photon GHZ states and fusion measurements are illustrated in Fig.~\ref{fig:cluster_percolation}(a)-(c). 

It was found by numerical simulation that a percolation threshold $p_{\rm th} \approx 0.625$ is possible by constructing a three-dimensional rectangular lattice~\cite{Gimeno2015From}. The success probability of the boosted type-II fusion measurement with four ancillary single photons or an entangled photon pair can reach 75\%~\cite{Ewert2014Efficient}, significantly higher than the percolation threshold. This implies that a renormalized universal cluster state can be constructed by utilizing a set of small entangled resource states and probabilistic fusion measurements. By further reusing the entangled resource states when the fusion fails, the percolation threshold can be reduced to $1/(n-1)$, where $n$ is the size of the small entangled resource states~\cite{Pant2019percolation}. The above schemes for generating cluster states require no active feedforward, representing a ballistic way of state generation. Most importantly, the resource overhead for ballistic generation of cluster state and subsequent measurement-based quantum computing is substantially lower than that of the KLM scheme. 

By incorporating the adaptive single-qubit measurements used to implement measurement-based gates into the fusion measurements, the resource overhead can be further reduced, forming a different computational paradigm called FBQC~\cite{Bartolucci2023fusion}. The FBQC utilizes small constant-sized entangled resource states, e.g., the four-ring or six-ring resource states, and projective entangling measurements such as type-II fusion measurements. Given the set of small entangled resource states, the application of a fusion network determines the implementation of specific quantum algorithms, and appropriate combinations of the fusion measurement outcomes give the outputs of the computation. 
The architecture for a DV FBQC consists of two main modules: the resource state generation (RSG) module and the measurement network module, as illustrated in Fig.~\ref{fig:cluster_percolation}(d). The RSG prepares single photons as photonic qubits and as ancilla for boosting the fusion probability and generates small constant-sized entangled resource states. The measurement network implements quantum algorithms and quantum error correction through fusion operations and/or adaptive single-qubit measurements. 
The computation proceeds in the time domain. The RSG sequentially generates single photons and small entangled resource states, while the measurement network constructs temporal cluster state and then measures it to implement gates and error correction. A classical processor processes the measurement outcomes and determines the measurement settings of next round of fusion and/or single-qubit measurements.

The dominant noise mechanisms in DV photonic architectures, most notably photon loss, measurement imperfections, and the probabilistic failure of fusion measurements, can be suppressed by employing three-dimensional (3D) topological cluster states, thereby enabling fault tolerance. In the 3D Raussendorf-Harrington-Goyal (RHG) lattice, qubit-loss rates as high as 24.9\% can be tolerated, whereas the threshold for measurement-error rates is significantly lower, at 0.63\%~\cite{Barrett2010Fault}. To compensate for the intrinsically non-deterministic nature of two-qubit entangling gates, additional redundancy must be incorporated to boost the effective success probability. This comes at the cost of substantial resource overhead, requiring large numbers of single photons, MZIs, optical switches, and single photon detectors. For such switch-based architectures, the per-component photon-loss and noise thresholds are approximately $10^{-3}$ and $10^{-5}$, respectively, and constructing a single qubit within the 3D RHG lattice requires at the order of $10^{5}$ detectors and comparable numbers of other components~\cite{Li2015Resource}, posing severe challenges for experimental realization. A complementary strategy avoids optical switches by directly assembling a faulty 3D RHG lattice using probabilistic fusion operations. This approach improves resource scaling and remains fault tolerant, with thresholds of around 2.9\% for measurement errors and 14.5\% for fusion failure, corresponding to a fusion success rate of 95\%~\cite{Herr2018LocalScalable, Auger2018Fault}. Recent advances within the FBQC paradigm further enhance these limits, achieving tolerance to 43.2\% fusion failure and photon-loss rates up to 2.7\% per photon~\cite{Bartolucci2023fusion}. These developments collectively map out the performance landscape for large-scale, fault-tolerant DV photonic quantum computation and clarify the architectural trade-offs that future implementations must navigate.

Existing architectures for DV photonic quantum computers are largely designed to suppress various types of errors using topological quantum error-correcting codes, notably the surface code. Recent progress in developing quantum low-density-parity-check (LDPC) codes~\cite{Breuckman2021, panteleev2022asymptotically, leverrier2022quantum, bravyi2024high, xu2024constant} with higher encoding rates than the surface code opens a promising pathway toward new, more resource-efficient architectures~\cite{chen2026fusion}. Implementing quantum LDPC codes typically requires long-range interactions, a feature naturally compatible with flying photonic qubits. Leveraging high-rate quantum LDPC codes could dramatically reduce the resource overhead required for achieving fault tolerance. Advancing this direction will require sustained effort in architecture design, numerical modeling, decoding algorithm development, and comprehensive resource estimation.

\section{CV Photonic Quantum Computation}

The probabilistic nature of entangling gates in DV photonic quantum computing results in substantial resource overhead, making it the primary obstacle to achieving large-scale quantum computation~\cite{Knill2001SchemeEfficient}. By encoding quantum information in the CV degrees of freedom of the quantum light field~\cite{Lloyd1999Quantum, Braunstein2005Quantum, Weedbrook2012Gaussian}, entangling gates can be implemented deterministically, enabling a deterministic generation of large-scale CV cluster states~\cite{Zhang2006continuous-variable, Menicucci2006universal}. The fundamental principle of CV MBQC is to employ entanglement as a computational resource, and implement quantum gates and error correction through adaptive homodyne measurements~\cite{Menicucci2014fault-tolerant}. 
Three key pillars for realizing fault-tolerant CV MBQC are: (a) the generation of essential resource states, including CV cluster states and non-Gaussian states; (b) the execution of homodyne measurements on the CV cluster states to implement a universal set of gates and arbitrary quantum circuits; and (c) the suppression of environmental noise through error correction, including the bosonic error correction and the concatenation of bosonic codes with high-performance qubit error correcting codes. 
The CV cluster state, also referred to as a qumode cluster state, is a multimode Gaussian entangled state that forms the fundamental substrate for CV MBQC~\cite{Zhang2006continuous-variable, Menicucci2006universal}. In analogy to the qubit cluster state, the ideal CV cluster state is constructed by preparing each mode in a momentum eigenstate followed by the application of CV controlled-$Z$ gates across appropriate pairs of modes, which are determined by a specific graph structure. The ideal CV cluster state, however, requires unbounded energy and is therefore physically unattainable. In practical implementations, CV cluster states are generated by initializing each mode in a finitely squeezed vacuum state and interfering them through networks of beam splitters~\cite{vanLoock2007building, Gu2009quantum}. Finite squeezing inevitably introduces deviations from the ideal state, leading to errors in subsequent computations. The degree of approximation is typically quantified through the variances of the nullifiers, which are linear combinations of quadrature operators. The smaller of the variances of the nullifiers, the better the approximation to the ideal CV cluster state. Squeezing of all nullifiers by at least 3 dB relative to shot noise is required to ensure full inseparability and genuine multipartite entanglement~\cite{Loock2003Detecting}. In contrast to DV cluster states, whose probabilistic generation imposes significant scaling constraints, CV cluster states can be prepared deterministically, offering a distinct advantage for scaling toward large-scale resource states.   

\begin{figure}[t]
\includegraphics[width=0.98 \columnwidth]{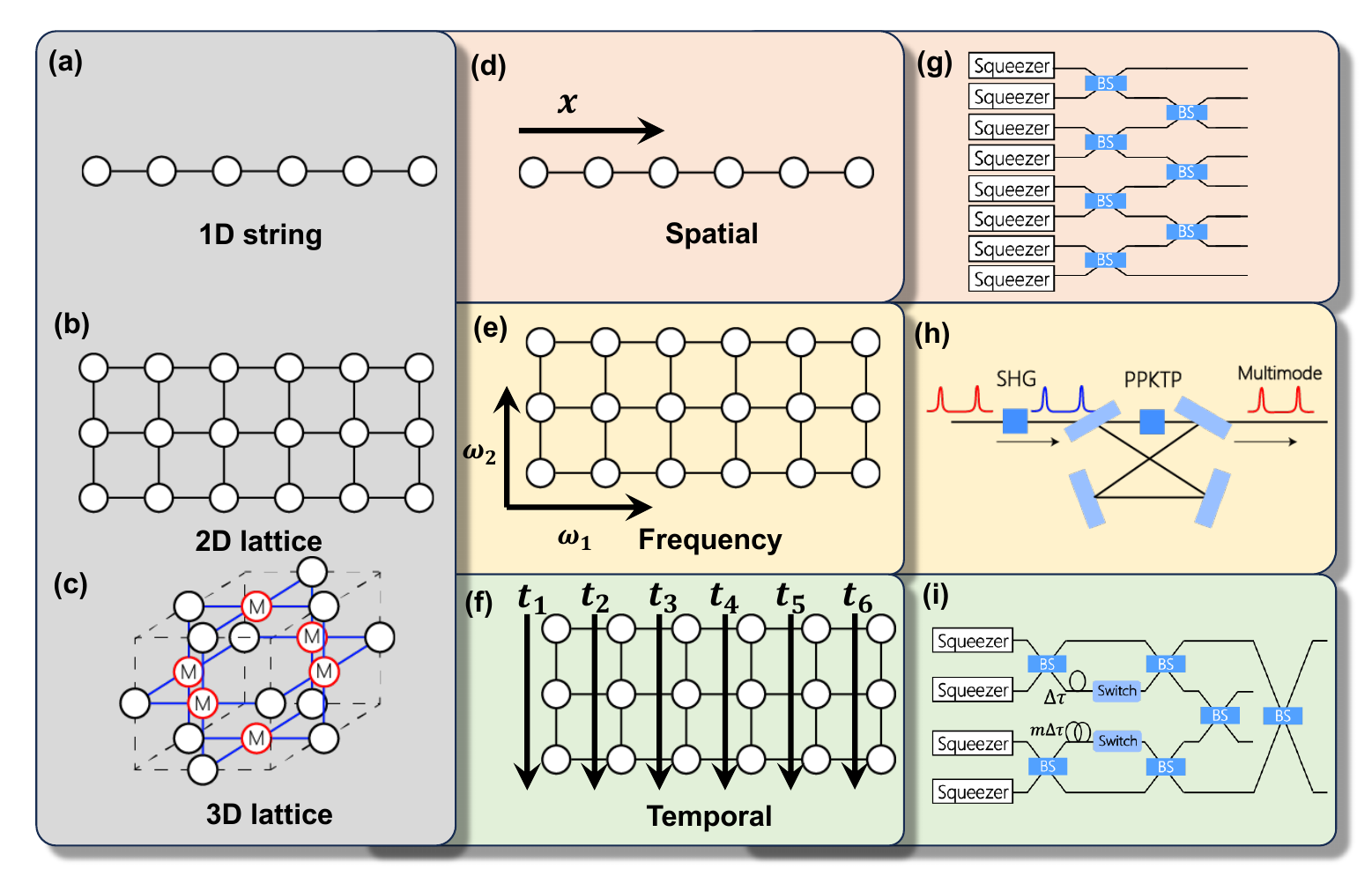}
\caption{ Schemes for generating spatial-, frequency- and temporal-domain CV cluster states. (a)-(c) CV cluster states with one-, two- and three-dimensional lattice structures. (d)-(f) CV cluster states encoded in the spatial, frequency and time degrees of freedom. (g)-(i) Experimental schemes for generating CV cluster states in the spatial, frequency and time domains. Each white circle represents a macronode that consists of either two quantum modes (one-dimensional cluster) or four quantum modes (two- and three-dimensional cluster). }
\label{fig:CVcluster}
\end{figure}

\begin{figure}[t]
\includegraphics[width=1.0 \columnwidth]{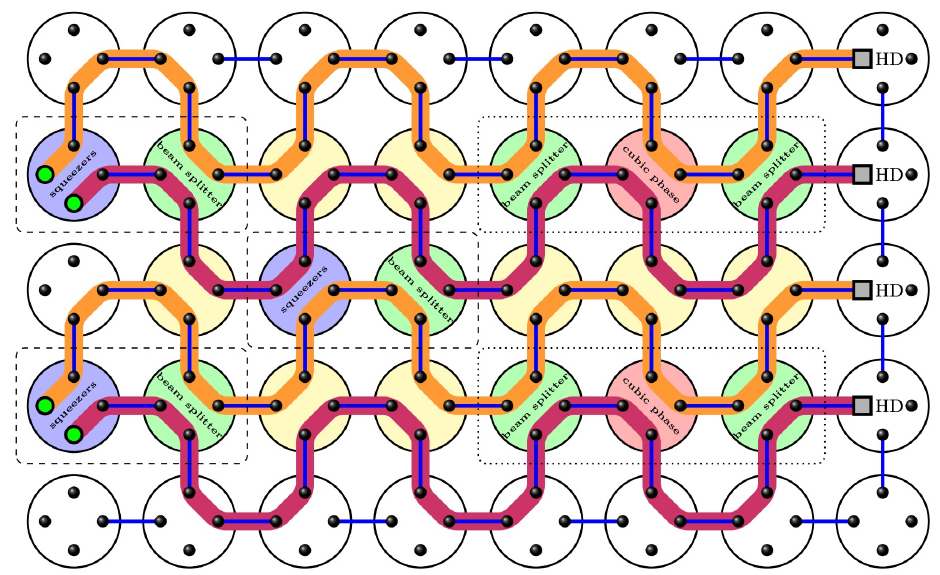}
\caption{ Example of measurement-based four-mode CV-IQP circuit~\cite{Su2018Implementing}, which can be implemented by the setup in Fig.~\ref{fig:CVcluster}(i). The blue shaded macronode implements two identical single-mode squeezers, the green shaded macronode implements a beam splitter, the pink shaded macronode implements a cubic phase gate and an identity gate, and the yellow shaded macronode implements two phase shifts which is not explicitly labelled. In the first column of macronodes, the four input modes (green circles) are prepared in the momentum squeezed states. In the first four columns, the two macronodes in a black dashed box together implements a CZ gate. From the fifth to the seventh column, the three macronodes in a black dotted box together implements a cubic phase gate and an identity gate. In the last column, the four output modes (grey squares) are detected by homodyne detectors. }
\label{fig:cluster_algorithm}
\end{figure}

\begin{figure*}[t]
\includegraphics[width=2.0 \columnwidth]{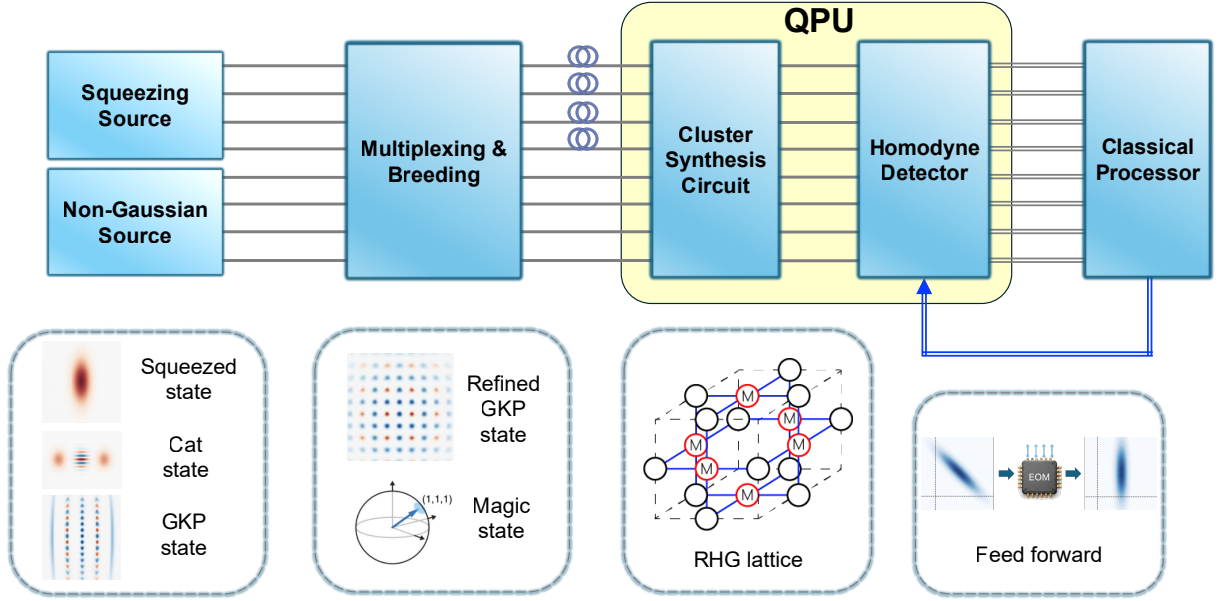}
\caption{ Schematic of CV fault-tolerant architecture. Squeezing and non-Gaussian light sources generate squeezed vacuum states, cat states, and GKP states. Multiplexing and breeding circuits breed the cat and GKP states to synthesize high-quality GKP states and magic states, and produce entangled GKP pairs. Cluster synthesis circuit generates high-dimensional CV cluster states such as RHG lattice and adaptive homodyne detection implements quantum algorithms and quantum error correction. Classical processor processes the measurement outcomes and decodes errors, and feed forward to the next cycle of measurements. }
\label{fig:cv_architecture}
\end{figure*}

On photonic platforms, a quantum mode can be encoded in various degrees of freedom of the quantum electromagnetic field, enabling CV cluster states across spatial, temporal, frequency, or hybrid domains, as illustrated in Fig.~\ref{fig:CVcluster}. Spatial cluster states provide simultaneous access to all modes and allow parallel measurements, significantly reducing circuit depth for certain quantum algorithms~\cite{Armstrong2012Programmable, Pooser2014Continuous}. However, the need for a large number of physical resources, e.g., squeezed light sources and beam splitters, poses significant experimental challenges. Temporal cluster states, enabled by time-domain multiplexing, reuse the same physical hardware to generate effectively unlimited numbers of modes but require long, low-loss delay lines~\cite{Yokoyama2013Ultra, Larsen2019Deterministic, Asavanant2019TimeDomain}. Frequency-domain cluster states exploit the structure of optical frequency combs, making it possible to generate high-dimensional entanglement from a single optical parametric oscillator (OPO)~\cite{Pysher2011Parallel, Chen2014experimental}. Their compatibility with flexible time-frequency mode selection in homodyne detection further enables engineering of cluster states with variable structures~\cite{Roslund2014Wavelength}. Hybrid cluster states that combine multiple degrees of freedom offer a promising route to complex, high-dimensional topologies, the kind required for implementing quantum error correction.

Non-Gaussian resources play a central role in extending Gaussian architectures to universal and fault-tolerant quantum computation~\cite{Lloyd1999Quantum}. Gaussian computations composed solely of Gaussian input states, Gaussian operations, and Gaussian measurements are efficiently classically simulable~\cite{Mari2012Positive} and thus cannot provide quantum advantage. Universality requires the incorporation of at least one non-Gaussian element, such as a non-Gaussian ancillary state, a non-Gaussian gate, or a non-Gaussian measurement. The need for non-Gaussianity becomes even more acute for error correction. Environmental noise inevitably degrades quantum coherence and compromises the target computational outcomes, thereby necessitating a proper design of CV quantum error-correcting codes. Realistic noise processes in CV quantum systems, including photon loss and random displacements, are Gaussian, and a no-go theorem states that purely Gaussian encodings cannot correct Gaussian noise~\cite{Niset2009No-go}. Consequently, effective CV quantum error-correcting codes must rely on non-Gaussian encoded states.

The fundamental primitive of CV MBQC is quantum teleportation, and in particular, the quantum gate teleportation~\cite{Walshe2020continuous}. Gaussian gates are implemented via homodyne measurements and classical feedforward on the CV cluster state. Non-Gaussian gates can be realized by injecting their corresponding non-Gaussian states into the cluster state and performing appropriate homodyne measurements, completing a universal set of gates. A two-dimensional (2D) CV cluster state, augmented with suitable non-Gaussian resources, suffices for universal CV MBQC. A specific quantum algorithm is implemented by first compiling a quantum circuit composed of a sequence of single- and two-mode gates, and then proceeding as follows: (1) select the computational modes within the cluster; (2) determine the measurement bases and patterns corresponding to a layer of gates; (3) perform homodyne detection and record the measurement outcomes; (4) apply feedforward to update the computational modes; and (5) iterate this process layer by layer until the entire quantum circuit has been executed. An example of implementing a CV instantaneous quantum polynomial (IQP) computation is shown in Fig.~\ref{fig:cluster_algorithm}~\cite{Su2018Implementing}. 

Non-ideal CV cluster states introduce excess noise to both the phase and momentum quadratures during the quantum gate teleportation, thereby distorting the encoded quantum information~\cite{Menicucci2006universal, Gu2009quantum} and fundamentally limiting scalability~\cite{Ohliger2010limitations}. Photon loss further reduces state energy and injects quadrature noise, degrading entanglement and gate fidelity. Quantum error correction is therefore indispensable for large-scale CV MBQC. Bosonic error-correcting codes, which encode quantum information into infinite-dimensional redundant Hilbert spaces~\cite{Cochrane1999Macroscopically, Ralph2003Quantum, Leghtas2013Hardware, Mirrahimi2014Danamically, Mechael2016New, Grimsmo2020Quantum}, are designed to suppress errors in CV quantum systems. Among these, the GKP code uniquely integrates with CV cluster states and is capable of correcting both random displacement noise originated from finite squeezing and photon loss. Fault tolerance can be achieved by concatenating the GKP code with qubit quantum error-correcting codes, provided that both the cluster-state squeezing and the GKP-state squeezing exceed approximately 20 dB~\cite{Menicucci2014fault-tolerant}. Using 3D cluster states based on the RHG lattice~\cite{Raussendorf2007faul-tolerant, Raussendorf2007topological} and topological protection, the squeezing requirement can be reduced between 10 dB and 13 dB~\cite{Fukui2017analog, Fukui2018highthreshold, Bourassa2021blueprintscalable, Tzitrin2021fault-tolerant, Larsen2021Fault-tolerant, Renault2025End}, a regime accessible to state-of-the-art photonic technologies. More recently, concatenation of the GKP code with quantum LDPC codes has emerged as a promising scheme to further reduce resource overhead~\cite{Walshe2025linear, Ide2025Fault-tolerant}.

Several fault-tolerant architectures for CV MBQC have been proposed~\cite{Bourassa2021blueprintscalable, Larsen2021Fault-tolerant, Tzitrin2021fault-tolerant, Renault2025End}, and their layout can be schematically illustrated in Fig.~\ref{fig:cv_architecture}. The squeezed-light sources generate single-mode squeezed vacuum states and the dedicated non-Gaussian sources probabilistically produce non-Gaussian states, such as cat states and primitive GKP states, via photon-number-resolving detection on suitably engineered multimode Gaussian states, followed by post selection. These resource states are subsequently processed in a multiplexing/breeding module, where multiple copies are bred into high-quality GKP states and magic states. This module also boosts the GKP generation rate through multiplexing and entangles the refined GKP states with squeezed-vacuum modes to generate EPR pairs. The quantum-processing unit (QPU) then assembles these ingredients into fault-tolerant CV cluster states, such as hybrid RHG lattices, and executes homodyne measurements according to the prescribed measurement pattern. The resulting outcomes are processed by classical controllers, which compute and apply the necessary feedforward operations. Through this continuous quantum-classical loop, the QPU implements measurement-based quantum gates and performs real-time quantum error correction. A key architectural consideration is the trade-off between the squeezing threshold and the success probability of GKP generation: lower GKP-generation success probabilities require higher squeezing levels to maintain fault tolerance~\cite{Bourassa2021blueprintscalable, Tzitrin2021fault-tolerant}. The adaptive breeding protocol~\cite{Weigand2018Generating}, which combines multiple cat states to synthesize high-fidelity GKP states, can substantially boost the success probability~\cite{Renault2025End}. As a result, the stringent squeezing requirements can be significantly relaxed, easing one of the central bottlenecks in scalable CV fault-tolerant architectures.

\section{Quantum Light Sources}

Photonic quantum computing relies on a broad spectrum of optical resource states, including single photons, entangled photons, single-mode squeezed vacuum states, entangled CV cluster states, and non-Gaussian states. These states serve distinct yet complementary functions within a quantum microprocessor: single photons and non-Gaussian states (e.g., GKP states) act as information carriers, whereas entangled photons and CV cluster states function as computational resources enabling universal quantum computation. This section reviews schemes for generating these resource states, their leading on-chip implementations, key performance metrics achieved to date and discusses the primary technical challenges associated with each resource class, highlighting the open problems that must be addressed.

\subsection{Single Photon Sources}

Single-photon sources aim to provide quantum states with high purity, high indistinguishability and near-deterministic emission. 
Purity reflects how accurately a source emits single photons without multiphoton contamination, indistinguishability is required for high-fidelity two-photon gates, whereas high efficiency directly impacts the attainable computational speed. Existing single-photon sources are broadly classified as probabilistic heralded sources and deterministic emitters. Probabilistic sources exploit nonlinear optical processes and heralded photon detection, while deterministic emitters leverage strong light-matter coupling to generate photons on demand.

Through SPDC or SFWM, EPR pairs can be generated by pumping a nonlinear optical medium with a laser. Detecting a single photon in the idler mode heralds the presence of a single photon in the signal mode. However, because the pair generation is inherently nondeterministic, heralded single-photon sources are intrinsically probabilistic. Nevertheless, this technique is currently the most extensively investigated and technologically mature route to single-photon generation. Crucially, the nonlinear optical processes are compatible with integrated photonic platforms, providing a promising pathway toward mass-manufacturable sources.

Although the heralded single-photon sources have demonstrated strong potential for achieving high purity and high indistinguishability, the key remaining challenge is to boost the generation efficiency toward unity. This can be potentially achieved by exploiting multiplexing techniques~\cite{Pittman2002Single, Migdall2002Tailoring}, including time, spatial and frequency multiplexing. Implementing these schemes typically requires multiple heralded single-photon sources, together with high-performance optical switches, adjustable delay lines, and/or conversion lasers. The achievable multiplexed heralding probability is $P_H = 1 - (1 - p)^N$, where $N$ is the number of multiplexed modes and $p$ is the probability of a trigger detector of the heralded single-photon source. An efficiency enhancement by a factor of 9.7 has been demonstrated using active time multiplexing of 40 heralded single photon sources~\cite{kaneda2019high}. 
Rapid advancement of integrated photonics enables co-integration of multiple heralded single-photon sources, high-performance SNSPDs and high-speed optical switches based on electro-optic modulation, and makes the implementation of on-chip multiplexing protocols increasingly feasible. 

The quantum dot embedded in a microcavity is a promising quantum emitter that can, in principle, generate single photons on demand~\cite{Lodahl2015Interfacing, senellart2017high}. Significant progress has been made during the past few years, with steady improvement in purity, indistinguishability and generation efficiency~\cite{He2013OnDemand, Wang2019Towards, Najer2019GatedQD, Tomm2021Bright}. The currently best quantum-dot single-photon source simultaneously achieves high purity of 0.98, indistinguishability of 0.985 and the overall system efficiency of 0.712~\cite{Ding2025HighEfficiency}. However, the high indistinguishability is achieved by performing the Hong-Ou-Mandel interference using single photons from a single quantum dot. Whilst photonic quantum computing requires distinguishable single photons from many separate single-photon sources. An indistinguishability of only 0.93 is attainable by interfering single photons from two completely separate quantum dots~\cite{zhai2022quantum}. Hybrid integrated photonic circuits provide a promising route to enhance the indistinguishability of single photons from separate sources. A lithium niobate photonic quantum platform that integrates arrays of quantum-dot-containing waveguides with 20 deterministic single-photon sources has been demonstrated, with capability of spectral tuning of individual quantum dot and achieving interference visibility of 0.73 from spatially separated quantum dots~\cite{Wang2025LargeScaleQD}.

\subsection{Entangled Photons}

There are two main schemes for generating entangled resource states: probabilistic and deterministic. The former utilizes probabilistic sources of entangled photon pairs, passive linear optical elements and post selection. Substantial experimental efforts have been devoted to increasing the size and efficiency, with the state of the art of generating 12-photon entangled states at a rate of one per hour~\cite{Zhong2018Photon}. The exponential decrease of success probability in the number of photons renders this sort of probabilistic protocols extremely hard to scale up. By introducing active feedforward and multiplexing, the scaling of the efficiency is improved and an enhancement of the efficiency for a factor of 35 has been experimentally demonstrated~\cite{Meyer-Scott2022Scalable}. However, the generation rate is still at the order of one per minute, far from the requirement of building large-scale cluster states. 

The deterministic scheme utilizes strong light-matter interaction in a cavity. Multi-photon entangled states can be generated deterministically through sequentially coupling a quantum dot to initially uncorrelated photonic qubits. This protocol was generalized to generate one-dimensional cluster states, which are particularly relevant to MBQC~\cite{Lindner2009Proposal}. Utilizing the interaction between two quantum dots, either through a conditional phase gate~\cite{Economou2010Optically} or the exchange interaction, two-dimensional cluster states can be generated. Based on this scheme, one-dimensional cluster states containing up to 10 photons have been experimentally demonstrated from quantum dots~\cite{schwartz2016deterministic, coste2023high, cogan2023deterministic} with generation rates up to GHz. 
Atom-cavity platforms have also demonstrated deterministic cluster-state generation, achieving cluster sizes of up to 14 photons, albeit at substantially lower generation rates of approximately one per three minutes~\cite{thomas2022efficient}. These deterministically generated photonic cluster states can serve as elementary resource units that are subsequently fused to construct larger and more complex cluster states~\cite{thomas2024fusion}. The main limitations are the low collection efficiency of photons emitted from the atomic system and the low indistinguishability of photons from different quantum emitters. Whether integrating deterministic emitters, such as quantum dots or atom-microcavity systems, directly onto photonic chips can substantially improve the generation efficiency of entangled photons remains an open question, necessitating extensive theoretical analysis and systematic experimental validation.

\subsection{Single- and Two-mode Squeezed Vacuum states}

\begin{figure}[h]
\includegraphics[width=0.95 \columnwidth]{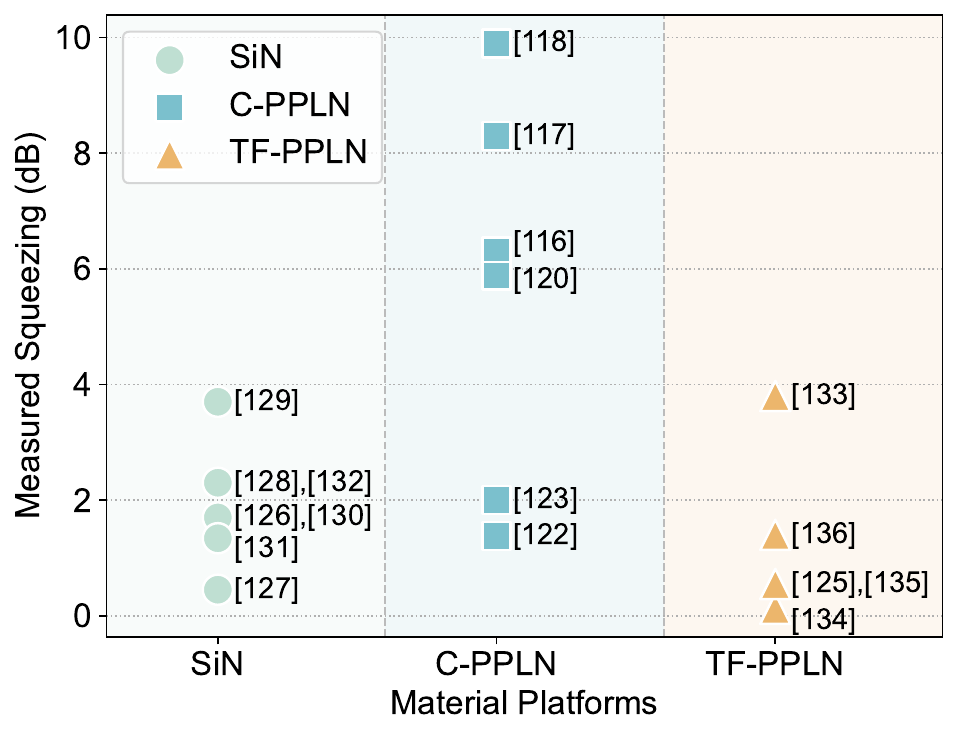}
\caption{ Squeezing levels of single-mode squeezed states generated on different integrated photonic platforms: silicon nitride (SiN)\cite{Dutt2015On-chip,cernansky2020nanophotonic,alfredo2024quantum,shen2025strong,zhang2021squeezed,zhao2020near-degenerate,ulanov2025quadrature}, conventional periodically poled lithium niobate (C-PPLN)\cite{Lenzini2018integrated,Mondain2019Lithium,Kashiwazaki2021LowLossPPLN,Kashiwazaki2023Over8dB,Hirota2025Generation10dB,Amari2023HighlyEfficient}, and thin-film periodically poled lithium niobate (TF-PPLN)\cite{Nehra2022Few,Stokowski2023Integrated,taewon2024singlemode,chen2022ultra,shi2025squeezed}.}
\label{fig:squeezing_platform}
\end{figure}

Preparing high-purity single-mode squeezed-vacuum states and EPR pairs constitutes the first step towards on-chip CV MBQC. Squeezed states should be generated with high squeezing level, broad bandwidth, low noise and well-controlled optical phases, and should be compatible with subsequent on-chip interference, multiplexing and homodyne detection. Generating squeezed states has been demonstrated in several integrated and waveguide platforms, including periodically poled lithium niobate (PPLN), silicon nitride and silica, as summarized in Fig.~\ref{fig:squeezing_platform}. 

Lithium niobate is one of the promising platforms for its large second-order optical nonlinearity, low propagation loss, strong electro-optic response and mature quasi-phase-matching technology. In particular, PPLN waveguides can realize efficient parametric down-conversion or optical parametric amplification with relatively low pump power, while thin-film lithium niobate (TFLN) further enables compact device footprints and monolithic integration with modulators, filters and interferometers. 
Although many high-performance demonstrations have been carried out in conventional PPLN (C-PPLN) waveguides rather than fully monolithic circuits, they establish an important benchmark for integrated lithium-niobate squeezed-light sources. 
Broadband continuous-wave squeezed-vacuum states with squeezing up to 10 dB have been generated in C-PPLN waveguides~\cite{Serkland1995Squeezing, Yoshino2007Generation, kashiwazaki2020continuous, Kashiwazaki2021LowLossPPLN, Kashiwazaki2023Over8dB, Hirota2025Generation10dB}, showing that quasi-phase-matched $\chi^{(2)}$ interactions can provide both high nonlinear efficiency and low-noise operation. Pulsed squeezed-vacuum states with squeezing up to 5.88 dB have also been demonstrated~\cite{Eto2011Efficient, Amari2023HighlyEfficient}, which is particularly relevant for temporal-mode multiplexing and pulsed CV quantum information protocols. Towards fully integrated implementations, on-chip squeezing has been explored in thin-film PPLN (TF-PPLN) waveguides~\cite{Kanter2002squeezing, Lenzini2018integrated, Mondain2019Lithium, Stefsky2017Waveguide}  and, more recently, in TFLN microresonators~\cite{taewon2024singlemode}. Waveguide-based devices are attractive for broadband traveling-wave squeezing and direct compatibility with time-domain multiplexing, whereas microresonators enhance the nonlinear interaction through a high-quality-factor cavity and can substantially reduce the required pump power. 
Remaining challenges include reducing propagation and coupling losses, suppressing photorefractive and pump-induced noise, improving poling quality in nanophotonic geometries and integrating low-loss pump rejection. Continued progress in TFLN fabrication and heterogeneous packaging may therefore turn lithium niobate from a high-performance off-chip squeezed-light platform into a compact source layer for on-chip quantum source generation and universal CV MBQC.

Another attractive platform is the silicon nitride, which has high third-order optical nonlinearity and low linear propagation loss and also lacks of two-photon absorption. More importantly, it is CMOS compatible and has a high degree of confinement, enabling large-scale integration on a monolithic chip. Typically, a high-quality factor microresonator is introduced to enhance the interaction strength and the resultant squeezing level. The intensity difference squeezing with 1.7 dB was generated in a silicon nitride microresonator integrated on chip through four-wave-mixing (FWM)~\cite{Dutt2015On-chip}. However, the quadrature squeezing in a single temporal mode is necessary in quantum sampling and other quantum information processing protocols. Through strongly driven spontaneous FWM below threshold in silicon nitride microresonator, two-mode quadrature squeezing in frequency domain was demonstrated~\cite{vaidya2020broadband}, which was subsequently used to demonstrate quantum sampling and related quantum applications~\cite{Arrazola2021QuantumCircuits}. The degenerate single-mode quadrature squeezing at one resonant frequency of the microcavity can be generated though FWM by using dual pump fields~\cite{Vernon2019Scalable}, and was experimentally demonstrated in~\cite{zhao2020near-degenerate}. However, the introduction of dual pump fields induces unwanted processes, the single-pump spontaneous FWM for each pump and Bragg-scattering FWM, thereby limiting the level of single-mode squeezing. These unwanted processes can be significantly suppressed by introducing an auxiliary microresonator that hybridize only the modes involved in the unwanted processes, enabling a generation of 1.65 dB of directly measured and 8 dB of inferred broadband single-mode squeezing on-chip~\cite{zhang2021squeezed}.


\begin{figure}[h]
\includegraphics[width=0.95 \columnwidth]{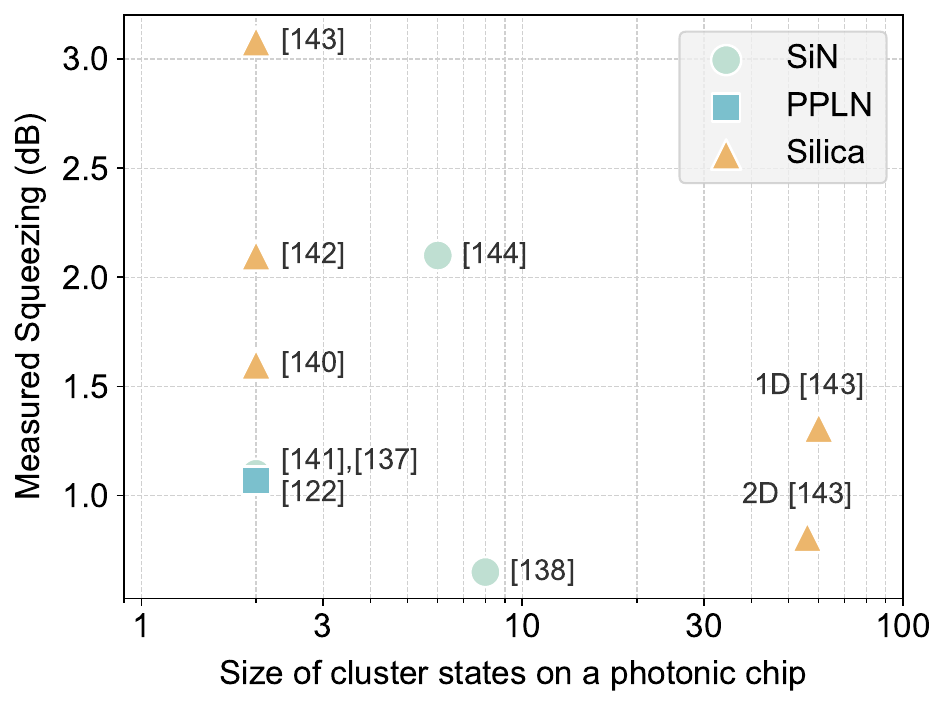}
\caption{ Squeezing levels of cluster states and their sizes\cite{Arrazola2021QuantumCircuits,yang2021squeezed,Jahanbozorgi2023generation,vaidya2020broadband,yang2022quantum,Wang2025LargeScale,Jia2025ContinuousVariable,Lenzini2018integrated}.}
\label{fig:squeezing_cluster}
\end{figure}

Extending from single- and two-mode squeezed vacuum states, the microresonator integrated on-chip enables simultaneous generation of many pairs of two-mode squeezed vacuum states by injecting a single pump field. This is due to the fact that the microresonator is a natural platform for generating frequency comb. Pump photons at a specific comb frequency can be converted into signal and idler photon pairs at lower and higher frequency modes, respectively. This was first demonstrated in a silica microresonator on a silicon chip and more than 20 frequency EPR pairs were generated simultaneously~\cite{yang2021squeezed, yang2022quantum}. A similar scheme for generating a set of two-mode squeezed vacuum states simultaneously has also been demonstrated in the more compatible silicon nitride platform~\cite{Jahanbozorgi2023generation}. 

Although on-chip squeezers have made substantial progress, the highest experimentally reported squeezing levels for on-chip squeezed vacuum states are still around 6 dB~\cite{Larsen2025Integrated, Shen2025HighlySqueezed}. This remains far below the squeezing level required for fault-tolerant CV quantum computation. The primary limitation arises from photon loss, including propagation loss, out-coupling loss, and the intrinsic loss within the on-chip squeezers. Therefore, optimizing the chip design to suppress photon loss is highly desirable. In parallel, strengthening the nonlinear coupling within the on-chip squeezer is also essential for increasing the achievable squeezing levels.

\subsection{CV cluster states}

Engineering on-chip CV cluster states is the next step to a fully integrated photonic quantum computing platform. One of the promising schemes to generate CV cluster states in frequency domain is to integrate a high-quality microresonator on chip, which generates quantum frequency combs. The Kerr-type nonlinearity in the microresonator enables the FWM processes that establish quantum correlations among quantum modes encoded by the comb lines. In particular, injecting a single primary pump line at one of the resonant frequencies of the microresonator can generate a set of EPR pairs via degenerate FWM process. Injecting a secondary pump line generates another set of EPR pairs and establishes quantum correlations between these two sets of EPR pairs through the non-degenerate FWM process. By utilizing the above mechanism, one- and two-dimensional cluster states can be generated~\cite{Wang2025LargeScale}. To generate multimode entangled states with various connectivity and topologies, a straightforward approach is to add additional pump lines to drive the microresonator. However, multiple pump lines would inevitably introduce Bragg scattering~\cite{Guidry2022QuantumOptics}, which ultimately destroys the cluster-type structure of the multimode entangled states. Nevertheless, the generated multipartite entanglement is still a useful quantum resource. By introducing, manipulating and detecting the supermodes, which are superpositions of the quantum frequency comb modes, one can effectively generate CV cluster states with various structures and connectivities~\cite{Jia2025ContinuousVariable}. Recent experimental demonstrations of generating multimode entangled states on chip are summarized in Fig.~\ref{fig:squeezing_cluster}.

The size and connectivity of existing CV cluster states remain limited. By employing time multiplexing, cluster states containing up to a billion modes have been generated~\cite{AghaeeRad2025Scaling}. However, scalability only in time may still fall short of supporting large-scale quantum computation. In contrast, cluster states generated through frequency multiplexing currently contain fewer than 100 modes~\cite{Wang2025LargeScale,Jia2025ContinuousVariable}, limiting their utility in large-scale CV MBQC. Increasing both the dimensionality and the size along each dimension of the CV cluster states is essential for enhancing their computational capabilities. Several proposals pursuing this direction have been explored recently~\cite{Wu2020quantum} and corresponding experimental demonstrations are highly anticipated.

\subsection{Non-Gaussian States}

\begin{figure*}[t]
\includegraphics[width=1.8\columnwidth]{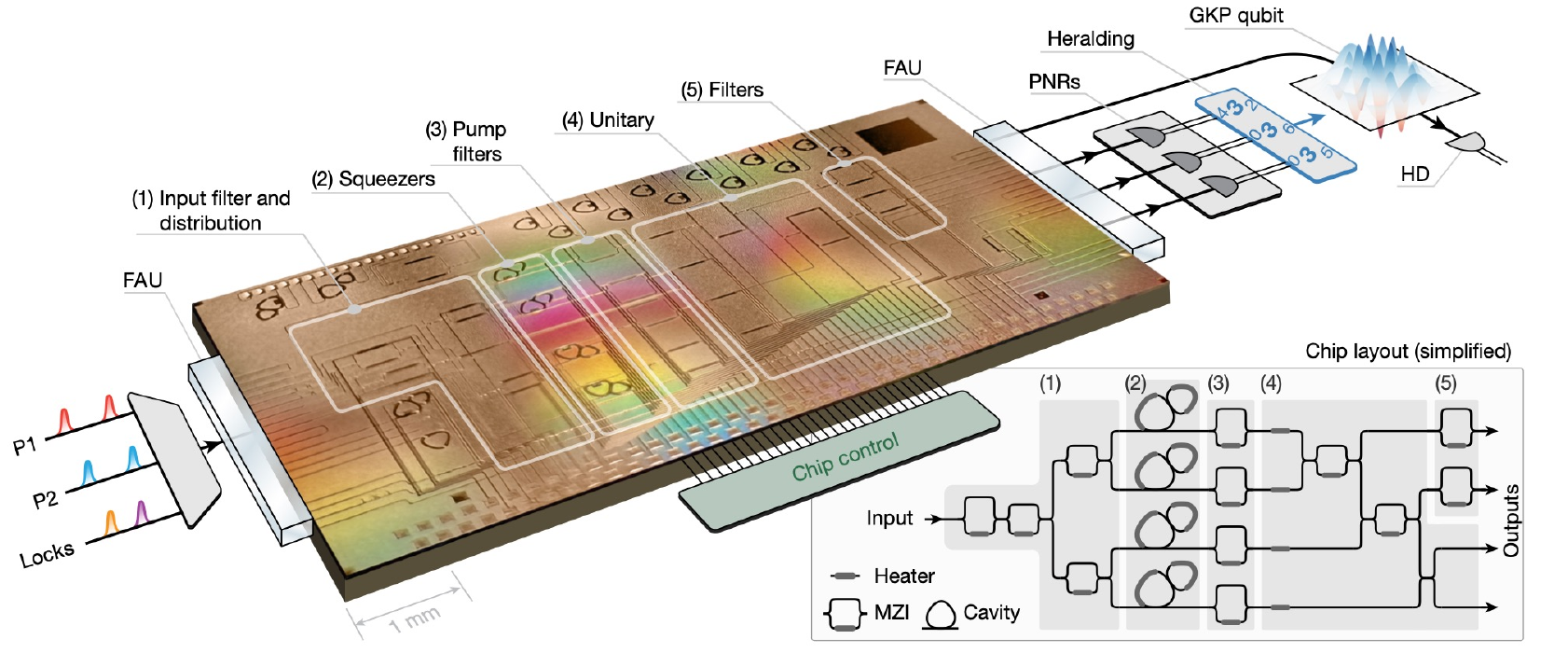}
\caption{ Generation of GKP states using a silicon nitride photonic chip and photon-number-resolving (PNR) detectors~\cite{Larsen2025Integrated}. Four single-mode squeezers and a four-mode linear optical network are integrated on chip, generating a four-mode Gaussian state. Three off-chip PNR detectors count the photon number of three output modes and the remaining mode is prepared in a GKP state, which is characterized via homodyne detection (HD). }
\label{fig:GKP_chip}
\end{figure*}

Recently, a generic framework for non-Gaussian state generation was proposed, in which a subset of modes of a multimode Gaussian state are measured by PNRDs and a non-Gaussian state is prepared by post-selecting a specific photon number pattern~\cite{Su2019conversion}. This non-Gaussian state generation framework can be experimentally implemented by employing squeezed vacuum states as inputs, a linear optical network, PNRDs and post selection, all of which can be achieved by the state-of-the-art photonic technologies. By preparing multimode Gaussian states on chip and performing photon-number-resolving detection off chip using transition edge sensor detectors, the generation of optical cat and GKP states was reported recently~\cite{AghaeeRad2025Scaling, Larsen2025Integrated}, as shown in Fig.~\ref{fig:GKP_chip}, which represents a major breakthrough towards a large-scale, fault-tolerant optical quantum computer on integrated photonic chips. 

The success rate and quality of the generated GKP states are limited. The existing proposals for generating GKP states are intrinsically probabilistic, leading to a trade-off between the state quality and success rate: higher-quality states are produced at lower rates. The best demonstrated generation rate to date is approximately 35 Hz~\cite{Larsen2025Integrated}, which is substantially slower than the generation rate of squeezed vacuum states. This severely limits the capability to demonstrate quantum algorithms and quantum error correction. The low generation rate arises primarily from the probabilistic nature of the preparation scheme and the long detection times of PNRDs. On the theoretical side, new schemes for generating non-Gaussian state are needed; on the hardware side, fast and high-efficiency PNRDs must be developed.


\section{Photon Detectors}

Three classes of photon detectors play central roles in photonic quantum computing: homodyne detectors, single-photon detectors, and photon-number-resolving detectors. These detectors are indispensable both for preparing photonic resource states and for implementing measurement-based gates. In the DV paradigm, single-photon detectors and PNRDs are required for state heralding, fusion measurements, and final readout. In contrast, CV computation relies primarily on high-performance homodyne detectors for quadrature measurements, supplemented by PNRDs for generating the non-Gaussian resource states required for universality and quantum error correction.

\subsection{Homodyne Detectors}

A homodyne detector (HD) is a phase-sensitive measurement device that enables precise characterization of weak signals by interfering them with a strong local oscillator (LO). In a balanced homodyne detection, the signal and LO are combined on a 50:50 beam splitter, and the resulting outputs are measured by two matched photodiodes. By subtracting the photocurrents, common-mode noise, primarily originating from the LO, is suppressed, allowing sensitive measurement of quantum fluctuations. Homodyne detection provides direct access to the field quadratures of the optical mode, with the measured quadrature determined by the relative phase between the signal and the LO. As such, HDs constitute a fundamental measurement primitive in CV MBQC, where phase-resolved quadrature measurements and feedforward operations implement quantum gates.

\begin{figure*}[t]
\includegraphics[width=1.9 \columnwidth]{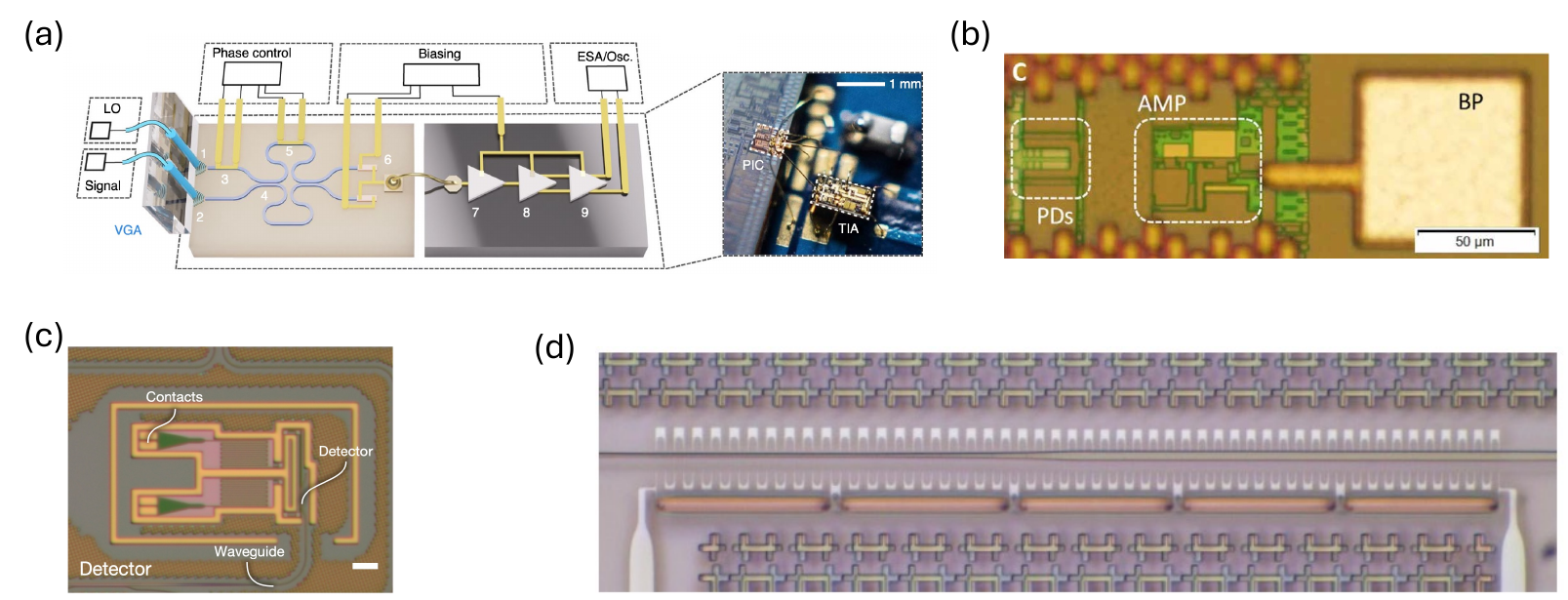}
\caption{ On-chip photon detectors. (a) Integrated photodiodes wire bonded to an unpackaged TIA die~\cite{Tasker2021SiliconPhotonics}; (b) homodyne detector with monolithic electronic–photonic integration~\cite{Tasker2024ABi-CMOS}; (c) superconducting nanowire single photon detector (SNSPD)~\cite{Alexander2025Manufacturable}; (d) photon-number-resolving detector through multiplexing many SNSPDs~\cite{Alexander2025Manufacturable}.  }
\label{fig:detector}
\end{figure*}

The first high-performance homodyne detectors with fully integrated optical components on a silicon-on-insulator (SOI) platform were demonstrated in Ref.~\cite{Raffaelli2018AHomodyne}, achieving a 3 dB bandwidth of 150 MHz, a shot-noise clearance of 11 dB, and a total detection efficiency of 0.59. Subsequent efforts have reported detector bandwidths in the MHz regime~\cite{Zhang2019Integrated}, although these implementations remain fundamentally limited by the interface between the photonic integrated circuit (PIC) and the transimpedance amplifier (TIA). In particular, the parasitic capacitance arising from the printed circuit board (PCB) interconnects and the input capacitance of packaged amplifier dies impose a severe constraint on bandwidth.
A major performance improvement was achieved by direct wire-bonding the PIC to an unpackaged TIA die (Fig.~\ref{fig:detector}(a)), thereby minimizing both the parasitic capacitance and inductance at the amplifier input. This hybrid integration strategy enabled a 3 dB bandwidth of 1.7 GHz~\cite{Tasker2021SiliconPhotonics}, comparable to the performance of homodyne detectors based on free-space photodiodes.
Further enhancement is made possible through monolithic electronic–photonic integration on a single substrate (Fig.~\ref{fig:detector}(b)), which eliminates the PIC-TIA interface altogether and reduces the parasitic capacitance to levels unattainable in multi-chip or discrete components. Using this approach, a 3 dB bandwidth of 15.3 GHz has been demonstrated~\cite{Tasker2024ABi-CMOS}, representing a significant milestone toward integrating high-speed quantum-limited detection directly on chip.

\subsection{Single-Photon Detectors}

High-efficiency single-photon detectors are indispensable for DV photonic quantum computation, which enable heralded generation of single photons and small entangled resource states, as well as single-qubit measurements and fusion measurements. The superconducting nanowire single-photon detectors (SNSPDs) are the most widely used and technically mature single-photon detectors, achieving the state-of-the-art detection efficiency of 98\%, operation speed on the order of a few nanoseconds and timing jitter of 3 ps~\cite{Reddy2020Superconducting, Korzh2020Sub3psSNSPD}. More importantly, the SNSPDs are compatible with integrated quantum photonics circuits, and they have been fabricated on various integrated photonic platforms, including GaAS, Si, SiN and LiNbO$_3$~\cite{Pernice2012TravellingWave, Schuck2013Waveguide}. The on-chip detection efficiency up to 93.4\% of the SNSPDs has been demonstrated, with the detectors operating at a temperature of approximately 2 K. The co-integration of SFWM sources of photon pairs, SNSPDs and tunable MZIs enables fully on-chip single-photon generation, and high-fidelity single-qubit state preparation, gate operation and measurements (Fig.~\ref{fig:detector}(c))~\cite{Alexander2025Manufacturable}. 

The PNRDs can discriminate the exact number of photons in a light field. The capability of counting the photon number helps identify the multi-photon components when heralding a single photon, thus increasing the purity of the single photon source; and identify the unwanted events in fusion-based entangled state generation and computation. One promising way to implement PNRDs is to spatially multiplex multiple SNSPDs (Fig.~\ref{fig:detector}(d)), which enables the discrimination of up to 100 photons on chip~\cite{Cheng2023Hundred}. A key advantage of this effective photon-number-resolving detection is the high detection speed, comparable to that of SNSPDs. Another way to implement PNRDs is to exploit the transition-edge sensor (TES), which can achieve high detection efficiency above 96\%~\cite{Larsen2025Integrated}. The PNRDs are also compatible with integrated quantum photonic circuits and have been integrated with GaAs, LiNbO$_3$ and SiO$_2$ waveguides. The  TES need to be operated at temperature below 50 mK, which imposes a stringent condition for experimental realization. Another disadvantage for PNRDs based on TES is that the detection speed is low, which is typically in the range of hundreds of kHz.

\section{Comparison of DV and CV paradigms}

\begin{table*}[t]
\centering
\small
\caption{Comparison of DV and CV photonic quantum computing paradigms.}
\label{tab:comparison}
\begin{tabular}{l  l  l  l}
\hline
\textbf{Feature} & \textbf{DV Paradigm} & \textbf{CV Paradigm} \\
\hline
\makecell[l]{Information \\ Encoding} & \makecell[l]{Single photons \\ path, polarization, time-bin} & \makecell[l]{Quadrature variables of \\ quantum optical modes}  \\
\hline
Resource State & \makecell[l]{Small DV cluster states \\ Large DV cluster states} & \makecell[l]{Large Gaussian cluster states \\ Non-Gaussian states}  \\
\hline
Key Operations & \makecell[l]{Phase shifters \\ Beam splitters \\ Probabilistic fusion} & \makecell[l]{Squeezing operation  \\ Linear optics \\ Homodyne detection }  \\
\hline
\makecell[l]{Non-Gaussian \\ Requirement} & \makecell[l]{Heralded single photon sources ~~~~~~~~~~ \\ PNRD} & GKP states, cubic phase  \\
\hline
Deterministic Gates?~~~~~~~~~~ & No (fusion gates $<100\%$) & \makecell[l]{Gaussian: yes \\ Non-Gaussian: no}  \\
\hline
FT Threshold & $\sim 2\%$ loss & $\sim10$--$15$ dB squeezing ~~  \\
\hline
\makecell[l]{Experimental \\ Maturity} & Mature at small scale & \makecell[l]{Mature Gaussian \\ Immature GKP}~~ \\
\hline
Scaling Method & \makecell[l]{Multiplexed photon sources  \\ Fusion measurement} & \makecell[l]{Time/frequency multiplexing \\ Delay loops}  \\
\hline
Main Bottleneck & \makecell[l]{Deterministic photons \\ Boosted fusion measurement} & High-quality GKP states \\
\hline
\end{tabular}
\end{table*}

The routes toward fault-tolerant photonic quantum computation have been deeply explored in both DV and CV paradigms. With rapid progress in integrated photonics, both approaches now have promising hardware paths toward large-scale implementation. Nevertheless, DV and CV photonic quantum computing rely on fundamentally different encoding strategies, resource states, operational primitives, and scaling pathways. These differences impose distinct technological requirements, which we summarize below and in Table~\ref{tab:comparison}.

In the DV paradigm, quantum information is encoded in the discrete degrees of freedom of single photons, which provides a clean notion of a qubit and enables straightforward state preparation, measurement, and verification. The discrete nature of the encoding makes the conceptual framework close to qubit-based models used on other quantum platforms. In contrast, the CV paradigm encodes quantum information in the continuous quadrature variables of an optical mode. The resulting infinite-dimensional Hilbert space offers a large redundant encoding space and enables deterministic entangling operations between modes. 

Both paradigms require nonclassical resource states, but with significant structural differences. The DV paradigm requires single photons, small entangled states (e.g., Bell pairs and GHZ states), and ultimately large-scale photonic cluster states. Since DV two-photon entangling gates are inherently nondeterministic, large clusters must be assembled via probabilistic fusion measurements. These fusion operations must exceed a percolation threshold to achieve scalable cluster generation, which places stringent requirements on source quality and detector performance. In the CV paradigm, the core resources are squeezed vacuum states, CV cluster states, and non-Gaussian states. A key advantage is that the two-mode entangling operations used to build large CV cluster states are deterministic, enabling scalable construction using only squeezed states and linear optics. However, universality requires non-Gaussian states, which can be generated only probabilistically within the framework of linear optics, thereby limiting practical scalability.

Despite significant architectural differences, both DV and CV photonic quantum computing face a common obstacle: the non-Gaussianity bottleneck, which fundamentally limits scalable fault-tolerant quantum computation. In the DV paradigm, the scalability depends on engineering probabilistic fusion operations with sufficiently high success probabilities. These can be boosted using high-purity, high-indistinguishability single photons, and high-efficiency SNSPDs and PNRDs. In the CV paradigm, non-Gaussian state generation is the main hurdle. Current schemes for non-Gaussian state generation rely on high-efficiency PNRDs and ultrafast optical switches. Strong light-matter interactions (e.g., in cavity QED systems or nonlinear nanophotonics) offer alternate deterministic pathways, but these approaches require substantial further development.

Photon loss constitutes the dominant error mechanism in both paradigms. Preliminary analyses indicate that the DV approach can tolerate about 2\% photon loss, assuming state-of-the-art detectors and encoding schemes. By concatenating GKP codes with high-performance qubit quantum error-correcting codes in the CV approach, photon loss can be mapped to finite squeezing noise. Fault tolerance requires achieving about 10 dB of squeezing, which is within reach of current state-of-the-art sources but remains technically demanding.

\section{Conclusions}\label{sec6}

We survey recent progress in photonic quantum computation implemented on integrated quantum photonic chips, which offer intrinsic scalability, compact footprints, enhanced phase stability, high-volume manufacturability, and programmability. Various measurement-based photonic architectures have been proposed to realize universal and fault-tolerant quantum computing, incorporating modular subsystems for resource-state generation, measurement-based gate implementation, and quantum error correction.
Substantial advances have been made in integrating the key building blocks required for large-scale photonic quantum computing onto integrated photonic platforms. Essential components, ranging from on-chip sources of single photons, squeezed states, non-Gaussian states, and cluster states, to high-performance MZIs with fast, low-loss phase modulation and broad bandwidth, as well as high-efficiency photon detectors for state preparation, characterization, and measurement-based gates, have now been demonstrated either individually or in combination. The rapid advancement of these technologies positions integrated photonic circuits as one of the most promising platforms for quantum computation and brings the vision of fully integrated, chip-scale quantum microprocessors within increasingly realistic reach.


Looking ahead, it remains crucial to explore new schemes for generating resource states with high efficiency and to develop architectures that incorporate quantum LDPC codes with higher encoding rates, thereby achieving substantial reductions in resource overhead. From a hardware perspective, continued improvements in the performance of integrated photonic building blocks are essential, including higher squeezing levels, faster and more efficient state-generation rates, and further enhancements in single-photon sources and detectors. Beyond individual components, the co-integration of all key building blocks on a monolithic chip, or alternatively, the development of ultra-low-loss interconnects linking modular subsystems, represents a major frontier. Meeting this challenge will likely require heterogeneous integration strategies that leverage the strengths of multiple material platforms, each optimized for specific functionalities. Collectively, these developments are key to enabling on-chip demonstrations of quantum algorithms and quantum error correction, and to pushing integrated photonic technologies toward the demanding thresholds required for fully fault-tolerant quantum computation.


{\bf Acknowledgements}: We would like to thank Timothy Ralph for insightful discussions. This research was conducted in the JC STEM Lab of Quantum Technology, funded by The Hong Kong Jockey Club Charities Trust. This work was also supported by Hong Kong Polytechnic University (1-CD3J, 1-BEAZ, 1-BDZ6, 1-BDWT, 1-BEDH, 1-BF06, 1-BEET, 4-ZZYB) and by Hong Kong Research Grant Council/University Grants Committee (grant no. 21203724). 

{\bf Conflict of interest}: The authors declare no competing interests. 

\bibliography{Photonic_QC}

\begin{thebibliography}{160}
\expandafter\ifx\csname natexlab\endcsname\relax\def\natexlab#1{#1}\fi
\expandafter\ifx\csname bibnamefont\endcsname\relax
  \def\bibnamefont#1{#1}\fi
\expandafter\ifx\csname bibfnamefont\endcsname\relax
  \def\bibfnamefont#1{#1}\fi
\expandafter\ifx\csname citenamefont\endcsname\relax
  \def\citenamefont#1{#1}\fi
\expandafter\ifx\csname url\endcsname\relax
  \def\url#1{\texttt{#1}}\fi
\expandafter\ifx\csname urlprefix\endcsname\relax\def\urlprefix{URL }\fi
\providecommand{\bibinfo}[2]{#2}
\providecommand{\eprint}[2][]{\url{#2}}

\bibitem[{\citenamefont{Devoret and
  Schoelkopf}(2013)}]{Devoret2013SuperconductingCircuits}
\bibinfo{author}{\bibfnamefont{M.~H.} \bibnamefont{Devoret}} \bibnamefont{and}
  \bibinfo{author}{\bibfnamefont{R.~J.} \bibnamefont{Schoelkopf}},
  \bibinfo{journal}{Science} \textbf{\bibinfo{volume}{339}},
  \bibinfo{pages}{1169} (\bibinfo{year}{2013}).

\bibitem[{\citenamefont{Krantz et~al.}(2019)\citenamefont{Krantz, Kjaergaard,
  Yan, Orlando, Gustavsson, and Oliver}}]{Krantz2019Quantum}
\bibinfo{author}{\bibfnamefont{P.}~\bibnamefont{Krantz}},
  \bibinfo{author}{\bibfnamefont{M.}~\bibnamefont{Kjaergaard}},
  \bibinfo{author}{\bibfnamefont{F.}~\bibnamefont{Yan}},
  \bibinfo{author}{\bibfnamefont{T.~P.} \bibnamefont{Orlando}},
  \bibinfo{author}{\bibfnamefont{S.}~\bibnamefont{Gustavsson}},
  \bibnamefont{and} \bibinfo{author}{\bibfnamefont{W.~D.}
  \bibnamefont{Oliver}}, \bibinfo{journal}{Applied Physics Reviews}
  \textbf{\bibinfo{volume}{6}}, \bibinfo{pages}{021318} (\bibinfo{year}{2019}).

\bibitem[{\citenamefont{Arute et~al.}(2019)\citenamefont{Arute, Arya, Babbush,
  Bacon, Bardin, Barends, Biswas, Boixo, ao, Buell
  et~al.}}]{Arute2019QuantumSupremacy}
\bibinfo{author}{\bibfnamefont{F.}~\bibnamefont{Arute}},
  \bibinfo{author}{\bibfnamefont{K.}~\bibnamefont{Arya}},
  \bibinfo{author}{\bibfnamefont{R.}~\bibnamefont{Babbush}},
  \bibinfo{author}{\bibfnamefont{D.}~\bibnamefont{Bacon}},
  \bibinfo{author}{\bibfnamefont{J.~C.} \bibnamefont{Bardin}},
  \bibinfo{author}{\bibfnamefont{R.}~\bibnamefont{Barends}},
  \bibinfo{author}{\bibfnamefont{R.}~\bibnamefont{Biswas}},
  \bibinfo{author}{\bibfnamefont{S.}~\bibnamefont{Boixo}},
  \bibinfo{author}{\bibfnamefont{F.~G. S. L.~B.} \bibnamefont{ao}},
  \bibinfo{author}{\bibfnamefont{D.~A.} \bibnamefont{Buell}},
  \bibnamefont{et~al.}, \bibinfo{journal}{Nature}
  \textbf{\bibinfo{volume}{574}}, \bibinfo{pages}{505} (\bibinfo{year}{2019}).

\bibitem[{\citenamefont{{Google Quantum AI and
  Collaborators}}(2025)}]{Acharya2025QuantumErrorCorrection}
\bibinfo{author}{\bibnamefont{{Google Quantum AI and Collaborators}}},
  \bibinfo{journal}{Nature} \textbf{\bibinfo{volume}{638}},
  \bibinfo{pages}{920} (\bibinfo{year}{2025}).

\bibitem[{\citenamefont{Cirac and Zoller}(1995)}]{Cirac1995Quantum}
\bibinfo{author}{\bibfnamefont{J.~I.} \bibnamefont{Cirac}} \bibnamefont{and}
  \bibinfo{author}{\bibfnamefont{P.}~\bibnamefont{Zoller}},
  \bibinfo{journal}{Phys. Rev. Lett.} \textbf{\bibinfo{volume}{74}},
  \bibinfo{pages}{4091} (\bibinfo{year}{1995}),
  \urlprefix\url{https://link.aps.org/doi/10.1103/PhysRevLett.74.4091}.

\bibitem[{\citenamefont{Monroe and Kim}(2013)}]{Monroe2013ScalingIonTrap}
\bibinfo{author}{\bibfnamefont{C.}~\bibnamefont{Monroe}} \bibnamefont{and}
  \bibinfo{author}{\bibfnamefont{J.}~\bibnamefont{Kim}},
  \bibinfo{journal}{Science} \textbf{\bibinfo{volume}{339}},
  \bibinfo{pages}{1164} (\bibinfo{year}{2013}).

\bibitem[{\citenamefont{Bruzewicz et~al.}(2019)\citenamefont{Bruzewicz,
  Chiaverini, McConnell, and Sage}}]{Bruzewicz2019TrappedIonQC}
\bibinfo{author}{\bibfnamefont{C.~D.} \bibnamefont{Bruzewicz}},
  \bibinfo{author}{\bibfnamefont{J.}~\bibnamefont{Chiaverini}},
  \bibinfo{author}{\bibfnamefont{R.}~\bibnamefont{McConnell}},
  \bibnamefont{and} \bibinfo{author}{\bibfnamefont{J.~M.} \bibnamefont{Sage}},
  \bibinfo{journal}{Applied Physics Reviews} \textbf{\bibinfo{volume}{6}},
  \bibinfo{pages}{021314} (\bibinfo{year}{2019}).

\bibitem[{\citenamefont{Evered et~al.}(2023)\citenamefont{Evered, Bluvstein,
  Kalinowski, Ebadi, Manovitz, Zhou, Li, Geim, Wang, Maskara
  et~al.}}]{Evered2023HighFidelity}
\bibinfo{author}{\bibfnamefont{S.~J.} \bibnamefont{Evered}},
  \bibinfo{author}{\bibfnamefont{D.}~\bibnamefont{Bluvstein}},
  \bibinfo{author}{\bibfnamefont{M.}~\bibnamefont{Kalinowski}},
  \bibinfo{author}{\bibfnamefont{S.}~\bibnamefont{Ebadi}},
  \bibinfo{author}{\bibfnamefont{T.}~\bibnamefont{Manovitz}},
  \bibinfo{author}{\bibfnamefont{H.}~\bibnamefont{Zhou}},
  \bibinfo{author}{\bibfnamefont{S.~H.} \bibnamefont{Li}},
  \bibinfo{author}{\bibfnamefont{A.~A.} \bibnamefont{Geim}},
  \bibinfo{author}{\bibfnamefont{T.~T.} \bibnamefont{Wang}},
  \bibinfo{author}{\bibfnamefont{N.}~\bibnamefont{Maskara}},
  \bibnamefont{et~al.}, \bibinfo{journal}{Nature}
  \textbf{\bibinfo{volume}{622}}, \bibinfo{pages}{268} (\bibinfo{year}{2023}).

\bibitem[{\citenamefont{Radnaev et~al.}(2025)\citenamefont{Radnaev, Chung,
  Cole, Mason, Ballance, Bedalov, Belknap, Berman, Blakely, Bloomfield
  et~al.}}]{Radnaev2025Universal}
\bibinfo{author}{\bibfnamefont{A.}~\bibnamefont{Radnaev}},
  \bibinfo{author}{\bibfnamefont{W.}~\bibnamefont{Chung}},
  \bibinfo{author}{\bibfnamefont{D.}~\bibnamefont{Cole}},
  \bibinfo{author}{\bibfnamefont{D.}~\bibnamefont{Mason}},
  \bibinfo{author}{\bibfnamefont{T.}~\bibnamefont{Ballance}},
  \bibinfo{author}{\bibfnamefont{M.}~\bibnamefont{Bedalov}},
  \bibinfo{author}{\bibfnamefont{D.}~\bibnamefont{Belknap}},
  \bibinfo{author}{\bibfnamefont{M.}~\bibnamefont{Berman}},
  \bibinfo{author}{\bibfnamefont{M.}~\bibnamefont{Blakely}},
  \bibinfo{author}{\bibfnamefont{I.}~\bibnamefont{Bloomfield}},
  \bibnamefont{et~al.}, \bibinfo{journal}{PRX Quantum}
  \textbf{\bibinfo{volume}{6}}, \bibinfo{pages}{030334} (\bibinfo{year}{2025}),
  \urlprefix\url{https://link.aps.org/doi/10.1103/66s8-jj18}.

\bibitem[{\citenamefont{Bluvstein et~al.}(2026)\citenamefont{Bluvstein, Geim,
  Li, Evered, Ataides, Baranes, Gu, Manovitz, Xu, Kalinowski
  et~al.}}]{Bluvstein2025FaultTolerant}
\bibinfo{author}{\bibfnamefont{D.}~\bibnamefont{Bluvstein}},
  \bibinfo{author}{\bibfnamefont{A.~A.} \bibnamefont{Geim}},
  \bibinfo{author}{\bibfnamefont{S.~H.} \bibnamefont{Li}},
  \bibinfo{author}{\bibfnamefont{S.~J.} \bibnamefont{Evered}},
  \bibinfo{author}{\bibfnamefont{J.~P.~B.} \bibnamefont{Ataides}},
  \bibinfo{author}{\bibfnamefont{G.}~\bibnamefont{Baranes}},
  \bibinfo{author}{\bibfnamefont{A.}~\bibnamefont{Gu}},
  \bibinfo{author}{\bibfnamefont{T.}~\bibnamefont{Manovitz}},
  \bibinfo{author}{\bibfnamefont{M.}~\bibnamefont{Xu}},
  \bibinfo{author}{\bibfnamefont{M.}~\bibnamefont{Kalinowski}},
  \bibnamefont{et~al.}, \bibinfo{journal}{Nature}
  \textbf{\bibinfo{volume}{649}}, \bibinfo{pages}{39} (\bibinfo{year}{2026}).

\bibitem[{\citenamefont{Weinstein et~al.}(2023)\citenamefont{Weinstein, Reed,
  Jones, Andrews, Barnes, Blumoff, Euliss, Eng, Fong, Ha
  et~al.}}]{Weinstein2023Universal}
\bibinfo{author}{\bibfnamefont{A.~J.} \bibnamefont{Weinstein}},
  \bibinfo{author}{\bibfnamefont{M.~D.} \bibnamefont{Reed}},
  \bibinfo{author}{\bibfnamefont{A.~M.} \bibnamefont{Jones}},
  \bibinfo{author}{\bibfnamefont{R.~W.} \bibnamefont{Andrews}},
  \bibinfo{author}{\bibfnamefont{D.}~\bibnamefont{Barnes}},
  \bibinfo{author}{\bibfnamefont{J.~Z.} \bibnamefont{Blumoff}},
  \bibinfo{author}{\bibfnamefont{L.~E.} \bibnamefont{Euliss}},
  \bibinfo{author}{\bibfnamefont{K.}~\bibnamefont{Eng}},
  \bibinfo{author}{\bibfnamefont{B.~H.} \bibnamefont{Fong}},
  \bibinfo{author}{\bibfnamefont{S.~D.} \bibnamefont{Ha}},
  \bibnamefont{et~al.}, \bibinfo{journal}{Nature}
  \textbf{\bibinfo{volume}{615}}, \bibinfo{pages}{817} (\bibinfo{year}{2023}).

\bibitem[{\citenamefont{Edlbauer et~al.}(2025)\citenamefont{Edlbauer, Wang,
  Huq, Thorvaldson, Jones, Misha, Pappas, Moehle, Hsueh, Bornemann
  et~al.}}]{Edlbauer2025An11Qubit}
\bibinfo{author}{\bibfnamefont{H.}~\bibnamefont{Edlbauer}},
  \bibinfo{author}{\bibfnamefont{J.}~\bibnamefont{Wang}},
  \bibinfo{author}{\bibfnamefont{A.~M.~S.} \bibnamefont{Huq}},
  \bibinfo{author}{\bibfnamefont{I.}~\bibnamefont{Thorvaldson}},
  \bibinfo{author}{\bibfnamefont{M.~T.} \bibnamefont{Jones}},
  \bibinfo{author}{\bibfnamefont{S.~H.} \bibnamefont{Misha}},
  \bibinfo{author}{\bibfnamefont{W.~J.} \bibnamefont{Pappas}},
  \bibinfo{author}{\bibfnamefont{C.~M.} \bibnamefont{Moehle}},
  \bibinfo{author}{\bibfnamefont{Y.}~\bibnamefont{Hsueh}},
  \bibinfo{author}{\bibfnamefont{H.}~\bibnamefont{Bornemann}},
  \bibnamefont{et~al.}, \bibinfo{journal}{Nature}
  \textbf{\bibinfo{volume}{648}}, \bibinfo{pages}{569} (\bibinfo{year}{2025}).

\bibitem[{\citenamefont{Kitaev}(2003)}]{Kitaev2003Fault}
\bibinfo{author}{\bibfnamefont{A.~Y.} \bibnamefont{Kitaev}},
  \bibinfo{journal}{Annals of Physics} \textbf{\bibinfo{volume}{303}},
  \bibinfo{pages}{2} (\bibinfo{year}{2003}).

\bibitem[{\citenamefont{Nayak et~al.}(2008)\citenamefont{Nayak, Simon, Stern,
  Freedman, and Das~Sarma}}]{Nayak2008NonAbelian}
\bibinfo{author}{\bibfnamefont{C.}~\bibnamefont{Nayak}},
  \bibinfo{author}{\bibfnamefont{S.~H.} \bibnamefont{Simon}},
  \bibinfo{author}{\bibfnamefont{A.}~\bibnamefont{Stern}},
  \bibinfo{author}{\bibfnamefont{M.}~\bibnamefont{Freedman}}, \bibnamefont{and}
  \bibinfo{author}{\bibfnamefont{S.}~\bibnamefont{Das~Sarma}},
  \bibinfo{journal}{Rev. Mod. Phys.} \textbf{\bibinfo{volume}{80}},
  \bibinfo{pages}{1083} (\bibinfo{year}{2008}),
  \urlprefix\url{https://link.aps.org/doi/10.1103/RevModPhys.80.1083}.

\bibitem[{\citenamefont{Knill et~al.}(2001)\citenamefont{Knill, Laflamme, and
  Milburn}}]{Knill2001SchemeEfficient}
\bibinfo{author}{\bibfnamefont{E.}~\bibnamefont{Knill}},
  \bibinfo{author}{\bibfnamefont{R.}~\bibnamefont{Laflamme}}, \bibnamefont{and}
  \bibinfo{author}{\bibfnamefont{G.~J.} \bibnamefont{Milburn}},
  \bibinfo{journal}{Nature} \textbf{\bibinfo{volume}{409}}, \bibinfo{pages}{46}
  (\bibinfo{year}{2001}).

\bibitem[{\citenamefont{Raussendorf and Briegel}(2001)}]{Raussendorf2001OneWay}
\bibinfo{author}{\bibfnamefont{R.}~\bibnamefont{Raussendorf}} \bibnamefont{and}
  \bibinfo{author}{\bibfnamefont{H.~J.} \bibnamefont{Briegel}},
  \bibinfo{journal}{Phys. Rev. Lett.} \textbf{\bibinfo{volume}{86}},
  \bibinfo{pages}{5188} (\bibinfo{year}{2001}),
  \urlprefix\url{https://link.aps.org/doi/10.1103/PhysRevLett.86.5188}.

\bibitem[{\citenamefont{Raussendorf et~al.}(2003)\citenamefont{Raussendorf,
  Browne, and Briegel}}]{Raussendorf2003Measurement}
\bibinfo{author}{\bibfnamefont{R.}~\bibnamefont{Raussendorf}},
  \bibinfo{author}{\bibfnamefont{D.~E.} \bibnamefont{Browne}},
  \bibnamefont{and} \bibinfo{author}{\bibfnamefont{H.~J.}
  \bibnamefont{Briegel}}, \bibinfo{journal}{Phys. Rev. A}
  \textbf{\bibinfo{volume}{68}}, \bibinfo{pages}{022312}
  (\bibinfo{year}{2003}),
  \urlprefix\url{https://link.aps.org/doi/10.1103/PhysRevA.68.022312}.

\bibitem[{\citenamefont{Nielsen}(2004)}]{Nielsen2004Optical}
\bibinfo{author}{\bibfnamefont{M.~A.} \bibnamefont{Nielsen}},
  \bibinfo{journal}{Phys. Rev. Lett.} \textbf{\bibinfo{volume}{93}},
  \bibinfo{pages}{040503} (\bibinfo{year}{2004}),
  \urlprefix\url{https://link.aps.org/doi/10.1103/PhysRevLett.93.040503}.

\bibitem[{\citenamefont{Browne and Rudolph}(2005)}]{Browne2005Resource}
\bibinfo{author}{\bibfnamefont{D.~E.} \bibnamefont{Browne}} \bibnamefont{and}
  \bibinfo{author}{\bibfnamefont{T.}~\bibnamefont{Rudolph}},
  \bibinfo{journal}{Phys. Rev. Lett.} \textbf{\bibinfo{volume}{95}},
  \bibinfo{pages}{010501} (\bibinfo{year}{2005}),
  \urlprefix\url{https://link.aps.org/doi/10.1103/PhysRevLett.95.010501}.

\bibitem[{\citenamefont{Kieling et~al.}(2007)\citenamefont{Kieling, Rudolph,
  and Eisert}}]{Kieling2007Percolation}
\bibinfo{author}{\bibfnamefont{K.}~\bibnamefont{Kieling}},
  \bibinfo{author}{\bibfnamefont{T.}~\bibnamefont{Rudolph}}, \bibnamefont{and}
  \bibinfo{author}{\bibfnamefont{J.}~\bibnamefont{Eisert}},
  \bibinfo{journal}{Phys. Rev. Lett.} \textbf{\bibinfo{volume}{99}},
  \bibinfo{pages}{130501} (\bibinfo{year}{2007}),
  \urlprefix\url{https://link.aps.org/doi/10.1103/PhysRevLett.99.130501}.

\bibitem[{\citenamefont{O'Brien et~al.}(2003)\citenamefont{O'Brien, Pryde,
  White, Ralph, and Branning}}]{OBrien2003ControlledNOT}
\bibinfo{author}{\bibfnamefont{J.~L.} \bibnamefont{O'Brien}},
  \bibinfo{author}{\bibfnamefont{G.~J.} \bibnamefont{Pryde}},
  \bibinfo{author}{\bibfnamefont{A.~G.} \bibnamefont{White}},
  \bibinfo{author}{\bibfnamefont{T.~C.} \bibnamefont{Ralph}}, \bibnamefont{and}
  \bibinfo{author}{\bibfnamefont{D.}~\bibnamefont{Branning}},
  \bibinfo{journal}{Nature} \textbf{\bibinfo{volume}{426}},
  \bibinfo{pages}{264} (\bibinfo{year}{2003}).

\bibitem[{\citenamefont{Walther et~al.}(2005)\citenamefont{Walther, Resch,
  Rudolph, Schenck, Weinfurter, Vedral, Aspelmeyer, and
  Zeilinger}}]{Walther2005Experimental}
\bibinfo{author}{\bibfnamefont{P.}~\bibnamefont{Walther}},
  \bibinfo{author}{\bibfnamefont{K.~J.} \bibnamefont{Resch}},
  \bibinfo{author}{\bibfnamefont{T.}~\bibnamefont{Rudolph}},
  \bibinfo{author}{\bibfnamefont{E.}~\bibnamefont{Schenck}},
  \bibinfo{author}{\bibfnamefont{H.}~\bibnamefont{Weinfurter}},
  \bibinfo{author}{\bibfnamefont{V.}~\bibnamefont{Vedral}},
  \bibinfo{author}{\bibfnamefont{M.}~\bibnamefont{Aspelmeyer}},
  \bibnamefont{and}
  \bibinfo{author}{\bibfnamefont{A.}~\bibnamefont{Zeilinger}},
  \bibinfo{journal}{Nature} \textbf{\bibinfo{volume}{434}},
  \bibinfo{pages}{169} (\bibinfo{year}{2005}).

\bibitem[{\citenamefont{Lu et~al.}(2007)\citenamefont{Lu, Zhou, G{\"u}hne, Gao,
  Zhang, Chen, Goebel, Yang, Kim, Zeilinger
  et~al.}}]{Lu2007ExperimentalEntanglement}
\bibinfo{author}{\bibfnamefont{C.-Y.} \bibnamefont{Lu}},
  \bibinfo{author}{\bibfnamefont{X.-Q.} \bibnamefont{Zhou}},
  \bibinfo{author}{\bibfnamefont{O.}~\bibnamefont{G{\"u}hne}},
  \bibinfo{author}{\bibfnamefont{W.-B.} \bibnamefont{Gao}},
  \bibinfo{author}{\bibfnamefont{J.}~\bibnamefont{Zhang}},
  \bibinfo{author}{\bibfnamefont{Z.-B.} \bibnamefont{Chen}},
  \bibinfo{author}{\bibfnamefont{A.}~\bibnamefont{Goebel}},
  \bibinfo{author}{\bibfnamefont{T.}~\bibnamefont{Yang}},
  \bibinfo{author}{\bibfnamefont{J.}~\bibnamefont{Kim}},
  \bibinfo{author}{\bibfnamefont{A.}~\bibnamefont{Zeilinger}},
  \bibnamefont{et~al.}, \bibinfo{journal}{Nature Physics}
  \textbf{\bibinfo{volume}{3}}, \bibinfo{pages}{91} (\bibinfo{year}{2007}).

\bibitem[{\citenamefont{Yokoyama et~al.}(2013)\citenamefont{Yokoyama, Ukai,
  Armstrong, Sornphiphatphong, Kaji, Suzuki, Yoshikawa, Yonezawa, Menicucci,
  and Furusawa}}]{Yokoyama2013Ultra}
\bibinfo{author}{\bibfnamefont{S.}~\bibnamefont{Yokoyama}},
  \bibinfo{author}{\bibfnamefont{R.}~\bibnamefont{Ukai}},
  \bibinfo{author}{\bibfnamefont{S.~C.} \bibnamefont{Armstrong}},
  \bibinfo{author}{\bibfnamefont{C.}~\bibnamefont{Sornphiphatphong}},
  \bibinfo{author}{\bibfnamefont{T.}~\bibnamefont{Kaji}},
  \bibinfo{author}{\bibfnamefont{S.}~\bibnamefont{Suzuki}},
  \bibinfo{author}{\bibfnamefont{J.}~\bibnamefont{Yoshikawa}},
  \bibinfo{author}{\bibfnamefont{H.}~\bibnamefont{Yonezawa}},
  \bibinfo{author}{\bibfnamefont{N.~C.} \bibnamefont{Menicucci}},
  \bibnamefont{and} \bibinfo{author}{\bibfnamefont{A.}~\bibnamefont{Furusawa}},
  \bibinfo{journal}{Nature Photonics} \textbf{\bibinfo{volume}{7}},
  \bibinfo{pages}{982} (\bibinfo{year}{2013}).

\bibitem[{\citenamefont{Larsen et~al.}(2019)\citenamefont{Larsen, Guo, Breum,
  Neergaard-Nielsen, and Andersen}}]{Larsen2019Deterministic}
\bibinfo{author}{\bibfnamefont{M.~V.} \bibnamefont{Larsen}},
  \bibinfo{author}{\bibfnamefont{X.}~\bibnamefont{Guo}},
  \bibinfo{author}{\bibfnamefont{C.~R.} \bibnamefont{Breum}},
  \bibinfo{author}{\bibfnamefont{J.~S.} \bibnamefont{Neergaard-Nielsen}},
  \bibnamefont{and} \bibinfo{author}{\bibfnamefont{U.~L.}
  \bibnamefont{Andersen}}, \bibinfo{journal}{Science}
  \textbf{\bibinfo{volume}{366}}, \bibinfo{pages}{369} (\bibinfo{year}{2019}).

\bibitem[{\citenamefont{Asavanant et~al.}(2019)\citenamefont{Asavanant,
  Shiozawa, Yokoyama, Charoensombutamon, Emura, Alexander, Takeda, Yoshikawa,
  Menicucci, Yonezawa et~al.}}]{Asavanant2019TimeDomain}
\bibinfo{author}{\bibfnamefont{W.}~\bibnamefont{Asavanant}},
  \bibinfo{author}{\bibfnamefont{Y.}~\bibnamefont{Shiozawa}},
  \bibinfo{author}{\bibfnamefont{S.}~\bibnamefont{Yokoyama}},
  \bibinfo{author}{\bibfnamefont{B.}~\bibnamefont{Charoensombutamon}},
  \bibinfo{author}{\bibfnamefont{H.}~\bibnamefont{Emura}},
  \bibinfo{author}{\bibfnamefont{R.~N.} \bibnamefont{Alexander}},
  \bibinfo{author}{\bibfnamefont{S.}~\bibnamefont{Takeda}},
  \bibinfo{author}{\bibfnamefont{J.-i.} \bibnamefont{Yoshikawa}},
  \bibinfo{author}{\bibfnamefont{N.~C.} \bibnamefont{Menicucci}},
  \bibinfo{author}{\bibfnamefont{H.}~\bibnamefont{Yonezawa}},
  \bibnamefont{et~al.}, \bibinfo{journal}{Science}
  \textbf{\bibinfo{volume}{366}}, \bibinfo{pages}{373–376}
  (\bibinfo{year}{2019}),
  \urlprefix\url{https://doi.org/10.1126/science.aay2645}.

\bibitem[{\citenamefont{Asavanant et~al.}(2021)\citenamefont{Asavanant,
  Charoensombutamon, Yokoyama, Ebihara, Nakamura, Alexander, Endo, Yoshikawa,
  Menicucci, Yonezawa et~al.}}]{Asavanant2021TimeDomain}
\bibinfo{author}{\bibfnamefont{W.}~\bibnamefont{Asavanant}},
  \bibinfo{author}{\bibfnamefont{B.}~\bibnamefont{Charoensombutamon}},
  \bibinfo{author}{\bibfnamefont{S.}~\bibnamefont{Yokoyama}},
  \bibinfo{author}{\bibfnamefont{T.}~\bibnamefont{Ebihara}},
  \bibinfo{author}{\bibfnamefont{T.}~\bibnamefont{Nakamura}},
  \bibinfo{author}{\bibfnamefont{R.~N.} \bibnamefont{Alexander}},
  \bibinfo{author}{\bibfnamefont{M.}~\bibnamefont{Endo}},
  \bibinfo{author}{\bibfnamefont{J.-i.} \bibnamefont{Yoshikawa}},
  \bibinfo{author}{\bibfnamefont{N.~C.} \bibnamefont{Menicucci}},
  \bibinfo{author}{\bibfnamefont{H.}~\bibnamefont{Yonezawa}},
  \bibnamefont{et~al.}, \bibinfo{journal}{Phys. Rev. Appl.}
  \textbf{\bibinfo{volume}{16}}, \bibinfo{pages}{034005}
  (\bibinfo{year}{2021}),
  \urlprefix\url{https://link.aps.org/doi/10.1103/PhysRevApplied.16.034005}.

\bibitem[{\citenamefont{Larsen et~al.}(2021{\natexlab{a}})\citenamefont{Larsen,
  Guo, Breum, Neergaard-Nielsen, and Andersen}}]{Larsen2021Deterministic}
\bibinfo{author}{\bibfnamefont{M.~V.} \bibnamefont{Larsen}},
  \bibinfo{author}{\bibfnamefont{X.}~\bibnamefont{Guo}},
  \bibinfo{author}{\bibfnamefont{C.~R.} \bibnamefont{Breum}},
  \bibinfo{author}{\bibfnamefont{J.~S.} \bibnamefont{Neergaard-Nielsen}},
  \bibnamefont{and} \bibinfo{author}{\bibfnamefont{U.~L.}
  \bibnamefont{Andersen}}, \bibinfo{journal}{Nature Physics}
  \textbf{\bibinfo{volume}{17}}, \bibinfo{pages}{1018}
  (\bibinfo{year}{2021}{\natexlab{a}}).

\bibitem[{\citenamefont{Gidney}(2025)}]{Gidney2025Factor2048LessMillion}
\bibinfo{author}{\bibfnamefont{C.}~\bibnamefont{Gidney}},
  \bibinfo{journal}{arXiv preprint} \textbf{\bibinfo{volume}{arXiv:2505.15917}}
  (\bibinfo{year}{2025}), \urlprefix\url{https://arxiv.org/abs/2505.15917}.

\bibitem[{\citenamefont{Webster et~al.}(2026)\citenamefont{Webster, Berent,
  Chandra, Hockings, Baspin, Thomsen, Smith, and
  Cohen}}]{Webster2026PinnacleArchitecture}
\bibinfo{author}{\bibfnamefont{P.}~\bibnamefont{Webster}},
  \bibinfo{author}{\bibfnamefont{L.}~\bibnamefont{Berent}},
  \bibinfo{author}{\bibfnamefont{O.}~\bibnamefont{Chandra}},
  \bibinfo{author}{\bibfnamefont{E.~T.} \bibnamefont{Hockings}},
  \bibinfo{author}{\bibfnamefont{N.}~\bibnamefont{Baspin}},
  \bibinfo{author}{\bibfnamefont{F.}~\bibnamefont{Thomsen}},
  \bibinfo{author}{\bibfnamefont{S.~C.} \bibnamefont{Smith}}, \bibnamefont{and}
  \bibinfo{author}{\bibfnamefont{L.~Z.} \bibnamefont{Cohen}},
  \bibinfo{journal}{arXiv preprint}
  \textbf{\bibinfo{volume}{quant‑ph/2602.11457}} (\bibinfo{year}{2026}),
  \urlprefix\url{https://arxiv.org/abs/2602.11457}.

\bibitem[{\citenamefont{Cain et~al.}(2026)\citenamefont{Cain, Xu, King, Picard,
  Levine, Endres, Preskill, Huang, and Bluvstein}}]{Cain2026Shor10kQuits}
\bibinfo{author}{\bibfnamefont{M.}~\bibnamefont{Cain}},
  \bibinfo{author}{\bibfnamefont{Q.}~\bibnamefont{Xu}},
  \bibinfo{author}{\bibfnamefont{R.}~\bibnamefont{King}},
  \bibinfo{author}{\bibfnamefont{L.~R.~B.} \bibnamefont{Picard}},
  \bibinfo{author}{\bibfnamefont{H.}~\bibnamefont{Levine}},
  \bibinfo{author}{\bibfnamefont{M.}~\bibnamefont{Endres}},
  \bibinfo{author}{\bibfnamefont{J.}~\bibnamefont{Preskill}},
  \bibinfo{author}{\bibfnamefont{H.}~\bibnamefont{Huang}}, \bibnamefont{and}
  \bibinfo{author}{\bibfnamefont{D.}~\bibnamefont{Bluvstein}},
  \bibinfo{journal}{arXiv preprint}
  \textbf{\bibinfo{volume}{quant‑ph/2603.28627}} (\bibinfo{year}{2026}),
  \urlprefix\url{https://arxiv.org/abs/2603.28627}.

\bibitem[{\citenamefont{Zhou et~al.}(2023)\citenamefont{Zhou, Ou, Fang,
  Alkhazraji, Xu, Wan, and Bowers}}]{zhou2023prospects}
\bibinfo{author}{\bibfnamefont{Z.}~\bibnamefont{Zhou}},
  \bibinfo{author}{\bibfnamefont{X.}~\bibnamefont{Ou}},
  \bibinfo{author}{\bibfnamefont{Y.}~\bibnamefont{Fang}},
  \bibinfo{author}{\bibfnamefont{E.}~\bibnamefont{Alkhazraji}},
  \bibinfo{author}{\bibfnamefont{R.}~\bibnamefont{Xu}},
  \bibinfo{author}{\bibfnamefont{Y.}~\bibnamefont{Wan}}, \bibnamefont{and}
  \bibinfo{author}{\bibfnamefont{J.~E.} \bibnamefont{Bowers}},
  \bibinfo{journal}{elight} \textbf{\bibinfo{volume}{3}}, \bibinfo{pages}{1}
  (\bibinfo{year}{2023}).

\bibitem[{\citenamefont{Dutt et~al.}(2024)\citenamefont{Dutt, Mohanty, Gaeta,
  and Lipson}}]{dutt2024nonlinear}
\bibinfo{author}{\bibfnamefont{A.}~\bibnamefont{Dutt}},
  \bibinfo{author}{\bibfnamefont{A.}~\bibnamefont{Mohanty}},
  \bibinfo{author}{\bibfnamefont{A.~L.} \bibnamefont{Gaeta}}, \bibnamefont{and}
  \bibinfo{author}{\bibfnamefont{M.}~\bibnamefont{Lipson}},
  \bibinfo{journal}{Nature Reviews Materials} \textbf{\bibinfo{volume}{9}},
  \bibinfo{pages}{321} (\bibinfo{year}{2024}).

\bibitem[{\citenamefont{Wang et~al.}(2020)\citenamefont{Wang, Sciarrino, Laing,
  and Thompson}}]{wang2020integrated}
\bibinfo{author}{\bibfnamefont{J.}~\bibnamefont{Wang}},
  \bibinfo{author}{\bibfnamefont{F.}~\bibnamefont{Sciarrino}},
  \bibinfo{author}{\bibfnamefont{A.}~\bibnamefont{Laing}}, \bibnamefont{and}
  \bibinfo{author}{\bibfnamefont{M.~G.} \bibnamefont{Thompson}},
  \bibinfo{journal}{Nature photonics} \textbf{\bibinfo{volume}{14}},
  \bibinfo{pages}{273} (\bibinfo{year}{2020}).

\bibitem[{\citenamefont{Kok et~al.}(2007)\citenamefont{Kok, Munro, Nemoto,
  Ralph, Dowling, and Milburn}}]{Kok2007Linear}
\bibinfo{author}{\bibfnamefont{P.}~\bibnamefont{Kok}},
  \bibinfo{author}{\bibfnamefont{W.~J.} \bibnamefont{Munro}},
  \bibinfo{author}{\bibfnamefont{K.}~\bibnamefont{Nemoto}},
  \bibinfo{author}{\bibfnamefont{T.~C.} \bibnamefont{Ralph}},
  \bibinfo{author}{\bibfnamefont{J.~P.} \bibnamefont{Dowling}},
  \bibnamefont{and} \bibinfo{author}{\bibfnamefont{G.~J.}
  \bibnamefont{Milburn}}, \bibinfo{journal}{Rev. Mod. Phys.}
  \textbf{\bibinfo{volume}{79}}, \bibinfo{pages}{135} (\bibinfo{year}{2007}),
  \urlprefix\url{https://link.aps.org/doi/10.1103/RevModPhys.79.135}.

\bibitem[{\citenamefont{O'Brien}(2007)}]{OBrien2007Optical}
\bibinfo{author}{\bibfnamefont{J.~L.} \bibnamefont{O'Brien}},
  \bibinfo{journal}{Science} \textbf{\bibinfo{volume}{318}},
  \bibinfo{pages}{1567} (\bibinfo{year}{2007}).

\bibitem[{\citenamefont{Weedbrook et~al.}(2012)\citenamefont{Weedbrook,
  Pirandola, Garc\'{\i}a-Patr\'on, Cerf, Ralph, Shapiro, and
  Lloyd}}]{Weedbrook2012Gaussian}
\bibinfo{author}{\bibfnamefont{C.}~\bibnamefont{Weedbrook}},
  \bibinfo{author}{\bibfnamefont{S.}~\bibnamefont{Pirandola}},
  \bibinfo{author}{\bibfnamefont{R.}~\bibnamefont{Garc\'{\i}a-Patr\'on}},
  \bibinfo{author}{\bibfnamefont{N.~J.} \bibnamefont{Cerf}},
  \bibinfo{author}{\bibfnamefont{T.~C.} \bibnamefont{Ralph}},
  \bibinfo{author}{\bibfnamefont{J.~H.} \bibnamefont{Shapiro}},
  \bibnamefont{and} \bibinfo{author}{\bibfnamefont{S.}~\bibnamefont{Lloyd}},
  \bibinfo{journal}{Rev. Mod. Phys.} \textbf{\bibinfo{volume}{84}},
  \bibinfo{pages}{621} (\bibinfo{year}{2012}),
  \urlprefix\url{https://link.aps.org/doi/10.1103/RevModPhys.84.621}.

\bibitem[{\citenamefont{O'Brien et~al.}(2009)\citenamefont{O'Brien, Furusawa,
  and V{\v{u}}kovi{\'c}}}]{OBrien2009PhotonicQuantum}
\bibinfo{author}{\bibfnamefont{J.~L.} \bibnamefont{O'Brien}},
  \bibinfo{author}{\bibfnamefont{A.}~\bibnamefont{Furusawa}}, \bibnamefont{and}
  \bibinfo{author}{\bibfnamefont{J.}~\bibnamefont{V{\v{u}}kovi{\'c}}},
  \bibinfo{journal}{Nature Photonics} \textbf{\bibinfo{volume}{3}},
  \bibinfo{pages}{687} (\bibinfo{year}{2009}).

\bibitem[{\citenamefont{Wang et~al.}(2025{\natexlab{a}})\citenamefont{Wang,
  Ralph, Renema, Lu, and Pan}}]{wang2025scalable}
\bibinfo{author}{\bibfnamefont{H.}~\bibnamefont{Wang}},
  \bibinfo{author}{\bibfnamefont{T.~C.} \bibnamefont{Ralph}},
  \bibinfo{author}{\bibfnamefont{J.~J.} \bibnamefont{Renema}},
  \bibinfo{author}{\bibfnamefont{C.-Y.} \bibnamefont{Lu}}, \bibnamefont{and}
  \bibinfo{author}{\bibfnamefont{J.-W.} \bibnamefont{Pan}},
  \bibinfo{journal}{Nature Materials} \textbf{\bibinfo{volume}{24}},
  \bibinfo{pages}{1883} (\bibinfo{year}{2025}{\natexlab{a}}).

\bibitem[{\citenamefont{Zhu et~al.}(2026)\citenamefont{Zhu, Chen, Ma, Zhao,
  Guo, Liang, Tu, He, Luo, Wang et~al.}}]{Zhu2026RecentProgress}
\bibinfo{author}{\bibfnamefont{H.}~\bibnamefont{Zhu}},
  \bibinfo{author}{\bibfnamefont{T.}~\bibnamefont{Chen}},
  \bibinfo{author}{\bibfnamefont{H.}~\bibnamefont{Ma}},
  \bibinfo{author}{\bibfnamefont{Z.}~\bibnamefont{Zhao}},
  \bibinfo{author}{\bibfnamefont{J.}~\bibnamefont{Guo}},
  \bibinfo{author}{\bibfnamefont{Q.}~\bibnamefont{Liang}},
  \bibinfo{author}{\bibfnamefont{Y.}~\bibnamefont{Tu}},
  \bibinfo{author}{\bibfnamefont{T.}~\bibnamefont{He}},
  \bibinfo{author}{\bibfnamefont{W.}~\bibnamefont{Luo}},
  \bibinfo{author}{\bibfnamefont{B.}~\bibnamefont{Wang}}, \bibnamefont{et~al.},
  \bibinfo{journal}{npj Nanophotonics} \textbf{\bibinfo{volume}{3}},
  \bibinfo{pages}{20} (\bibinfo{year}{2026}).

\bibitem[{\citenamefont{Bartolucci et~al.}(2023)\citenamefont{Bartolucci,
  Birchall, Bombin, Cable, Dawson, Gimeno-Segovia, Johnston, Kieling,
  Nickerson, Pant et~al.}}]{Bartolucci2023fusion}
\bibinfo{author}{\bibfnamefont{S.}~\bibnamefont{Bartolucci}},
  \bibinfo{author}{\bibfnamefont{P.}~\bibnamefont{Birchall}},
  \bibinfo{author}{\bibfnamefont{H.}~\bibnamefont{Bombin}},
  \bibinfo{author}{\bibfnamefont{H.}~\bibnamefont{Cable}},
  \bibinfo{author}{\bibfnamefont{C.}~\bibnamefont{Dawson}},
  \bibinfo{author}{\bibfnamefont{M.}~\bibnamefont{Gimeno-Segovia}},
  \bibinfo{author}{\bibfnamefont{E.}~\bibnamefont{Johnston}},
  \bibinfo{author}{\bibfnamefont{K.}~\bibnamefont{Kieling}},
  \bibinfo{author}{\bibfnamefont{N.}~\bibnamefont{Nickerson}},
  \bibinfo{author}{\bibfnamefont{M.}~\bibnamefont{Pant}}, \bibnamefont{et~al.},
  \bibinfo{journal}{Nature Communications} \textbf{\bibinfo{volume}{14}},
  \bibinfo{pages}{912} (\bibinfo{year}{2023}).

\bibitem[{\citenamefont{Bartlett and Fan}(2020)}]{Bartlett2020Universal}
\bibinfo{author}{\bibfnamefont{B.}~\bibnamefont{Bartlett}} \bibnamefont{and}
  \bibinfo{author}{\bibfnamefont{S.}~\bibnamefont{Fan}},
  \bibinfo{journal}{Phys. Rev. A} \textbf{\bibinfo{volume}{101}},
  \bibinfo{pages}{042319} (\bibinfo{year}{2020}),
  \urlprefix\url{https://link.aps.org/doi/10.1103/PhysRevA.101.042319}.

\bibitem[{\citenamefont{Simon and Mukunda}(1990)}]{Simon1990Minimal}
\bibinfo{author}{\bibfnamefont{R.}~\bibnamefont{Simon}} \bibnamefont{and}
  \bibinfo{author}{\bibfnamefont{N.}~\bibnamefont{Mukunda}},
  \bibinfo{journal}{Physics Letters A} \textbf{\bibinfo{volume}{143}},
  \bibinfo{pages}{165} (\bibinfo{year}{1990}), ISSN \bibinfo{issn}{0375-9601},
  \urlprefix\url{https://www.sciencedirect.com/science/article/pii/0375960190907324}.

\bibitem[{\citenamefont{Pant et~al.}(2019)\citenamefont{Pant, Towsley, Englund,
  and Guha}}]{Pant2019percolation}
\bibinfo{author}{\bibfnamefont{M.}~\bibnamefont{Pant}},
  \bibinfo{author}{\bibfnamefont{D.}~\bibnamefont{Towsley}},
  \bibinfo{author}{\bibfnamefont{D.}~\bibnamefont{Englund}}, \bibnamefont{and}
  \bibinfo{author}{\bibfnamefont{S.}~\bibnamefont{Guha}},
  \bibinfo{journal}{Nature communications} \textbf{\bibinfo{volume}{10}},
  \bibinfo{pages}{1070} (\bibinfo{year}{2019}).

\bibitem[{\citenamefont{Gimeno-Segovia
  et~al.}(2015)\citenamefont{Gimeno-Segovia, Shadbolt, Browne, and
  Rudolph}}]{Gimeno2015From}
\bibinfo{author}{\bibfnamefont{M.}~\bibnamefont{Gimeno-Segovia}},
  \bibinfo{author}{\bibfnamefont{P.}~\bibnamefont{Shadbolt}},
  \bibinfo{author}{\bibfnamefont{D.~E.} \bibnamefont{Browne}},
  \bibnamefont{and} \bibinfo{author}{\bibfnamefont{T.}~\bibnamefont{Rudolph}},
  \bibinfo{journal}{Phys. Rev. Lett.} \textbf{\bibinfo{volume}{115}},
  \bibinfo{pages}{020502} (\bibinfo{year}{2015}),
  \urlprefix\url{https://link.aps.org/doi/10.1103/PhysRevLett.115.020502}.

\bibitem[{\citenamefont{Ewert and van Loock}(2014)}]{Ewert2014Efficient}
\bibinfo{author}{\bibfnamefont{F.}~\bibnamefont{Ewert}} \bibnamefont{and}
  \bibinfo{author}{\bibfnamefont{P.}~\bibnamefont{van Loock}},
  \bibinfo{journal}{Phys. Rev. Lett.} \textbf{\bibinfo{volume}{113}},
  \bibinfo{pages}{140403} (\bibinfo{year}{2014}),
  \urlprefix\url{https://link.aps.org/doi/10.1103/PhysRevLett.113.140403}.

\bibitem[{\citenamefont{Barrett and Stace}(2010)}]{Barrett2010Fault}
\bibinfo{author}{\bibfnamefont{S.~D.} \bibnamefont{Barrett}} \bibnamefont{and}
  \bibinfo{author}{\bibfnamefont{T.~M.} \bibnamefont{Stace}},
  \bibinfo{journal}{Phys. Rev. Lett.} \textbf{\bibinfo{volume}{105}},
  \bibinfo{pages}{200502} (\bibinfo{year}{2010}),
  \urlprefix\url{https://link.aps.org/doi/10.1103/PhysRevLett.105.200502}.

\bibitem[{\citenamefont{Li et~al.}(2015)\citenamefont{Li, Humphreys, Mendoza,
  and Benjamin}}]{Li2015Resource}
\bibinfo{author}{\bibfnamefont{Y.}~\bibnamefont{Li}},
  \bibinfo{author}{\bibfnamefont{P.~C.} \bibnamefont{Humphreys}},
  \bibinfo{author}{\bibfnamefont{G.~J.} \bibnamefont{Mendoza}},
  \bibnamefont{and} \bibinfo{author}{\bibfnamefont{S.~C.}
  \bibnamefont{Benjamin}}, \bibinfo{journal}{Phys. Rev. X}
  \textbf{\bibinfo{volume}{5}}, \bibinfo{pages}{041007} (\bibinfo{year}{2015}),
  \urlprefix\url{https://link.aps.org/doi/10.1103/PhysRevX.5.041007}.

\bibitem[{\citenamefont{Herr et~al.}(2018)\citenamefont{Herr, Paler, Devitt,
  and Nori}}]{Herr2018LocalScalable}
\bibinfo{author}{\bibfnamefont{D.}~\bibnamefont{Herr}},
  \bibinfo{author}{\bibfnamefont{A.}~\bibnamefont{Paler}},
  \bibinfo{author}{\bibfnamefont{S.~J.} \bibnamefont{Devitt}},
  \bibnamefont{and} \bibinfo{author}{\bibfnamefont{F.}~\bibnamefont{Nori}},
  \bibinfo{journal}{npj Quantum Information} \textbf{\bibinfo{volume}{4}},
  \bibinfo{pages}{27} (\bibinfo{year}{2018}).

\bibitem[{\citenamefont{Auger et~al.}(2018)\citenamefont{Auger, Anwar,
  Gimeno-Segovia, Stace, and Browne}}]{Auger2018Fault}
\bibinfo{author}{\bibfnamefont{J.~M.} \bibnamefont{Auger}},
  \bibinfo{author}{\bibfnamefont{H.}~\bibnamefont{Anwar}},
  \bibinfo{author}{\bibfnamefont{M.}~\bibnamefont{Gimeno-Segovia}},
  \bibinfo{author}{\bibfnamefont{T.~M.} \bibnamefont{Stace}}, \bibnamefont{and}
  \bibinfo{author}{\bibfnamefont{D.~E.} \bibnamefont{Browne}},
  \bibinfo{journal}{Phys. Rev. A} \textbf{\bibinfo{volume}{97}},
  \bibinfo{pages}{030301} (\bibinfo{year}{2018}),
  \urlprefix\url{https://link.aps.org/doi/10.1103/PhysRevA.97.030301}.

\bibitem[{\citenamefont{Breuckmann and Eberhardt}(2021)}]{Breuckman2021}
\bibinfo{author}{\bibfnamefont{N.~P.} \bibnamefont{Breuckmann}}
  \bibnamefont{and} \bibinfo{author}{\bibfnamefont{J.~N.}
  \bibnamefont{Eberhardt}}, \bibinfo{journal}{PRX Quantum}
  \textbf{\bibinfo{volume}{2}}, \bibinfo{pages}{040101} (\bibinfo{year}{2021}),
  \urlprefix\url{https://link.aps.org/doi/10.1103/PRXQuantum.2.040101}.

\bibitem[{\citenamefont{Panteleev and
  Kalachev}(2022)}]{panteleev2022asymptotically}
\bibinfo{author}{\bibfnamefont{P.}~\bibnamefont{Panteleev}} \bibnamefont{and}
  \bibinfo{author}{\bibfnamefont{G.}~\bibnamefont{Kalachev}}, in
  \emph{\bibinfo{booktitle}{Proceedings of the 54th Annual ACM SIGACT Symposium
  on Theory of Computing}} (\bibinfo{year}{2022}), pp.
  \bibinfo{pages}{375--388}.

\bibitem[{\citenamefont{Leverrier and Z{\'e}mor}(2022)}]{leverrier2022quantum}
\bibinfo{author}{\bibfnamefont{A.}~\bibnamefont{Leverrier}} \bibnamefont{and}
  \bibinfo{author}{\bibfnamefont{G.}~\bibnamefont{Z{\'e}mor}}, in
  \emph{\bibinfo{booktitle}{2022 IEEE 63rd Annual Symposium on Foundations of
  Computer Science (FOCS)}} (\bibinfo{organization}{IEEE},
  \bibinfo{year}{2022}), pp. \bibinfo{pages}{872--883}.

\bibitem[{\citenamefont{Bravyi et~al.}(2024)\citenamefont{Bravyi, Cross,
  Gambetta, Maslov, Rall, and Yoder}}]{bravyi2024high}
\bibinfo{author}{\bibfnamefont{S.}~\bibnamefont{Bravyi}},
  \bibinfo{author}{\bibfnamefont{A.~W.} \bibnamefont{Cross}},
  \bibinfo{author}{\bibfnamefont{J.~M.} \bibnamefont{Gambetta}},
  \bibinfo{author}{\bibfnamefont{D.}~\bibnamefont{Maslov}},
  \bibinfo{author}{\bibfnamefont{P.}~\bibnamefont{Rall}}, \bibnamefont{and}
  \bibinfo{author}{\bibfnamefont{T.~J.} \bibnamefont{Yoder}},
  \bibinfo{journal}{Nature} \textbf{\bibinfo{volume}{627}},
  \bibinfo{pages}{778} (\bibinfo{year}{2024}).

\bibitem[{\citenamefont{Xu et~al.}(2024)\citenamefont{Xu, Bonilla~Ataides,
  Pattison, Raveendran, Bluvstein, Wurtz, Vasi{\'c}, Lukin, Jiang, and
  Zhou}}]{xu2024constant}
\bibinfo{author}{\bibfnamefont{Q.}~\bibnamefont{Xu}},
  \bibinfo{author}{\bibfnamefont{J.~P.} \bibnamefont{Bonilla~Ataides}},
  \bibinfo{author}{\bibfnamefont{C.~A.} \bibnamefont{Pattison}},
  \bibinfo{author}{\bibfnamefont{N.}~\bibnamefont{Raveendran}},
  \bibinfo{author}{\bibfnamefont{D.}~\bibnamefont{Bluvstein}},
  \bibinfo{author}{\bibfnamefont{J.}~\bibnamefont{Wurtz}},
  \bibinfo{author}{\bibfnamefont{B.}~\bibnamefont{Vasi{\'c}}},
  \bibinfo{author}{\bibfnamefont{M.~D.} \bibnamefont{Lukin}},
  \bibinfo{author}{\bibfnamefont{L.}~\bibnamefont{Jiang}}, \bibnamefont{and}
  \bibinfo{author}{\bibfnamefont{H.}~\bibnamefont{Zhou}},
  \bibinfo{journal}{Nature Physics} \textbf{\bibinfo{volume}{20}},
  \bibinfo{pages}{1084} (\bibinfo{year}{2024}).

\bibitem[{\citenamefont{Chen et~al.}(2026)\citenamefont{Chen, L{\"o}bl, Chan,
  S{\o}rensen, and Paesani}}]{chen2026fusion}
\bibinfo{author}{\bibfnamefont{S.~X.} \bibnamefont{Chen}},
  \bibinfo{author}{\bibfnamefont{M.~C.} \bibnamefont{L{\"o}bl}},
  \bibinfo{author}{\bibfnamefont{M.~L.} \bibnamefont{Chan}},
  \bibinfo{author}{\bibfnamefont{A.~S.} \bibnamefont{S{\o}rensen}},
  \bibnamefont{and} \bibinfo{author}{\bibfnamefont{S.}~\bibnamefont{Paesani}},
  \bibinfo{journal}{npj Quantum Information}  (\bibinfo{year}{2026}).

\bibitem[{\citenamefont{Lloyd and Braunstein}(1999)}]{Lloyd1999Quantum}
\bibinfo{author}{\bibfnamefont{S.}~\bibnamefont{Lloyd}} \bibnamefont{and}
  \bibinfo{author}{\bibfnamefont{S.~L.} \bibnamefont{Braunstein}},
  \bibinfo{journal}{Phys. Rev. Lett.} \textbf{\bibinfo{volume}{82}},
  \bibinfo{pages}{1784} (\bibinfo{year}{1999}),
  \urlprefix\url{https://link.aps.org/doi/10.1103/PhysRevLett.82.1784}.

\bibitem[{\citenamefont{Braunstein and van
  Loock}(2005)}]{Braunstein2005Quantum}
\bibinfo{author}{\bibfnamefont{S.~L.} \bibnamefont{Braunstein}}
  \bibnamefont{and} \bibinfo{author}{\bibfnamefont{P.}~\bibnamefont{van
  Loock}}, \bibinfo{journal}{Rev. Mod. Phys.} \textbf{\bibinfo{volume}{77}},
  \bibinfo{pages}{513} (\bibinfo{year}{2005}),
  \urlprefix\url{https://link.aps.org/doi/10.1103/RevModPhys.77.513}.

\bibitem[{\citenamefont{Zhang and
  Braunstein}(2006)}]{Zhang2006continuous-variable}
\bibinfo{author}{\bibfnamefont{J.}~\bibnamefont{Zhang}} \bibnamefont{and}
  \bibinfo{author}{\bibfnamefont{S.~L.} \bibnamefont{Braunstein}},
  \bibinfo{journal}{Phys. Rev. A} \textbf{\bibinfo{volume}{73}},
  \bibinfo{pages}{032318} (\bibinfo{year}{2006}),
  \urlprefix\url{https://link.aps.org/doi/10.1103/PhysRevA.73.032318}.

\bibitem[{\citenamefont{Menicucci et~al.}(2006)\citenamefont{Menicucci, van
  Loock, Gu, Weedbrook, Ralph, and Nielsen}}]{Menicucci2006universal}
\bibinfo{author}{\bibfnamefont{N.~C.} \bibnamefont{Menicucci}},
  \bibinfo{author}{\bibfnamefont{P.}~\bibnamefont{van Loock}},
  \bibinfo{author}{\bibfnamefont{M.}~\bibnamefont{Gu}},
  \bibinfo{author}{\bibfnamefont{C.}~\bibnamefont{Weedbrook}},
  \bibinfo{author}{\bibfnamefont{T.~C.} \bibnamefont{Ralph}}, \bibnamefont{and}
  \bibinfo{author}{\bibfnamefont{M.~A.} \bibnamefont{Nielsen}},
  \bibinfo{journal}{Phys. Rev. Lett.} \textbf{\bibinfo{volume}{97}},
  \bibinfo{pages}{110501} (\bibinfo{year}{2006}),
  \urlprefix\url{https://link.aps.org/doi/10.1103/PhysRevLett.97.110501}.

\bibitem[{\citenamefont{Menicucci}(2014)}]{Menicucci2014fault-tolerant}
\bibinfo{author}{\bibfnamefont{N.~C.} \bibnamefont{Menicucci}},
  \bibinfo{journal}{Phys. Rev. Lett.} \textbf{\bibinfo{volume}{112}},
  \bibinfo{pages}{120504} (\bibinfo{year}{2014}),
  \urlprefix\url{https://link.aps.org/doi/10.1103/PhysRevLett.112.120504}.

\bibitem[{\citenamefont{van Loock et~al.}(2007)\citenamefont{van Loock,
  Weedbrook, and Gu}}]{vanLoock2007building}
\bibinfo{author}{\bibfnamefont{P.}~\bibnamefont{van Loock}},
  \bibinfo{author}{\bibfnamefont{C.}~\bibnamefont{Weedbrook}},
  \bibnamefont{and} \bibinfo{author}{\bibfnamefont{M.}~\bibnamefont{Gu}},
  \bibinfo{journal}{Phys. Rev. A} \textbf{\bibinfo{volume}{76}},
  \bibinfo{pages}{032321} (\bibinfo{year}{2007}),
  \urlprefix\url{https://link.aps.org/doi/10.1103/PhysRevA.76.032321}.

\bibitem[{\citenamefont{Gu et~al.}(2009)\citenamefont{Gu, Weedbrook, Menicucci,
  Ralph, and van Loock}}]{Gu2009quantum}
\bibinfo{author}{\bibfnamefont{M.}~\bibnamefont{Gu}},
  \bibinfo{author}{\bibfnamefont{C.}~\bibnamefont{Weedbrook}},
  \bibinfo{author}{\bibfnamefont{N.~C.} \bibnamefont{Menicucci}},
  \bibinfo{author}{\bibfnamefont{T.~C.} \bibnamefont{Ralph}}, \bibnamefont{and}
  \bibinfo{author}{\bibfnamefont{P.}~\bibnamefont{van Loock}},
  \bibinfo{journal}{Phys. Rev. A} \textbf{\bibinfo{volume}{79}},
  \bibinfo{pages}{062318} (\bibinfo{year}{2009}),
  \urlprefix\url{https://link.aps.org/doi/10.1103/PhysRevA.79.062318}.

\bibitem[{\citenamefont{van Loock and Furusawa}(2003)}]{Loock2003Detecting}
\bibinfo{author}{\bibfnamefont{P.}~\bibnamefont{van Loock}} \bibnamefont{and}
  \bibinfo{author}{\bibfnamefont{A.}~\bibnamefont{Furusawa}},
  \bibinfo{journal}{Phys. Rev. A} \textbf{\bibinfo{volume}{67}},
  \bibinfo{pages}{052315} (\bibinfo{year}{2003}),
  \urlprefix\url{https://link.aps.org/doi/10.1103/PhysRevA.67.052315}.

\bibitem[{\citenamefont{Su et~al.}(2018)\citenamefont{Su, Sabapathy, Myers, Qi,
  Weedbrook, and Br\'adler}}]{Su2018Implementing}
\bibinfo{author}{\bibfnamefont{D.}~\bibnamefont{Su}},
  \bibinfo{author}{\bibfnamefont{K.~K.} \bibnamefont{Sabapathy}},
  \bibinfo{author}{\bibfnamefont{C.~R.} \bibnamefont{Myers}},
  \bibinfo{author}{\bibfnamefont{H.}~\bibnamefont{Qi}},
  \bibinfo{author}{\bibfnamefont{C.}~\bibnamefont{Weedbrook}},
  \bibnamefont{and}
  \bibinfo{author}{\bibfnamefont{K.}~\bibnamefont{Br\'adler}},
  \bibinfo{journal}{Phys. Rev. A} \textbf{\bibinfo{volume}{98}},
  \bibinfo{pages}{032316} (\bibinfo{year}{2018}),
  \urlprefix\url{https://link.aps.org/doi/10.1103/PhysRevA.98.032316}.

\bibitem[{\citenamefont{Armstrong et~al.}(2012)\citenamefont{Armstrong,
  Morizur, Janousek, Hage, Treps, Lam, and Bachor}}]{Armstrong2012Programmable}
\bibinfo{author}{\bibfnamefont{S.}~\bibnamefont{Armstrong}},
  \bibinfo{author}{\bibfnamefont{J.-F.} \bibnamefont{Morizur}},
  \bibinfo{author}{\bibfnamefont{J.}~\bibnamefont{Janousek}},
  \bibinfo{author}{\bibfnamefont{B.}~\bibnamefont{Hage}},
  \bibinfo{author}{\bibfnamefont{N.}~\bibnamefont{Treps}},
  \bibinfo{author}{\bibfnamefont{P.~K.} \bibnamefont{Lam}}, \bibnamefont{and}
  \bibinfo{author}{\bibfnamefont{H.-A.} \bibnamefont{Bachor}},
  \bibinfo{journal}{Nature Communications} \textbf{\bibinfo{volume}{3}},
  \bibinfo{pages}{1026} (\bibinfo{year}{2012}).

\bibitem[{\citenamefont{Pooser and Jing}(2014)}]{Pooser2014Continuous}
\bibinfo{author}{\bibfnamefont{R.}~\bibnamefont{Pooser}} \bibnamefont{and}
  \bibinfo{author}{\bibfnamefont{J.}~\bibnamefont{Jing}},
  \bibinfo{journal}{Phys. Rev. A} \textbf{\bibinfo{volume}{90}},
  \bibinfo{pages}{043841} (\bibinfo{year}{2014}),
  \urlprefix\url{https://link.aps.org/doi/10.1103/PhysRevA.90.043841}.

\bibitem[{\citenamefont{Pysher et~al.}(2011)\citenamefont{Pysher, Miwa,
  Shahrokhshahi, Bloomer, and Pfister}}]{Pysher2011Parallel}
\bibinfo{author}{\bibfnamefont{M.}~\bibnamefont{Pysher}},
  \bibinfo{author}{\bibfnamefont{Y.}~\bibnamefont{Miwa}},
  \bibinfo{author}{\bibfnamefont{R.}~\bibnamefont{Shahrokhshahi}},
  \bibinfo{author}{\bibfnamefont{R.}~\bibnamefont{Bloomer}}, \bibnamefont{and}
  \bibinfo{author}{\bibfnamefont{O.}~\bibnamefont{Pfister}},
  \bibinfo{journal}{Phys. Rev. Lett.} \textbf{\bibinfo{volume}{107}},
  \bibinfo{pages}{030505} (\bibinfo{year}{2011}),
  \urlprefix\url{https://link.aps.org/doi/10.1103/PhysRevLett.107.030505}.

\bibitem[{\citenamefont{Chen et~al.}(2014)\citenamefont{Chen, Menicucci, and
  Pfister}}]{Chen2014experimental}
\bibinfo{author}{\bibfnamefont{M.}~\bibnamefont{Chen}},
  \bibinfo{author}{\bibfnamefont{N.~C.} \bibnamefont{Menicucci}},
  \bibnamefont{and} \bibinfo{author}{\bibfnamefont{O.}~\bibnamefont{Pfister}},
  \bibinfo{journal}{Phys. Rev. Lett.} \textbf{\bibinfo{volume}{112}},
  \bibinfo{pages}{120505} (\bibinfo{year}{2014}),
  \urlprefix\url{https://link.aps.org/doi/10.1103/PhysRevLett.112.120505}.

\bibitem[{\citenamefont{Roslund et~al.}(2014)\citenamefont{Roslund, De~Araujo,
  Jiang, Fabre, and Treps}}]{Roslund2014Wavelength}
\bibinfo{author}{\bibfnamefont{J.}~\bibnamefont{Roslund}},
  \bibinfo{author}{\bibfnamefont{R.~M.} \bibnamefont{De~Araujo}},
  \bibinfo{author}{\bibfnamefont{S.}~\bibnamefont{Jiang}},
  \bibinfo{author}{\bibfnamefont{C.}~\bibnamefont{Fabre}}, \bibnamefont{and}
  \bibinfo{author}{\bibfnamefont{N.}~\bibnamefont{Treps}},
  \bibinfo{journal}{Nature Photonics} \textbf{\bibinfo{volume}{8}},
  \bibinfo{pages}{109} (\bibinfo{year}{2014}).

\bibitem[{\citenamefont{Mari and Eisert}(2012)}]{Mari2012Positive}
\bibinfo{author}{\bibfnamefont{A.}~\bibnamefont{Mari}} \bibnamefont{and}
  \bibinfo{author}{\bibfnamefont{J.}~\bibnamefont{Eisert}},
  \bibinfo{journal}{Phys. Rev. Lett.} \textbf{\bibinfo{volume}{109}},
  \bibinfo{pages}{230503} (\bibinfo{year}{2012}),
  \urlprefix\url{https://link.aps.org/doi/10.1103/PhysRevLett.109.230503}.

\bibitem[{\citenamefont{Niset et~al.}(2009)\citenamefont{Niset,
  Fiur\'a\ifmmode~\check{s}\else \v{s}\fi{}ek, and Cerf}}]{Niset2009No-go}
\bibinfo{author}{\bibfnamefont{J.}~\bibnamefont{Niset}},
  \bibinfo{author}{\bibfnamefont{J.}~\bibnamefont{Fiur\'a\ifmmode~\check{s}\else
  \v{s}\fi{}ek}}, \bibnamefont{and} \bibinfo{author}{\bibfnamefont{N.~J.}
  \bibnamefont{Cerf}}, \bibinfo{journal}{Phys. Rev. Lett.}
  \textbf{\bibinfo{volume}{102}}, \bibinfo{pages}{120501}
  (\bibinfo{year}{2009}),
  \urlprefix\url{https://link.aps.org/doi/10.1103/PhysRevLett.102.120501}.

\bibitem[{\citenamefont{Walshe et~al.}(2020)\citenamefont{Walshe, Baragiola,
  Alexander, and Menicucci}}]{Walshe2020continuous}
\bibinfo{author}{\bibfnamefont{B.~W.} \bibnamefont{Walshe}},
  \bibinfo{author}{\bibfnamefont{B.~Q.} \bibnamefont{Baragiola}},
  \bibinfo{author}{\bibfnamefont{R.~N.} \bibnamefont{Alexander}},
  \bibnamefont{and} \bibinfo{author}{\bibfnamefont{N.~C.}
  \bibnamefont{Menicucci}}, \bibinfo{journal}{Phys. Rev. A}
  \textbf{\bibinfo{volume}{102}}, \bibinfo{pages}{062411}
  (\bibinfo{year}{2020}),
  \urlprefix\url{https://link.aps.org/doi/10.1103/PhysRevA.102.062411}.

\bibitem[{\citenamefont{Ohliger et~al.}(2010)\citenamefont{Ohliger, Kieling,
  and Eisert}}]{Ohliger2010limitations}
\bibinfo{author}{\bibfnamefont{M.}~\bibnamefont{Ohliger}},
  \bibinfo{author}{\bibfnamefont{K.}~\bibnamefont{Kieling}}, \bibnamefont{and}
  \bibinfo{author}{\bibfnamefont{J.}~\bibnamefont{Eisert}},
  \bibinfo{journal}{Phys. Rev. A} \textbf{\bibinfo{volume}{82}},
  \bibinfo{pages}{042336} (\bibinfo{year}{2010}),
  \urlprefix\url{https://link.aps.org/doi/10.1103/PhysRevA.82.042336}.

\bibitem[{\citenamefont{Cochrane et~al.}(1999)\citenamefont{Cochrane, Milburn,
  and Munro}}]{Cochrane1999Macroscopically}
\bibinfo{author}{\bibfnamefont{P.~T.} \bibnamefont{Cochrane}},
  \bibinfo{author}{\bibfnamefont{G.~J.} \bibnamefont{Milburn}},
  \bibnamefont{and} \bibinfo{author}{\bibfnamefont{W.~J.} \bibnamefont{Munro}},
  \bibinfo{journal}{Phys. Rev. A} \textbf{\bibinfo{volume}{59}},
  \bibinfo{pages}{2631} (\bibinfo{year}{1999}),
  \urlprefix\url{https://link.aps.org/doi/10.1103/PhysRevA.59.2631}.

\bibitem[{\citenamefont{Ralph et~al.}(2003)\citenamefont{Ralph, Gilchrist,
  Milburn, Munro, and Glancy}}]{Ralph2003Quantum}
\bibinfo{author}{\bibfnamefont{T.~C.} \bibnamefont{Ralph}},
  \bibinfo{author}{\bibfnamefont{A.}~\bibnamefont{Gilchrist}},
  \bibinfo{author}{\bibfnamefont{G.~J.} \bibnamefont{Milburn}},
  \bibinfo{author}{\bibfnamefont{W.~J.} \bibnamefont{Munro}}, \bibnamefont{and}
  \bibinfo{author}{\bibfnamefont{S.}~\bibnamefont{Glancy}},
  \bibinfo{journal}{Phys. Rev. A} \textbf{\bibinfo{volume}{68}},
  \bibinfo{pages}{042319} (\bibinfo{year}{2003}),
  \urlprefix\url{https://link.aps.org/doi/10.1103/PhysRevA.68.042319}.

\bibitem[{\citenamefont{Leghtas et~al.}(2013)\citenamefont{Leghtas, Kirchmair,
  Vlastakis, Schoelkopf, Devoret, and Mirrahimi}}]{Leghtas2013Hardware}
\bibinfo{author}{\bibfnamefont{Z.}~\bibnamefont{Leghtas}},
  \bibinfo{author}{\bibfnamefont{G.}~\bibnamefont{Kirchmair}},
  \bibinfo{author}{\bibfnamefont{B.}~\bibnamefont{Vlastakis}},
  \bibinfo{author}{\bibfnamefont{R.~J.} \bibnamefont{Schoelkopf}},
  \bibinfo{author}{\bibfnamefont{M.~H.} \bibnamefont{Devoret}},
  \bibnamefont{and}
  \bibinfo{author}{\bibfnamefont{M.}~\bibnamefont{Mirrahimi}},
  \bibinfo{journal}{Phys. Rev. Lett.} \textbf{\bibinfo{volume}{111}},
  \bibinfo{pages}{120501} (\bibinfo{year}{2013}),
  \urlprefix\url{https://link.aps.org/doi/10.1103/PhysRevLett.111.120501}.

\bibitem[{\citenamefont{Mirrahimi et~al.}(2014)\citenamefont{Mirrahimi,
  Leghtas, Albert, Touzard, Schoelkopf, Jiang, and
  Devoret}}]{Mirrahimi2014Danamically}
\bibinfo{author}{\bibfnamefont{M.}~\bibnamefont{Mirrahimi}},
  \bibinfo{author}{\bibfnamefont{Z.}~\bibnamefont{Leghtas}},
  \bibinfo{author}{\bibfnamefont{V.~V.} \bibnamefont{Albert}},
  \bibinfo{author}{\bibfnamefont{S.}~\bibnamefont{Touzard}},
  \bibinfo{author}{\bibfnamefont{R.~J.} \bibnamefont{Schoelkopf}},
  \bibinfo{author}{\bibfnamefont{L.}~\bibnamefont{Jiang}}, \bibnamefont{and}
  \bibinfo{author}{\bibfnamefont{M.~H.} \bibnamefont{Devoret}},
  \bibinfo{journal}{New Journal of Physics} \textbf{\bibinfo{volume}{16}},
  \bibinfo{pages}{045014} (\bibinfo{year}{2014}),
  \urlprefix\url{https://doi.org/10.1088/1367-2630/16/4/045014}.

\bibitem[{\citenamefont{Michael et~al.}(2016)\citenamefont{Michael, Silveri,
  Brierley, Albert, Salmilehto, Jiang, and Girvin}}]{Mechael2016New}
\bibinfo{author}{\bibfnamefont{M.~H.} \bibnamefont{Michael}},
  \bibinfo{author}{\bibfnamefont{M.}~\bibnamefont{Silveri}},
  \bibinfo{author}{\bibfnamefont{R.~T.} \bibnamefont{Brierley}},
  \bibinfo{author}{\bibfnamefont{V.~V.} \bibnamefont{Albert}},
  \bibinfo{author}{\bibfnamefont{J.}~\bibnamefont{Salmilehto}},
  \bibinfo{author}{\bibfnamefont{L.}~\bibnamefont{Jiang}}, \bibnamefont{and}
  \bibinfo{author}{\bibfnamefont{S.~M.} \bibnamefont{Girvin}},
  \bibinfo{journal}{Phys. Rev. X} \textbf{\bibinfo{volume}{6}},
  \bibinfo{pages}{031006} (\bibinfo{year}{2016}),
  \urlprefix\url{https://link.aps.org/doi/10.1103/PhysRevX.6.031006}.

\bibitem[{\citenamefont{Grimsmo et~al.}(2020)\citenamefont{Grimsmo, Combes, and
  Baragiola}}]{Grimsmo2020Quantum}
\bibinfo{author}{\bibfnamefont{A.~L.} \bibnamefont{Grimsmo}},
  \bibinfo{author}{\bibfnamefont{J.}~\bibnamefont{Combes}}, \bibnamefont{and}
  \bibinfo{author}{\bibfnamefont{B.~Q.} \bibnamefont{Baragiola}},
  \bibinfo{journal}{Phys. Rev. X} \textbf{\bibinfo{volume}{10}},
  \bibinfo{pages}{011058} (\bibinfo{year}{2020}),
  \urlprefix\url{https://link.aps.org/doi/10.1103/PhysRevX.10.011058}.

\bibitem[{\citenamefont{Raussendorf and
  Harrington}(2007)}]{Raussendorf2007faul-tolerant}
\bibinfo{author}{\bibfnamefont{R.}~\bibnamefont{Raussendorf}} \bibnamefont{and}
  \bibinfo{author}{\bibfnamefont{J.}~\bibnamefont{Harrington}},
  \bibinfo{journal}{Phys. Rev. Lett.} \textbf{\bibinfo{volume}{98}},
  \bibinfo{pages}{190504} (\bibinfo{year}{2007}),
  \urlprefix\url{https://link.aps.org/doi/10.1103/PhysRevLett.98.190504}.

\bibitem[{\citenamefont{Raussendorf et~al.}(2007)\citenamefont{Raussendorf,
  Harrington, and Goyal}}]{Raussendorf2007topological}
\bibinfo{author}{\bibfnamefont{R.}~\bibnamefont{Raussendorf}},
  \bibinfo{author}{\bibfnamefont{J.}~\bibnamefont{Harrington}},
  \bibnamefont{and} \bibinfo{author}{\bibfnamefont{K.}~\bibnamefont{Goyal}},
  \bibinfo{journal}{New Journal of Physics} \textbf{\bibinfo{volume}{9}},
  \bibinfo{pages}{199} (\bibinfo{year}{2007}),
  \urlprefix\url{https://doi.org/10.1088/1367-2630/9/6/199}.

\bibitem[{\citenamefont{Fukui et~al.}(2017)\citenamefont{Fukui, Tomita, and
  Okamoto}}]{Fukui2017analog}
\bibinfo{author}{\bibfnamefont{K.}~\bibnamefont{Fukui}},
  \bibinfo{author}{\bibfnamefont{A.}~\bibnamefont{Tomita}}, \bibnamefont{and}
  \bibinfo{author}{\bibfnamefont{A.}~\bibnamefont{Okamoto}},
  \bibinfo{journal}{Phys. Rev. Lett.} \textbf{\bibinfo{volume}{119}},
  \bibinfo{pages}{180507} (\bibinfo{year}{2017}),
  \urlprefix\url{https://link.aps.org/doi/10.1103/PhysRevLett.119.180507}.

\bibitem[{\citenamefont{Fukui et~al.}(2018)\citenamefont{Fukui, Tomita,
  Okamoto, and Fujii}}]{Fukui2018highthreshold}
\bibinfo{author}{\bibfnamefont{K.}~\bibnamefont{Fukui}},
  \bibinfo{author}{\bibfnamefont{A.}~\bibnamefont{Tomita}},
  \bibinfo{author}{\bibfnamefont{A.}~\bibnamefont{Okamoto}}, \bibnamefont{and}
  \bibinfo{author}{\bibfnamefont{K.}~\bibnamefont{Fujii}},
  \bibinfo{journal}{Phys. Rev. X} \textbf{\bibinfo{volume}{8}},
  \bibinfo{pages}{021054} (\bibinfo{year}{2018}),
  \urlprefix\url{https://link.aps.org/doi/10.1103/PhysRevX.8.021054}.

\bibitem[{\citenamefont{Bourassa et~al.}(2021)\citenamefont{Bourassa,
  Alexander, Vasmer, Patil, Tzitrin, Matsuura, Su, Baragiola, Guha, Dauphinais
  et~al.}}]{Bourassa2021blueprintscalable}
\bibinfo{author}{\bibfnamefont{J.~E.} \bibnamefont{Bourassa}},
  \bibinfo{author}{\bibfnamefont{R.~N.} \bibnamefont{Alexander}},
  \bibinfo{author}{\bibfnamefont{M.}~\bibnamefont{Vasmer}},
  \bibinfo{author}{\bibfnamefont{A.}~\bibnamefont{Patil}},
  \bibinfo{author}{\bibfnamefont{I.}~\bibnamefont{Tzitrin}},
  \bibinfo{author}{\bibfnamefont{T.}~\bibnamefont{Matsuura}},
  \bibinfo{author}{\bibfnamefont{D.}~\bibnamefont{Su}},
  \bibinfo{author}{\bibfnamefont{B.~Q.} \bibnamefont{Baragiola}},
  \bibinfo{author}{\bibfnamefont{S.}~\bibnamefont{Guha}},
  \bibinfo{author}{\bibfnamefont{G.}~\bibnamefont{Dauphinais}},
  \bibnamefont{et~al.}, \bibinfo{journal}{{Quantum}}
  \textbf{\bibinfo{volume}{5}}, \bibinfo{pages}{392} (\bibinfo{year}{2021}),
  ISSN \bibinfo{issn}{2521-327X},
  \urlprefix\url{https://doi.org/10.22331/q-2021-02-04-392}.

\bibitem[{\citenamefont{Tzitrin et~al.}(2021)\citenamefont{Tzitrin, Matsuura,
  Alexander, Dauphinais, Bourassa, Sabapathy, Menicucci, and
  Dhand}}]{Tzitrin2021fault-tolerant}
\bibinfo{author}{\bibfnamefont{I.}~\bibnamefont{Tzitrin}},
  \bibinfo{author}{\bibfnamefont{T.}~\bibnamefont{Matsuura}},
  \bibinfo{author}{\bibfnamefont{R.~N.} \bibnamefont{Alexander}},
  \bibinfo{author}{\bibfnamefont{G.}~\bibnamefont{Dauphinais}},
  \bibinfo{author}{\bibfnamefont{J.~E.} \bibnamefont{Bourassa}},
  \bibinfo{author}{\bibfnamefont{K.~K.} \bibnamefont{Sabapathy}},
  \bibinfo{author}{\bibfnamefont{N.~C.} \bibnamefont{Menicucci}},
  \bibnamefont{and} \bibinfo{author}{\bibfnamefont{I.}~\bibnamefont{Dhand}},
  \bibinfo{journal}{PRX Quantum} \textbf{\bibinfo{volume}{2}},
  \bibinfo{pages}{040353} (\bibinfo{year}{2021}),
  \urlprefix\url{https://link.aps.org/doi/10.1103/PRXQuantum.2.040353}.

\bibitem[{\citenamefont{Larsen et~al.}(2021{\natexlab{b}})\citenamefont{Larsen,
  Chamberland, Noh, Neergaard-Nielsen, and
  Andersen}}]{Larsen2021Fault-tolerant}
\bibinfo{author}{\bibfnamefont{M.~V.} \bibnamefont{Larsen}},
  \bibinfo{author}{\bibfnamefont{C.}~\bibnamefont{Chamberland}},
  \bibinfo{author}{\bibfnamefont{K.}~\bibnamefont{Noh}},
  \bibinfo{author}{\bibfnamefont{J.~S.} \bibnamefont{Neergaard-Nielsen}},
  \bibnamefont{and} \bibinfo{author}{\bibfnamefont{U.~L.}
  \bibnamefont{Andersen}}, \bibinfo{journal}{PRX Quantum}
  \textbf{\bibinfo{volume}{2}}, \bibinfo{pages}{030325}
  (\bibinfo{year}{2021}{\natexlab{b}}),
  \urlprefix\url{https://link.aps.org/doi/10.1103/PRXQuantum.2.030325}.

\bibitem[{\citenamefont{Renault et~al.}(2025)\citenamefont{Renault, Yard,
  Pooser, Eaton, and Zaidi}}]{Renault2025End}
\bibinfo{author}{\bibfnamefont{P.}~\bibnamefont{Renault}},
  \bibinfo{author}{\bibfnamefont{P.}~\bibnamefont{Yard}},
  \bibinfo{author}{\bibfnamefont{R.~C.} \bibnamefont{Pooser}},
  \bibinfo{author}{\bibfnamefont{M.}~\bibnamefont{Eaton}}, \bibnamefont{and}
  \bibinfo{author}{\bibfnamefont{H.~A.} \bibnamefont{Zaidi}},
  \bibinfo{journal}{Quantum} \textbf{\bibinfo{volume}{9}},
  \bibinfo{pages}{1796} (\bibinfo{year}{2025}),
  \urlprefix\url{https://quantum-journal.org/papers/q-2025-07-14-1796/}.

\bibitem[{\citenamefont{Walshe et~al.}(2025)\citenamefont{Walshe, Baragiola,
  Ferretti, Gefaell, Vasmer, Weil, Matsuura, Jaeken, Pantaleoni, Han
  et~al.}}]{Walshe2025linear}
\bibinfo{author}{\bibfnamefont{B.~W.} \bibnamefont{Walshe}},
  \bibinfo{author}{\bibfnamefont{B.~Q.} \bibnamefont{Baragiola}},
  \bibinfo{author}{\bibfnamefont{H.}~\bibnamefont{Ferretti}},
  \bibinfo{author}{\bibfnamefont{J.}~\bibnamefont{Gefaell}},
  \bibinfo{author}{\bibfnamefont{M.}~\bibnamefont{Vasmer}},
  \bibinfo{author}{\bibfnamefont{R.}~\bibnamefont{Weil}},
  \bibinfo{author}{\bibfnamefont{T.}~\bibnamefont{Matsuura}},
  \bibinfo{author}{\bibfnamefont{T.}~\bibnamefont{Jaeken}},
  \bibinfo{author}{\bibfnamefont{G.}~\bibnamefont{Pantaleoni}},
  \bibinfo{author}{\bibfnamefont{Z.}~\bibnamefont{Han}}, \bibnamefont{et~al.},
  \bibinfo{journal}{Phys. Rev. Lett.} \textbf{\bibinfo{volume}{134}},
  \bibinfo{pages}{100602} (\bibinfo{year}{2025}),
  \urlprefix\url{https://link.aps.org/doi/10.1103/PhysRevLett.134.100602}.

\bibitem[{\citenamefont{Ide et~al.}(2025)\citenamefont{Ide, Gowda, Nadkarni,
  and Dauphinais}}]{Ide2025Fault-tolerant}
\bibinfo{author}{\bibfnamefont{B.}~\bibnamefont{Ide}},
  \bibinfo{author}{\bibfnamefont{M.~G.} \bibnamefont{Gowda}},
  \bibinfo{author}{\bibfnamefont{P.~J.} \bibnamefont{Nadkarni}},
  \bibnamefont{and}
  \bibinfo{author}{\bibfnamefont{G.}~\bibnamefont{Dauphinais}},
  \bibinfo{journal}{Phys. Rev. X} \textbf{\bibinfo{volume}{15}},
  \bibinfo{pages}{021088} (\bibinfo{year}{2025}),
  \urlprefix\url{https://link.aps.org/doi/10.1103/PhysRevX.15.021088}.

\bibitem[{\citenamefont{Weigand and Terhal}(2018)}]{Weigand2018Generating}
\bibinfo{author}{\bibfnamefont{D.~J.} \bibnamefont{Weigand}} \bibnamefont{and}
  \bibinfo{author}{\bibfnamefont{B.~M.} \bibnamefont{Terhal}},
  \bibinfo{journal}{Phys. Rev. A} \textbf{\bibinfo{volume}{97}},
  \bibinfo{pages}{022341} (\bibinfo{year}{2018}),
  \urlprefix\url{https://link.aps.org/doi/10.1103/PhysRevA.97.022341}.

\bibitem[{\citenamefont{Pittman et~al.}(2002)\citenamefont{Pittman, Jacobs, and
  Franson}}]{Pittman2002Single}
\bibinfo{author}{\bibfnamefont{T.~B.} \bibnamefont{Pittman}},
  \bibinfo{author}{\bibfnamefont{B.~C.} \bibnamefont{Jacobs}},
  \bibnamefont{and} \bibinfo{author}{\bibfnamefont{J.~D.}
  \bibnamefont{Franson}}, \bibinfo{journal}{Phys. Rev. A}
  \textbf{\bibinfo{volume}{66}}, \bibinfo{pages}{042303}
  (\bibinfo{year}{2002}),
  \urlprefix\url{https://link.aps.org/doi/10.1103/PhysRevA.66.042303}.

\bibitem[{\citenamefont{Migdall et~al.}(2002)\citenamefont{Migdall, Branning,
  and Castelletto}}]{Migdall2002Tailoring}
\bibinfo{author}{\bibfnamefont{A.~L.} \bibnamefont{Migdall}},
  \bibinfo{author}{\bibfnamefont{D.}~\bibnamefont{Branning}}, \bibnamefont{and}
  \bibinfo{author}{\bibfnamefont{S.}~\bibnamefont{Castelletto}},
  \bibinfo{journal}{Phys. Rev. A} \textbf{\bibinfo{volume}{66}},
  \bibinfo{pages}{053805} (\bibinfo{year}{2002}),
  \urlprefix\url{https://link.aps.org/doi/10.1103/PhysRevA.66.053805}.

\bibitem[{\citenamefont{Kaneda and Kwiat}(2019)}]{kaneda2019high}
\bibinfo{author}{\bibfnamefont{F.}~\bibnamefont{Kaneda}} \bibnamefont{and}
  \bibinfo{author}{\bibfnamefont{P.~G.} \bibnamefont{Kwiat}},
  \bibinfo{journal}{Science advances} \textbf{\bibinfo{volume}{5}},
  \bibinfo{pages}{eaaw8586} (\bibinfo{year}{2019}).

\bibitem[{\citenamefont{Lodahl et~al.}(2015)\citenamefont{Lodahl, Mahmoodian,
  and Stobbe}}]{Lodahl2015Interfacing}
\bibinfo{author}{\bibfnamefont{P.}~\bibnamefont{Lodahl}},
  \bibinfo{author}{\bibfnamefont{S.}~\bibnamefont{Mahmoodian}},
  \bibnamefont{and} \bibinfo{author}{\bibfnamefont{S.}~\bibnamefont{Stobbe}},
  \bibinfo{journal}{Rev. Mod. Phys.} \textbf{\bibinfo{volume}{87}},
  \bibinfo{pages}{347} (\bibinfo{year}{2015}),
  \urlprefix\url{https://link.aps.org/doi/10.1103/RevModPhys.87.347}.

\bibitem[{\citenamefont{Senellart et~al.}(2017)\citenamefont{Senellart,
  Solomon, and White}}]{senellart2017high}
\bibinfo{author}{\bibfnamefont{P.}~\bibnamefont{Senellart}},
  \bibinfo{author}{\bibfnamefont{G.}~\bibnamefont{Solomon}}, \bibnamefont{and}
  \bibinfo{author}{\bibfnamefont{A.}~\bibnamefont{White}},
  \bibinfo{journal}{Nature nanotechnology} \textbf{\bibinfo{volume}{12}},
  \bibinfo{pages}{1026} (\bibinfo{year}{2017}).

\bibitem[{\citenamefont{He et~al.}(2013)\citenamefont{He, He, Wei, Wu, Gong,
  Schneider, Kamp, H{\"o}fling, Lu, and Pan}}]{He2013OnDemand}
\bibinfo{author}{\bibfnamefont{Y.-M.} \bibnamefont{He}},
  \bibinfo{author}{\bibfnamefont{Y.}~\bibnamefont{He}},
  \bibinfo{author}{\bibfnamefont{Y.-J.} \bibnamefont{Wei}},
  \bibinfo{author}{\bibfnamefont{D.}~\bibnamefont{Wu}},
  \bibinfo{author}{\bibfnamefont{M.-L.} \bibnamefont{Gong}},
  \bibinfo{author}{\bibfnamefont{C.}~\bibnamefont{Schneider}},
  \bibinfo{author}{\bibfnamefont{M.}~\bibnamefont{Kamp}},
  \bibinfo{author}{\bibfnamefont{S.}~\bibnamefont{H{\"o}fling}},
  \bibinfo{author}{\bibfnamefont{C.~Y.} \bibnamefont{Lu}}, \bibnamefont{and}
  \bibinfo{author}{\bibfnamefont{J.-W.} \bibnamefont{Pan}},
  \bibinfo{journal}{Nature Nanotechnology} \textbf{\bibinfo{volume}{8}},
  \bibinfo{pages}{213} (\bibinfo{year}{2013}).

\bibitem[{\citenamefont{Wang et~al.}(2019)\citenamefont{Wang, He, Chung, Hu,
  Yu, Chen, Ding, Chen, Qin, Yang et~al.}}]{Wang2019Towards}
\bibinfo{author}{\bibfnamefont{H.}~\bibnamefont{Wang}},
  \bibinfo{author}{\bibfnamefont{Y.-M.} \bibnamefont{He}},
  \bibinfo{author}{\bibfnamefont{T.~H.} \bibnamefont{Chung}},
  \bibinfo{author}{\bibfnamefont{H.}~\bibnamefont{Hu}},
  \bibinfo{author}{\bibfnamefont{Y.}~\bibnamefont{Yu}},
  \bibinfo{author}{\bibfnamefont{S.}~\bibnamefont{Chen}},
  \bibinfo{author}{\bibfnamefont{X.}~\bibnamefont{Ding}},
  \bibinfo{author}{\bibfnamefont{M.-C.} \bibnamefont{Chen}},
  \bibinfo{author}{\bibfnamefont{J.}~\bibnamefont{Qin}},
  \bibinfo{author}{\bibfnamefont{X.}~\bibnamefont{Yang}}, \bibnamefont{et~al.},
  \bibinfo{journal}{Nature Photonics} \textbf{\bibinfo{volume}{13}},
  \bibinfo{pages}{770} (\bibinfo{year}{2019}).

\bibitem[{\citenamefont{Najer et~al.}(2019)\citenamefont{Najer, L{\"o}bl,
  R{\"u}tter, Kuhlmann, and Warburton}}]{Najer2019GatedQD}
\bibinfo{author}{\bibfnamefont{D.}~\bibnamefont{Najer}},
  \bibinfo{author}{\bibfnamefont{M.~C.} \bibnamefont{L{\"o}bl}},
  \bibinfo{author}{\bibfnamefont{D.~R.} \bibnamefont{R{\"u}tter}},
  \bibinfo{author}{\bibfnamefont{A.~V.} \bibnamefont{Kuhlmann}},
  \bibnamefont{and} \bibinfo{author}{\bibfnamefont{R.~J.}
  \bibnamefont{Warburton}}, \bibinfo{journal}{Nature}
  \textbf{\bibinfo{volume}{575}}, \bibinfo{pages}{622} (\bibinfo{year}{2019}).

\bibitem[{\citenamefont{Tomm et~al.}(2021)\citenamefont{Tomm, Javadi,
  Antoniadis, Najer, L{\"o}bl, Korsch, Schott, Valentin, Wieck, Ludwig
  et~al.}}]{Tomm2021Bright}
\bibinfo{author}{\bibfnamefont{N.}~\bibnamefont{Tomm}},
  \bibinfo{author}{\bibfnamefont{A.}~\bibnamefont{Javadi}},
  \bibinfo{author}{\bibfnamefont{N.~O.} \bibnamefont{Antoniadis}},
  \bibinfo{author}{\bibfnamefont{D.}~\bibnamefont{Najer}},
  \bibinfo{author}{\bibfnamefont{M.~C.} \bibnamefont{L{\"o}bl}},
  \bibinfo{author}{\bibfnamefont{A.~R.} \bibnamefont{Korsch}},
  \bibinfo{author}{\bibfnamefont{R.}~\bibnamefont{Schott}},
  \bibinfo{author}{\bibfnamefont{S.~R.} \bibnamefont{Valentin}},
  \bibinfo{author}{\bibfnamefont{A.~D.} \bibnamefont{Wieck}},
  \bibinfo{author}{\bibfnamefont{A.}~\bibnamefont{Ludwig}},
  \bibnamefont{et~al.}, \bibinfo{journal}{Nature Nanotechnology}
  \textbf{\bibinfo{volume}{16}}, \bibinfo{pages}{399} (\bibinfo{year}{2021}).

\bibitem[{\citenamefont{Ding et~al.}(2025)\citenamefont{Ding, Guo, Xu, Liu,
  Zou, Zhao, Ge, Zhang, Liu, Wang et~al.}}]{Ding2025HighEfficiency}
\bibinfo{author}{\bibfnamefont{X.}~\bibnamefont{Ding}},
  \bibinfo{author}{\bibfnamefont{Y.-P.} \bibnamefont{Guo}},
  \bibinfo{author}{\bibfnamefont{M.-C.} \bibnamefont{Xu}},
  \bibinfo{author}{\bibfnamefont{R.-Z.} \bibnamefont{Liu}},
  \bibinfo{author}{\bibfnamefont{G.-Y.} \bibnamefont{Zou}},
  \bibinfo{author}{\bibfnamefont{J.-Y.} \bibnamefont{Zhao}},
  \bibinfo{author}{\bibfnamefont{Z.-X.} \bibnamefont{Ge}},
  \bibinfo{author}{\bibfnamefont{Q.-H.} \bibnamefont{Zhang}},
  \bibinfo{author}{\bibfnamefont{H.-L.} \bibnamefont{Liu}},
  \bibinfo{author}{\bibfnamefont{L.-J.} \bibnamefont{Wang}},
  \bibnamefont{et~al.}, \bibinfo{journal}{Nature Photonics}
  \textbf{\bibinfo{volume}{19}}, \bibinfo{pages}{387} (\bibinfo{year}{2025}).

\bibitem[{\citenamefont{Zhai et~al.}(2022)\citenamefont{Zhai, Nguyen, Spinnler,
  Ritzmann, L{\"o}bl, Wieck, Ludwig, Javadi, and Warburton}}]{zhai2022quantum}
\bibinfo{author}{\bibfnamefont{L.}~\bibnamefont{Zhai}},
  \bibinfo{author}{\bibfnamefont{G.~N.} \bibnamefont{Nguyen}},
  \bibinfo{author}{\bibfnamefont{C.}~\bibnamefont{Spinnler}},
  \bibinfo{author}{\bibfnamefont{J.}~\bibnamefont{Ritzmann}},
  \bibinfo{author}{\bibfnamefont{M.~C.} \bibnamefont{L{\"o}bl}},
  \bibinfo{author}{\bibfnamefont{A.~D.} \bibnamefont{Wieck}},
  \bibinfo{author}{\bibfnamefont{A.}~\bibnamefont{Ludwig}},
  \bibinfo{author}{\bibfnamefont{A.}~\bibnamefont{Javadi}}, \bibnamefont{and}
  \bibinfo{author}{\bibfnamefont{R.~J.} \bibnamefont{Warburton}},
  \bibinfo{journal}{Nature nanotechnology} \textbf{\bibinfo{volume}{17}},
  \bibinfo{pages}{829} (\bibinfo{year}{2022}).

\bibitem[{\citenamefont{Wang et~al.}(2025{\natexlab{b}})\citenamefont{Wang,
  Zhang, Chen, Zhu, Qin, Dong, Cai, Sui, Wu, Zhang
  et~al.}}]{Wang2025LargeScaleQD}
\bibinfo{author}{\bibfnamefont{X.}~\bibnamefont{Wang}},
  \bibinfo{author}{\bibfnamefont{X.}~\bibnamefont{Zhang}},
  \bibinfo{author}{\bibfnamefont{B.}~\bibnamefont{Chen}},
  \bibinfo{author}{\bibfnamefont{Y.}~\bibnamefont{Zhu}},
  \bibinfo{author}{\bibfnamefont{Y.}~\bibnamefont{Qin}},
  \bibinfo{author}{\bibfnamefont{L.}~\bibnamefont{Dong}},
  \bibinfo{author}{\bibfnamefont{J.}~\bibnamefont{Cai}},
  \bibinfo{author}{\bibfnamefont{D.}~\bibnamefont{Sui}},
  \bibinfo{author}{\bibfnamefont{J.}~\bibnamefont{Wu}},
  \bibinfo{author}{\bibfnamefont{Q.}~\bibnamefont{Zhang}},
  \bibnamefont{et~al.}, \bibinfo{journal}{Nature Materials}
  \textbf{\bibinfo{volume}{24}}, \bibinfo{pages}{1898}
  (\bibinfo{year}{2025}{\natexlab{b}}).

\bibitem[{\citenamefont{Zhong et~al.}(2018)\citenamefont{Zhong, Li, Li, Peng,
  Su, Hu, He, Ding, Zhang, Li et~al.}}]{Zhong2018Photon}
\bibinfo{author}{\bibfnamefont{H.-S.} \bibnamefont{Zhong}},
  \bibinfo{author}{\bibfnamefont{Y.}~\bibnamefont{Li}},
  \bibinfo{author}{\bibfnamefont{W.}~\bibnamefont{Li}},
  \bibinfo{author}{\bibfnamefont{L.-C.} \bibnamefont{Peng}},
  \bibinfo{author}{\bibfnamefont{Z.-E.} \bibnamefont{Su}},
  \bibinfo{author}{\bibfnamefont{Y.}~\bibnamefont{Hu}},
  \bibinfo{author}{\bibfnamefont{Y.-M.} \bibnamefont{He}},
  \bibinfo{author}{\bibfnamefont{X.}~\bibnamefont{Ding}},
  \bibinfo{author}{\bibfnamefont{W.}~\bibnamefont{Zhang}},
  \bibinfo{author}{\bibfnamefont{H.}~\bibnamefont{Li}}, \bibnamefont{et~al.},
  \bibinfo{journal}{Phys. Rev. Lett.} \textbf{\bibinfo{volume}{121}},
  \bibinfo{pages}{250505} (\bibinfo{year}{2018}),
  \urlprefix\url{https://link.aps.org/doi/10.1103/PhysRevLett.121.250505}.

\bibitem[{\citenamefont{Meyer-Scott et~al.}(2022)\citenamefont{Meyer-Scott,
  Prasannan, Dhand, Eigner, Quiring, Barkhofen, Brecht, Plenio, and
  Silberhorn}}]{Meyer-Scott2022Scalable}
\bibinfo{author}{\bibfnamefont{E.}~\bibnamefont{Meyer-Scott}},
  \bibinfo{author}{\bibfnamefont{N.}~\bibnamefont{Prasannan}},
  \bibinfo{author}{\bibfnamefont{I.}~\bibnamefont{Dhand}},
  \bibinfo{author}{\bibfnamefont{C.}~\bibnamefont{Eigner}},
  \bibinfo{author}{\bibfnamefont{V.}~\bibnamefont{Quiring}},
  \bibinfo{author}{\bibfnamefont{S.}~\bibnamefont{Barkhofen}},
  \bibinfo{author}{\bibfnamefont{B.}~\bibnamefont{Brecht}},
  \bibinfo{author}{\bibfnamefont{M.~B.} \bibnamefont{Plenio}},
  \bibnamefont{and}
  \bibinfo{author}{\bibfnamefont{C.}~\bibnamefont{Silberhorn}},
  \bibinfo{journal}{Phys. Rev. Lett.} \textbf{\bibinfo{volume}{129}},
  \bibinfo{pages}{150501} (\bibinfo{year}{2022}),
  \urlprefix\url{https://link.aps.org/doi/10.1103/PhysRevLett.129.150501}.

\bibitem[{\citenamefont{Lindner and Rudolph}(2009)}]{Lindner2009Proposal}
\bibinfo{author}{\bibfnamefont{N.~H.} \bibnamefont{Lindner}} \bibnamefont{and}
  \bibinfo{author}{\bibfnamefont{T.}~\bibnamefont{Rudolph}},
  \bibinfo{journal}{Phys. Rev. Lett.} \textbf{\bibinfo{volume}{103}},
  \bibinfo{pages}{113602} (\bibinfo{year}{2009}),
  \urlprefix\url{https://link.aps.org/doi/10.1103/PhysRevLett.103.113602}.

\bibitem[{\citenamefont{Economou et~al.}(2010)\citenamefont{Economou, Lindner,
  and Rudolph}}]{Economou2010Optically}
\bibinfo{author}{\bibfnamefont{S.~E.} \bibnamefont{Economou}},
  \bibinfo{author}{\bibfnamefont{N.}~\bibnamefont{Lindner}}, \bibnamefont{and}
  \bibinfo{author}{\bibfnamefont{T.}~\bibnamefont{Rudolph}},
  \bibinfo{journal}{Phys. Rev. Lett.} \textbf{\bibinfo{volume}{105}},
  \bibinfo{pages}{093601} (\bibinfo{year}{2010}),
  \urlprefix\url{https://link.aps.org/doi/10.1103/PhysRevLett.105.093601}.

\bibitem[{\citenamefont{Schwartz et~al.}(2016)\citenamefont{Schwartz, Cogan,
  Schmidgall, Don, Gantz, Kenneth, Lindner, and
  Gershoni}}]{schwartz2016deterministic}
\bibinfo{author}{\bibfnamefont{I.}~\bibnamefont{Schwartz}},
  \bibinfo{author}{\bibfnamefont{D.}~\bibnamefont{Cogan}},
  \bibinfo{author}{\bibfnamefont{E.~R.} \bibnamefont{Schmidgall}},
  \bibinfo{author}{\bibfnamefont{Y.}~\bibnamefont{Don}},
  \bibinfo{author}{\bibfnamefont{L.}~\bibnamefont{Gantz}},
  \bibinfo{author}{\bibfnamefont{O.}~\bibnamefont{Kenneth}},
  \bibinfo{author}{\bibfnamefont{N.~H.} \bibnamefont{Lindner}},
  \bibnamefont{and} \bibinfo{author}{\bibfnamefont{D.}~\bibnamefont{Gershoni}},
  \bibinfo{journal}{Science} \textbf{\bibinfo{volume}{354}},
  \bibinfo{pages}{434} (\bibinfo{year}{2016}).

\bibitem[{\citenamefont{Coste et~al.}(2023)\citenamefont{Coste, Fioretto,
  Belabas, Wein, Hilaire, Frantzeskakis, Gundin, Goes, Somaschi, Morassi
  et~al.}}]{coste2023high}
\bibinfo{author}{\bibfnamefont{N.}~\bibnamefont{Coste}},
  \bibinfo{author}{\bibfnamefont{D.}~\bibnamefont{Fioretto}},
  \bibinfo{author}{\bibfnamefont{N.}~\bibnamefont{Belabas}},
  \bibinfo{author}{\bibfnamefont{S.}~\bibnamefont{Wein}},
  \bibinfo{author}{\bibfnamefont{P.}~\bibnamefont{Hilaire}},
  \bibinfo{author}{\bibfnamefont{R.}~\bibnamefont{Frantzeskakis}},
  \bibinfo{author}{\bibfnamefont{M.}~\bibnamefont{Gundin}},
  \bibinfo{author}{\bibfnamefont{B.}~\bibnamefont{Goes}},
  \bibinfo{author}{\bibfnamefont{N.}~\bibnamefont{Somaschi}},
  \bibinfo{author}{\bibfnamefont{M.}~\bibnamefont{Morassi}},
  \bibnamefont{et~al.}, \bibinfo{journal}{Nature Photonics}
  \textbf{\bibinfo{volume}{17}}, \bibinfo{pages}{582} (\bibinfo{year}{2023}).

\bibitem[{\citenamefont{Cogan et~al.}(2023)\citenamefont{Cogan, Su, Kenneth,
  and Gershoni}}]{cogan2023deterministic}
\bibinfo{author}{\bibfnamefont{D.}~\bibnamefont{Cogan}},
  \bibinfo{author}{\bibfnamefont{Z.-E.} \bibnamefont{Su}},
  \bibinfo{author}{\bibfnamefont{O.}~\bibnamefont{Kenneth}}, \bibnamefont{and}
  \bibinfo{author}{\bibfnamefont{D.}~\bibnamefont{Gershoni}},
  \bibinfo{journal}{Nature Photonics} \textbf{\bibinfo{volume}{17}},
  \bibinfo{pages}{324} (\bibinfo{year}{2023}).

\bibitem[{\citenamefont{Thomas et~al.}(2022)\citenamefont{Thomas, Ruscio,
  Morin, and Rempe}}]{thomas2022efficient}
\bibinfo{author}{\bibfnamefont{P.}~\bibnamefont{Thomas}},
  \bibinfo{author}{\bibfnamefont{L.}~\bibnamefont{Ruscio}},
  \bibinfo{author}{\bibfnamefont{O.}~\bibnamefont{Morin}}, \bibnamefont{and}
  \bibinfo{author}{\bibfnamefont{G.}~\bibnamefont{Rempe}},
  \bibinfo{journal}{Nature} \textbf{\bibinfo{volume}{608}},
  \bibinfo{pages}{677} (\bibinfo{year}{2022}).

\bibitem[{\citenamefont{Thomas et~al.}(2024)\citenamefont{Thomas, Ruscio,
  Morin, and Rempe}}]{thomas2024fusion}
\bibinfo{author}{\bibfnamefont{P.}~\bibnamefont{Thomas}},
  \bibinfo{author}{\bibfnamefont{L.}~\bibnamefont{Ruscio}},
  \bibinfo{author}{\bibfnamefont{O.}~\bibnamefont{Morin}}, \bibnamefont{and}
  \bibinfo{author}{\bibfnamefont{G.}~\bibnamefont{Rempe}},
  \bibinfo{journal}{Nature} \textbf{\bibinfo{volume}{629}},
  \bibinfo{pages}{567} (\bibinfo{year}{2024}).

\bibitem[{\citenamefont{Serkland et~al.}(1995)\citenamefont{Serkland, Fejer,
  Byer, and Yamamoto}}]{Serkland1995Squeezing}
\bibinfo{author}{\bibfnamefont{D.~K.} \bibnamefont{Serkland}},
  \bibinfo{author}{\bibfnamefont{M.~M.} \bibnamefont{Fejer}},
  \bibinfo{author}{\bibfnamefont{R.~L.} \bibnamefont{Byer}}, \bibnamefont{and}
  \bibinfo{author}{\bibfnamefont{Y.}~\bibnamefont{Yamamoto}},
  \bibinfo{journal}{Opt. Lett.} \textbf{\bibinfo{volume}{20}},
  \bibinfo{pages}{1649} (\bibinfo{year}{1995}),
  \urlprefix\url{https://opg.optica.org/ol/abstract.cfm?URI=ol-20-15-1649}.

\bibitem[{\citenamefont{Yoshino et~al.}(2007)\citenamefont{Yoshino, Aoki, and
  Furusawa}}]{Yoshino2007Generation}
\bibinfo{author}{\bibfnamefont{K.~I.} \bibnamefont{Yoshino}},
  \bibinfo{author}{\bibfnamefont{T.}~\bibnamefont{Aoki}}, \bibnamefont{and}
  \bibinfo{author}{\bibfnamefont{A.}~\bibnamefont{Furusawa}},
  \bibinfo{journal}{Applied Physics Letters} \textbf{\bibinfo{volume}{90}},
  \bibinfo{pages}{041111} (\bibinfo{year}{2007}).

\bibitem[{\citenamefont{Kashiwazaki et~al.}(2020)\citenamefont{Kashiwazaki,
  Takanashi, Yamashima, Kazama, Enbutsu, Kasahara, Umeki, and
  Furusawa}}]{kashiwazaki2020continuous}
\bibinfo{author}{\bibfnamefont{T.}~\bibnamefont{Kashiwazaki}},
  \bibinfo{author}{\bibfnamefont{N.}~\bibnamefont{Takanashi}},
  \bibinfo{author}{\bibfnamefont{T.}~\bibnamefont{Yamashima}},
  \bibinfo{author}{\bibfnamefont{T.}~\bibnamefont{Kazama}},
  \bibinfo{author}{\bibfnamefont{K.}~\bibnamefont{Enbutsu}},
  \bibinfo{author}{\bibfnamefont{R.}~\bibnamefont{Kasahara}},
  \bibinfo{author}{\bibfnamefont{T.}~\bibnamefont{Umeki}}, \bibnamefont{and}
  \bibinfo{author}{\bibfnamefont{A.}~\bibnamefont{Furusawa}},
  \bibinfo{journal}{APL Photonics} \textbf{\bibinfo{volume}{5}}
  (\bibinfo{year}{2020}).

\bibitem[{\citenamefont{Kashiwazaki et~al.}(2021)\citenamefont{Kashiwazaki,
  Yamashima, Takanashi, Inoue, Umeki, and
  Furusawa}}]{Kashiwazaki2021LowLossPPLN}
\bibinfo{author}{\bibfnamefont{T.}~\bibnamefont{Kashiwazaki}},
  \bibinfo{author}{\bibfnamefont{T.}~\bibnamefont{Yamashima}},
  \bibinfo{author}{\bibfnamefont{N.}~\bibnamefont{Takanashi}},
  \bibinfo{author}{\bibfnamefont{A.}~\bibnamefont{Inoue}},
  \bibinfo{author}{\bibfnamefont{T.}~\bibnamefont{Umeki}}, \bibnamefont{and}
  \bibinfo{author}{\bibfnamefont{A.}~\bibnamefont{Furusawa}},
  \bibinfo{journal}{Applied Physics Letters} \textbf{\bibinfo{volume}{119}},
  \bibinfo{pages}{251104} (\bibinfo{year}{2021}).

\bibitem[{\citenamefont{Kashiwazaki et~al.}(2023)\citenamefont{Kashiwazaki,
  Yamashima, Enbutsu, Kazama, Inoue, Fukui, Endo, Umeki, and
  Furusawa}}]{Kashiwazaki2023Over8dB}
\bibinfo{author}{\bibfnamefont{T.}~\bibnamefont{Kashiwazaki}},
  \bibinfo{author}{\bibfnamefont{T.}~\bibnamefont{Yamashima}},
  \bibinfo{author}{\bibfnamefont{K.}~\bibnamefont{Enbutsu}},
  \bibinfo{author}{\bibfnamefont{T.}~\bibnamefont{Kazama}},
  \bibinfo{author}{\bibfnamefont{A.}~\bibnamefont{Inoue}},
  \bibinfo{author}{\bibfnamefont{K.}~\bibnamefont{Fukui}},
  \bibinfo{author}{\bibfnamefont{M.}~\bibnamefont{Endo}},
  \bibinfo{author}{\bibfnamefont{T.}~\bibnamefont{Umeki}}, \bibnamefont{and}
  \bibinfo{author}{\bibfnamefont{A.}~\bibnamefont{Furusawa}},
  \bibinfo{journal}{Applied Physics Letters} \textbf{\bibinfo{volume}{122}},
  \bibinfo{pages}{234003} (\bibinfo{year}{2023}).

\bibitem[{\citenamefont{Hirota et~al.}(2025)\citenamefont{Hirota, Kashiwazaki,
  Ha, Yamashima, Jaturaphagorn, Suzuki, Takahashi, Kawasaki, Inoue, Asavanant
  et~al.}}]{Hirota2025Generation10dB}
\bibinfo{author}{\bibfnamefont{K.}~\bibnamefont{Hirota}},
  \bibinfo{author}{\bibfnamefont{T.}~\bibnamefont{Kashiwazaki}},
  \bibinfo{author}{\bibfnamefont{G.}~\bibnamefont{Ha}},
  \bibinfo{author}{\bibfnamefont{T.}~\bibnamefont{Yamashima}},
  \bibinfo{author}{\bibfnamefont{P.}~\bibnamefont{Jaturaphagorn}},
  \bibinfo{author}{\bibfnamefont{T.}~\bibnamefont{Suzuki}},
  \bibinfo{author}{\bibfnamefont{K.}~\bibnamefont{Takahashi}},
  \bibinfo{author}{\bibfnamefont{A.}~\bibnamefont{Kawasaki}},
  \bibinfo{author}{\bibfnamefont{A.}~\bibnamefont{Inoue}},
  \bibinfo{author}{\bibfnamefont{W.}~\bibnamefont{Asavanant}},
  \bibnamefont{et~al.}, \bibinfo{journal}{arXiv}
  \textbf{\bibinfo{volume}{abs/2511.15082}} (\bibinfo{year}{2025}),
  \bibinfo{note}{preprint}, \urlprefix\url{https://arxiv.org/abs/2511.15082}.

\bibitem[{\citenamefont{Eto et~al.}(2011)\citenamefont{Eto, Koshio, Ohshiro,
  Sakurai, Horie, Hirano, and Sasaki}}]{Eto2011Efficient}
\bibinfo{author}{\bibfnamefont{Y.}~\bibnamefont{Eto}},
  \bibinfo{author}{\bibfnamefont{A.}~\bibnamefont{Koshio}},
  \bibinfo{author}{\bibfnamefont{A.}~\bibnamefont{Ohshiro}},
  \bibinfo{author}{\bibfnamefont{J.}~\bibnamefont{Sakurai}},
  \bibinfo{author}{\bibfnamefont{K.}~\bibnamefont{Horie}},
  \bibinfo{author}{\bibfnamefont{T.}~\bibnamefont{Hirano}}, \bibnamefont{and}
  \bibinfo{author}{\bibfnamefont{M.}~\bibnamefont{Sasaki}},
  \bibinfo{journal}{Opt. Lett.} \textbf{\bibinfo{volume}{36}},
  \bibinfo{pages}{4653} (\bibinfo{year}{2011}),
  \urlprefix\url{https://opg.optica.org/ol/abstract.cfm?URI=ol-36-23-4653}.

\bibitem[{\citenamefont{Amari et~al.}(2023)\citenamefont{Amari, Takai, and
  Hirano}}]{Amari2023HighlyEfficient}
\bibinfo{author}{\bibfnamefont{J.}~\bibnamefont{Amari}},
  \bibinfo{author}{\bibfnamefont{J.}~\bibnamefont{Takai}}, \bibnamefont{and}
  \bibinfo{author}{\bibfnamefont{T.}~\bibnamefont{Hirano}},
  \bibinfo{journal}{Opt. Continuum} \textbf{\bibinfo{volume}{2}},
  \bibinfo{pages}{933} (\bibinfo{year}{2023}),
  \urlprefix\url{https://opg.optica.org/optcon/abstract.cfm?URI=optcon-2-4-933}.

\bibitem[{\citenamefont{Kanter et~al.}(2002)\citenamefont{Kanter, Kumar,
  Roussev, Kurz, Parameswaran, and Fejer}}]{Kanter2002squeezing}
\bibinfo{author}{\bibfnamefont{G.~S.} \bibnamefont{Kanter}},
  \bibinfo{author}{\bibfnamefont{P.}~\bibnamefont{Kumar}},
  \bibinfo{author}{\bibfnamefont{R.~V.} \bibnamefont{Roussev}},
  \bibinfo{author}{\bibfnamefont{J.}~\bibnamefont{Kurz}},
  \bibinfo{author}{\bibfnamefont{K.~R.} \bibnamefont{Parameswaran}},
  \bibnamefont{and} \bibinfo{author}{\bibfnamefont{M.~M.} \bibnamefont{Fejer}},
  \bibinfo{journal}{Opt. Express} \textbf{\bibinfo{volume}{10}},
  \bibinfo{pages}{177} (\bibinfo{year}{2002}),
  \urlprefix\url{https://opg.optica.org/oe/abstract.cfm?URI=oe-10-3-177}.

\bibitem[{\citenamefont{Lenzini et~al.}(2018)\citenamefont{Lenzini, Janousek,
  Thearle, Villa, Haylock, Kasture, Cui, Phan, Dao, Yonezawa
  et~al.}}]{Lenzini2018integrated}
\bibinfo{author}{\bibfnamefont{F.}~\bibnamefont{Lenzini}},
  \bibinfo{author}{\bibfnamefont{J.}~\bibnamefont{Janousek}},
  \bibinfo{author}{\bibfnamefont{O.}~\bibnamefont{Thearle}},
  \bibinfo{author}{\bibfnamefont{M.}~\bibnamefont{Villa}},
  \bibinfo{author}{\bibfnamefont{B.}~\bibnamefont{Haylock}},
  \bibinfo{author}{\bibfnamefont{S.}~\bibnamefont{Kasture}},
  \bibinfo{author}{\bibfnamefont{L.}~\bibnamefont{Cui}},
  \bibinfo{author}{\bibfnamefont{H.-P.} \bibnamefont{Phan}},
  \bibinfo{author}{\bibfnamefont{D.~V.} \bibnamefont{Dao}},
  \bibinfo{author}{\bibfnamefont{H.}~\bibnamefont{Yonezawa}},
  \bibnamefont{et~al.}, \bibinfo{journal}{Science Advances}
  \textbf{\bibinfo{volume}{4}}, \bibinfo{pages}{eaat9331}
  (\bibinfo{year}{2018}),
  \eprint{https://www.science.org/doi/pdf/10.1126/sciadv.aat9331},
  \urlprefix\url{https://www.science.org/doi/abs/10.1126/sciadv.aat9331}.

\bibitem[{\citenamefont{Mondain et~al.}(2019)\citenamefont{Mondain, Lunghi,
  Zavatta, Gouzien, Doutre, Micheli, Tanzilli, and
  D'Auria}}]{Mondain2019Lithium}
\bibinfo{author}{\bibfnamefont{F.}~\bibnamefont{Mondain}},
  \bibinfo{author}{\bibfnamefont{T.}~\bibnamefont{Lunghi}},
  \bibinfo{author}{\bibfnamefont{A.}~\bibnamefont{Zavatta}},
  \bibinfo{author}{\bibfnamefont{E.}~\bibnamefont{Gouzien}},
  \bibinfo{author}{\bibfnamefont{F.}~\bibnamefont{Doutre}},
  \bibinfo{author}{\bibfnamefont{M.~D.} \bibnamefont{Micheli}},
  \bibinfo{author}{\bibfnamefont{S.}~\bibnamefont{Tanzilli}}, \bibnamefont{and}
  \bibinfo{author}{\bibfnamefont{V.}~\bibnamefont{D'Auria}},
  \bibinfo{journal}{Photon. Res.} \textbf{\bibinfo{volume}{7}},
  \bibinfo{pages}{A36} (\bibinfo{year}{2019}),
  \urlprefix\url{https://opg.optica.org/prj/abstract.cfm?URI=prj-7-7-A36}.

\bibitem[{\citenamefont{Stefszky et~al.}(2017)\citenamefont{Stefszky, Ricken,
  Eigner, Quiring, Herrmann, and Silberhorn}}]{Stefsky2017Waveguide}
\bibinfo{author}{\bibfnamefont{M.}~\bibnamefont{Stefszky}},
  \bibinfo{author}{\bibfnamefont{R.}~\bibnamefont{Ricken}},
  \bibinfo{author}{\bibfnamefont{C.}~\bibnamefont{Eigner}},
  \bibinfo{author}{\bibfnamefont{V.}~\bibnamefont{Quiring}},
  \bibinfo{author}{\bibfnamefont{H.}~\bibnamefont{Herrmann}}, \bibnamefont{and}
  \bibinfo{author}{\bibfnamefont{C.}~\bibnamefont{Silberhorn}},
  \bibinfo{journal}{Phys. Rev. Appl.} \textbf{\bibinfo{volume}{7}},
  \bibinfo{pages}{044026} (\bibinfo{year}{2017}),
  \urlprefix\url{https://link.aps.org/doi/10.1103/PhysRevApplied.7.044026}.

\bibitem[{\citenamefont{Park et~al.}(2024)\citenamefont{Park, Stokowski,
  Ansari, Gyger, Multani, Celik, Hwang, Dean, Mayor, McKenna
  et~al.}}]{taewon2024singlemode}
\bibinfo{author}{\bibfnamefont{T.}~\bibnamefont{Park}},
  \bibinfo{author}{\bibfnamefont{H.}~\bibnamefont{Stokowski}},
  \bibinfo{author}{\bibfnamefont{V.}~\bibnamefont{Ansari}},
  \bibinfo{author}{\bibfnamefont{S.}~\bibnamefont{Gyger}},
  \bibinfo{author}{\bibfnamefont{K.~K.~S.} \bibnamefont{Multani}},
  \bibinfo{author}{\bibfnamefont{O.~T.} \bibnamefont{Celik}},
  \bibinfo{author}{\bibfnamefont{A.~Y.} \bibnamefont{Hwang}},
  \bibinfo{author}{\bibfnamefont{D.~J.} \bibnamefont{Dean}},
  \bibinfo{author}{\bibfnamefont{F.}~\bibnamefont{Mayor}},
  \bibinfo{author}{\bibfnamefont{T.~P.} \bibnamefont{McKenna}},
  \bibnamefont{et~al.}, \bibinfo{journal}{Science Advances}
  \textbf{\bibinfo{volume}{10}}, \bibinfo{pages}{eadl1814}
  (\bibinfo{year}{2024}),
  \eprint{https://www.science.org/doi/pdf/10.1126/sciadv.adl1814},
  \urlprefix\url{https://www.science.org/doi/abs/10.1126/sciadv.adl1814}.

\bibitem[{\citenamefont{Dutt et~al.}(2015)\citenamefont{Dutt, Luke,
  Manipatruni, Gaeta, Nussenzveig, and Lipson}}]{Dutt2015On-chip}
\bibinfo{author}{\bibfnamefont{A.}~\bibnamefont{Dutt}},
  \bibinfo{author}{\bibfnamefont{K.}~\bibnamefont{Luke}},
  \bibinfo{author}{\bibfnamefont{S.}~\bibnamefont{Manipatruni}},
  \bibinfo{author}{\bibfnamefont{A.~L.} \bibnamefont{Gaeta}},
  \bibinfo{author}{\bibfnamefont{P.}~\bibnamefont{Nussenzveig}},
  \bibnamefont{and} \bibinfo{author}{\bibfnamefont{M.}~\bibnamefont{Lipson}},
  \bibinfo{journal}{Phys. Rev. Appl.} \textbf{\bibinfo{volume}{3}},
  \bibinfo{pages}{044005} (\bibinfo{year}{2015}),
  \urlprefix\url{https://link.aps.org/doi/10.1103/PhysRevApplied.3.044005}.

\bibitem[{\citenamefont{Cernansky and
  Politi}(2020)}]{cernansky2020nanophotonic}
\bibinfo{author}{\bibfnamefont{R.}~\bibnamefont{Cernansky}} \bibnamefont{and}
  \bibinfo{author}{\bibfnamefont{A.}~\bibnamefont{Politi}},
  \bibinfo{journal}{APL Photonics} \textbf{\bibinfo{volume}{5}}
  (\bibinfo{year}{2020}).

\bibitem[{\citenamefont{Alfredo~K{\"o}gler
  et~al.}(2024)\citenamefont{Alfredo~K{\"o}gler, Couto~Rickli,
  Ribeiro~Domeneguetti, Ji, Gaeta, Lipson, Martinelli, and
  Nussenzveig}}]{alfredo2024quantum}
\bibinfo{author}{\bibfnamefont{R.}~\bibnamefont{Alfredo~K{\"o}gler}},
  \bibinfo{author}{\bibfnamefont{G.}~\bibnamefont{Couto~Rickli}},
  \bibinfo{author}{\bibfnamefont{R.}~\bibnamefont{Ribeiro~Domeneguetti}},
  \bibinfo{author}{\bibfnamefont{X.}~\bibnamefont{Ji}},
  \bibinfo{author}{\bibfnamefont{A.~L.} \bibnamefont{Gaeta}},
  \bibinfo{author}{\bibfnamefont{M.}~\bibnamefont{Lipson}},
  \bibinfo{author}{\bibfnamefont{M.}~\bibnamefont{Martinelli}},
  \bibnamefont{and}
  \bibinfo{author}{\bibfnamefont{P.}~\bibnamefont{Nussenzveig}},
  \bibinfo{journal}{Optics Letters} \textbf{\bibinfo{volume}{49}},
  \bibinfo{pages}{3150} (\bibinfo{year}{2024}).

\bibitem[{\citenamefont{Shen et~al.}(2025{\natexlab{a}})\citenamefont{Shen,
  Hsieh, Sridhar, Feldman, Chang, Smith, and Dutt}}]{shen2025strong}
\bibinfo{author}{\bibfnamefont{Y.}~\bibnamefont{Shen}},
  \bibinfo{author}{\bibfnamefont{P.-Y.} \bibnamefont{Hsieh}},
  \bibinfo{author}{\bibfnamefont{S.~K.} \bibnamefont{Sridhar}},
  \bibinfo{author}{\bibfnamefont{S.}~\bibnamefont{Feldman}},
  \bibinfo{author}{\bibfnamefont{Y.-C.} \bibnamefont{Chang}},
  \bibinfo{author}{\bibfnamefont{T.~A.} \bibnamefont{Smith}}, \bibnamefont{and}
  \bibinfo{author}{\bibfnamefont{A.}~\bibnamefont{Dutt}},
  \bibinfo{journal}{Optica} \textbf{\bibinfo{volume}{12}}, \bibinfo{pages}{302}
  (\bibinfo{year}{2025}{\natexlab{a}}).

\bibitem[{\citenamefont{Zhang et~al.}(2021)\citenamefont{Zhang, Menotti, Tan,
  Vaidya, Mahler, Helt, Zatti, Liscidini, Morrison, and
  Vernon}}]{zhang2021squeezed}
\bibinfo{author}{\bibfnamefont{Y.}~\bibnamefont{Zhang}},
  \bibinfo{author}{\bibfnamefont{M.}~\bibnamefont{Menotti}},
  \bibinfo{author}{\bibfnamefont{K.}~\bibnamefont{Tan}},
  \bibinfo{author}{\bibfnamefont{V.}~\bibnamefont{Vaidya}},
  \bibinfo{author}{\bibfnamefont{D.}~\bibnamefont{Mahler}},
  \bibinfo{author}{\bibfnamefont{L.}~\bibnamefont{Helt}},
  \bibinfo{author}{\bibfnamefont{L.}~\bibnamefont{Zatti}},
  \bibinfo{author}{\bibfnamefont{M.}~\bibnamefont{Liscidini}},
  \bibinfo{author}{\bibfnamefont{B.}~\bibnamefont{Morrison}}, \bibnamefont{and}
  \bibinfo{author}{\bibfnamefont{Z.}~\bibnamefont{Vernon}},
  \bibinfo{journal}{Nature communications} \textbf{\bibinfo{volume}{12}},
  \bibinfo{pages}{2233} (\bibinfo{year}{2021}).

\bibitem[{\citenamefont{Zhao et~al.}(2020)\citenamefont{Zhao, Okawachi, Jang,
  Ji, Lipson, and Gaeta}}]{zhao2020near-degenerate}
\bibinfo{author}{\bibfnamefont{Y.}~\bibnamefont{Zhao}},
  \bibinfo{author}{\bibfnamefont{Y.}~\bibnamefont{Okawachi}},
  \bibinfo{author}{\bibfnamefont{J.~K.} \bibnamefont{Jang}},
  \bibinfo{author}{\bibfnamefont{X.}~\bibnamefont{Ji}},
  \bibinfo{author}{\bibfnamefont{M.}~\bibnamefont{Lipson}}, \bibnamefont{and}
  \bibinfo{author}{\bibfnamefont{A.~L.} \bibnamefont{Gaeta}},
  \bibinfo{journal}{Phys. Rev. Lett.} \textbf{\bibinfo{volume}{124}},
  \bibinfo{pages}{193601} (\bibinfo{year}{2020}),
  \urlprefix\url{https://link.aps.org/doi/10.1103/PhysRevLett.124.193601}.

\bibitem[{\citenamefont{Ulanov et~al.}(2025)\citenamefont{Ulanov, Ruhnke,
  Wildi, and Herr}}]{ulanov2025quadrature}
\bibinfo{author}{\bibfnamefont{A.~E.} \bibnamefont{Ulanov}},
  \bibinfo{author}{\bibfnamefont{B.}~\bibnamefont{Ruhnke}},
  \bibinfo{author}{\bibfnamefont{T.}~\bibnamefont{Wildi}}, \bibnamefont{and}
  \bibinfo{author}{\bibfnamefont{T.}~\bibnamefont{Herr}},
  \bibinfo{journal}{Nature Communications}  (\bibinfo{year}{2025}).

\bibitem[{\citenamefont{Nehra et~al.}(2022)\citenamefont{Nehra, Sekine,
  Ledezma, Guo, Gray, Roy, and Marandi}}]{Nehra2022Few}
\bibinfo{author}{\bibfnamefont{R.}~\bibnamefont{Nehra}},
  \bibinfo{author}{\bibfnamefont{R.}~\bibnamefont{Sekine}},
  \bibinfo{author}{\bibfnamefont{L.}~\bibnamefont{Ledezma}},
  \bibinfo{author}{\bibfnamefont{Q.}~\bibnamefont{Guo}},
  \bibinfo{author}{\bibfnamefont{R.~M.} \bibnamefont{Gray}},
  \bibinfo{author}{\bibfnamefont{A.}~\bibnamefont{Roy}}, \bibnamefont{and}
  \bibinfo{author}{\bibfnamefont{A.}~\bibnamefont{Marandi}},
  \bibinfo{journal}{Science} \textbf{\bibinfo{volume}{377}},
  \bibinfo{pages}{1333} (\bibinfo{year}{2022}),
  \eprint{https://www.science.org/doi/pdf/10.1126/science.abo6213},
  \urlprefix\url{https://www.science.org/doi/abs/10.1126/science.abo6213}.

\bibitem[{\citenamefont{Stokowski et~al.}(2023)\citenamefont{Stokowski,
  McKenna, Park, Hwang, Dean, Celik, Ansari, Fejer, and
  Safavi‑Naeini}}]{Stokowski2023Integrated}
\bibinfo{author}{\bibfnamefont{H.~S.} \bibnamefont{Stokowski}},
  \bibinfo{author}{\bibfnamefont{T.~P.} \bibnamefont{McKenna}},
  \bibinfo{author}{\bibfnamefont{T.}~\bibnamefont{Park}},
  \bibinfo{author}{\bibfnamefont{A.~Y.} \bibnamefont{Hwang}},
  \bibinfo{author}{\bibfnamefont{D.~J.} \bibnamefont{Dean}},
  \bibinfo{author}{\bibfnamefont{O.~T.} \bibnamefont{Celik}},
  \bibinfo{author}{\bibfnamefont{V.}~\bibnamefont{Ansari}},
  \bibinfo{author}{\bibfnamefont{M.~M.} \bibnamefont{Fejer}}, \bibnamefont{and}
  \bibinfo{author}{\bibfnamefont{A.~H.} \bibnamefont{Safavi‑Naeini}},
  \bibinfo{journal}{Nature Communications} \textbf{\bibinfo{volume}{14}},
  \bibinfo{pages}{3355} (\bibinfo{year}{2023}).

\bibitem[{\citenamefont{Chen et~al.}(2022)\citenamefont{Chen, Briggs, Hou, and
  Fan}}]{chen2022ultra}
\bibinfo{author}{\bibfnamefont{P.-K.} \bibnamefont{Chen}},
  \bibinfo{author}{\bibfnamefont{I.}~\bibnamefont{Briggs}},
  \bibinfo{author}{\bibfnamefont{S.}~\bibnamefont{Hou}}, \bibnamefont{and}
  \bibinfo{author}{\bibfnamefont{L.}~\bibnamefont{Fan}},
  \bibinfo{journal}{Optics Letters} \textbf{\bibinfo{volume}{47}},
  \bibinfo{pages}{1506} (\bibinfo{year}{2022}).

\bibitem[{\citenamefont{Shi et~al.}(2025)\citenamefont{Shi, Baiju, Chen,
  Mohanraj, Wang, Dhyani, Shajilal, Zhao, Yang, Li et~al.}}]{shi2025squeezed}
\bibinfo{author}{\bibfnamefont{X.}~\bibnamefont{Shi}},
  \bibinfo{author}{\bibfnamefont{A.~A.} \bibnamefont{Baiju}},
  \bibinfo{author}{\bibfnamefont{X.}~\bibnamefont{Chen}},
  \bibinfo{author}{\bibfnamefont{S.~S.} \bibnamefont{Mohanraj}},
  \bibinfo{author}{\bibfnamefont{S.}~\bibnamefont{Wang}},
  \bibinfo{author}{\bibfnamefont{V.}~\bibnamefont{Dhyani}},
  \bibinfo{author}{\bibfnamefont{B.}~\bibnamefont{Shajilal}},
  \bibinfo{author}{\bibfnamefont{M.}~\bibnamefont{Zhao}},
  \bibinfo{author}{\bibfnamefont{R.}~\bibnamefont{Yang}},
  \bibinfo{author}{\bibfnamefont{Y.}~\bibnamefont{Li}}, \bibnamefont{et~al.},
  \bibinfo{journal}{Nanophotonics} \textbf{\bibinfo{volume}{14}},
  \bibinfo{pages}{4721} (\bibinfo{year}{2025}).

\bibitem[{\citenamefont{Vaidya et~al.}(2020)\citenamefont{Vaidya, Morrison,
  Helt, Shahrokshahi, Mahler, Collins, Tan, Lavoie, Repingon, Menotti
  et~al.}}]{vaidya2020broadband}
\bibinfo{author}{\bibfnamefont{V.~D.} \bibnamefont{Vaidya}},
  \bibinfo{author}{\bibfnamefont{B.}~\bibnamefont{Morrison}},
  \bibinfo{author}{\bibfnamefont{L.~G.} \bibnamefont{Helt}},
  \bibinfo{author}{\bibfnamefont{R.}~\bibnamefont{Shahrokshahi}},
  \bibinfo{author}{\bibfnamefont{D.~H.} \bibnamefont{Mahler}},
  \bibinfo{author}{\bibfnamefont{M.~J.} \bibnamefont{Collins}},
  \bibinfo{author}{\bibfnamefont{K.}~\bibnamefont{Tan}},
  \bibinfo{author}{\bibfnamefont{J.}~\bibnamefont{Lavoie}},
  \bibinfo{author}{\bibfnamefont{A.}~\bibnamefont{Repingon}},
  \bibinfo{author}{\bibfnamefont{M.}~\bibnamefont{Menotti}},
  \bibnamefont{et~al.}, \bibinfo{journal}{Science Advances}
  \textbf{\bibinfo{volume}{6}}, \bibinfo{pages}{eaba9186}
  (\bibinfo{year}{2020}),
  \eprint{https://www.science.org/doi/pdf/10.1126/sciadv.aba9186},
  \urlprefix\url{https://www.science.org/doi/abs/10.1126/sciadv.aba9186}.

\bibitem[{\citenamefont{Arrazola et~al.}(2021)\citenamefont{Arrazola, Bergholm,
  Br{\'a}dler, Bromley, Collins, Dhand, Fumagalli, Gerrits, Goussev, Helt
  et~al.}}]{Arrazola2021QuantumCircuits}
\bibinfo{author}{\bibfnamefont{J.~M.} \bibnamefont{Arrazola}},
  \bibinfo{author}{\bibfnamefont{V.}~\bibnamefont{Bergholm}},
  \bibinfo{author}{\bibfnamefont{K.}~\bibnamefont{Br{\'a}dler}},
  \bibinfo{author}{\bibfnamefont{T.~R.} \bibnamefont{Bromley}},
  \bibinfo{author}{\bibfnamefont{M.~J.} \bibnamefont{Collins}},
  \bibinfo{author}{\bibfnamefont{I.}~\bibnamefont{Dhand}},
  \bibinfo{author}{\bibfnamefont{A.}~\bibnamefont{Fumagalli}},
  \bibinfo{author}{\bibfnamefont{T.}~\bibnamefont{Gerrits}},
  \bibinfo{author}{\bibfnamefont{A.}~\bibnamefont{Goussev}},
  \bibinfo{author}{\bibfnamefont{L.~G.} \bibnamefont{Helt}},
  \bibnamefont{et~al.}, \bibinfo{journal}{Nature}
  \textbf{\bibinfo{volume}{591}}, \bibinfo{pages}{54} (\bibinfo{year}{2021}).

\bibitem[{\citenamefont{Vernon et~al.}(2019)\citenamefont{Vernon, Quesada,
  Liscidini, Morrison, Menotti, Tan, and Sipe}}]{Vernon2019Scalable}
\bibinfo{author}{\bibfnamefont{Z.}~\bibnamefont{Vernon}},
  \bibinfo{author}{\bibfnamefont{N.}~\bibnamefont{Quesada}},
  \bibinfo{author}{\bibfnamefont{M.}~\bibnamefont{Liscidini}},
  \bibinfo{author}{\bibfnamefont{B.}~\bibnamefont{Morrison}},
  \bibinfo{author}{\bibfnamefont{M.}~\bibnamefont{Menotti}},
  \bibinfo{author}{\bibfnamefont{K.}~\bibnamefont{Tan}}, \bibnamefont{and}
  \bibinfo{author}{\bibfnamefont{J.}~\bibnamefont{Sipe}},
  \bibinfo{journal}{Phys. Rev. Appl.} \textbf{\bibinfo{volume}{12}},
  \bibinfo{pages}{064024} (\bibinfo{year}{2019}),
  \urlprefix\url{https://link.aps.org/doi/10.1103/PhysRevApplied.12.064024}.

\bibitem[{\citenamefont{Yang et~al.}(2021)\citenamefont{Yang, Jahanbozorgi,
  Jeong, Sun, Pfister, Lee, and Yi}}]{yang2021squeezed}
\bibinfo{author}{\bibfnamefont{Z.}~\bibnamefont{Yang}},
  \bibinfo{author}{\bibfnamefont{M.}~\bibnamefont{Jahanbozorgi}},
  \bibinfo{author}{\bibfnamefont{D.}~\bibnamefont{Jeong}},
  \bibinfo{author}{\bibfnamefont{S.}~\bibnamefont{Sun}},
  \bibinfo{author}{\bibfnamefont{O.}~\bibnamefont{Pfister}},
  \bibinfo{author}{\bibfnamefont{H.}~\bibnamefont{Lee}}, \bibnamefont{and}
  \bibinfo{author}{\bibfnamefont{X.}~\bibnamefont{Yi}},
  \bibinfo{journal}{Nature Communications} \textbf{\bibinfo{volume}{12}},
  \bibinfo{pages}{4781} (\bibinfo{year}{2021}).

\bibitem[{\citenamefont{Jahanbozorgi et~al.}(2023)\citenamefont{Jahanbozorgi,
  Yang, Sun, Chen, Liu, Wang, and Yi}}]{Jahanbozorgi2023generation}
\bibinfo{author}{\bibfnamefont{M.}~\bibnamefont{Jahanbozorgi}},
  \bibinfo{author}{\bibfnamefont{Z.}~\bibnamefont{Yang}},
  \bibinfo{author}{\bibfnamefont{S.}~\bibnamefont{Sun}},
  \bibinfo{author}{\bibfnamefont{H.}~\bibnamefont{Chen}},
  \bibinfo{author}{\bibfnamefont{R.}~\bibnamefont{Liu}},
  \bibinfo{author}{\bibfnamefont{B.}~\bibnamefont{Wang}}, \bibnamefont{and}
  \bibinfo{author}{\bibfnamefont{X.}~\bibnamefont{Yi}},
  \bibinfo{journal}{Optica} \textbf{\bibinfo{volume}{10}},
  \bibinfo{pages}{1100} (\bibinfo{year}{2023}),
  \urlprefix\url{https://opg.optica.org/optica/abstract.cfm?URI=optica-10-8-1100}.

\bibitem[{\citenamefont{Yang et~al.}(2022)\citenamefont{Yang, Jahanbozorgi,
  Jeong, Sun, Pfister, Lee, and Yi}}]{yang2022quantum}
\bibinfo{author}{\bibfnamefont{Z.}~\bibnamefont{Yang}},
  \bibinfo{author}{\bibfnamefont{M.}~\bibnamefont{Jahanbozorgi}},
  \bibinfo{author}{\bibfnamefont{D.}~\bibnamefont{Jeong}},
  \bibinfo{author}{\bibfnamefont{S.}~\bibnamefont{Sun}},
  \bibinfo{author}{\bibfnamefont{O.}~\bibnamefont{Pfister}},
  \bibinfo{author}{\bibfnamefont{H.}~\bibnamefont{Lee}}, \bibnamefont{and}
  \bibinfo{author}{\bibfnamefont{X.}~\bibnamefont{Yi}}, in
  \emph{\bibinfo{booktitle}{Conference on Lasers and Electro-Optics}}
  (\bibinfo{publisher}{Optica Publishing Group}, \bibinfo{year}{2022}), p.
  \bibinfo{pages}{FW1A.7},
  \urlprefix\url{https://opg.optica.org/abstract.cfm?URI=CLEO_QELS-2022-FW1A.7}.

\bibitem[{\citenamefont{Wang et~al.}(2025{\natexlab{c}})\citenamefont{Wang, Li,
  Wang, Zhou, Cheng, Jing, Sun, Li, Li, Wu et~al.}}]{Wang2025LargeScale}
\bibinfo{author}{\bibfnamefont{Z.}~\bibnamefont{Wang}},
  \bibinfo{author}{\bibfnamefont{K.}~\bibnamefont{Li}},
  \bibinfo{author}{\bibfnamefont{Y.}~\bibnamefont{Wang}},
  \bibinfo{author}{\bibfnamefont{X.}~\bibnamefont{Zhou}},
  \bibinfo{author}{\bibfnamefont{Y.}~\bibnamefont{Cheng}},
  \bibinfo{author}{\bibfnamefont{B.}~\bibnamefont{Jing}},
  \bibinfo{author}{\bibfnamefont{F.}~\bibnamefont{Sun}},
  \bibinfo{author}{\bibfnamefont{J.}~\bibnamefont{Li}},
  \bibinfo{author}{\bibfnamefont{Z.}~\bibnamefont{Li}},
  \bibinfo{author}{\bibfnamefont{B.}~\bibnamefont{Wu}}, \bibnamefont{et~al.},
  \bibinfo{journal}{Light: Science \& Applications}
  \textbf{\bibinfo{volume}{14}}, \bibinfo{pages}{164}
  (\bibinfo{year}{2025}{\natexlab{c}}).

\bibitem[{\citenamefont{Jia et~al.}(2025)\citenamefont{Jia, Zhai, Zhu, You,
  Cao, Zhang, Zheng, Fu, Mao, Dai et~al.}}]{Jia2025ContinuousVariable}
\bibinfo{author}{\bibfnamefont{X.}~\bibnamefont{Jia}},
  \bibinfo{author}{\bibfnamefont{C.}~\bibnamefont{Zhai}},
  \bibinfo{author}{\bibfnamefont{X.}~\bibnamefont{Zhu}},
  \bibinfo{author}{\bibfnamefont{C.}~\bibnamefont{You}},
  \bibinfo{author}{\bibfnamefont{Y.}~\bibnamefont{Cao}},
  \bibinfo{author}{\bibfnamefont{X.}~\bibnamefont{Zhang}},
  \bibinfo{author}{\bibfnamefont{Y.}~\bibnamefont{Zheng}},
  \bibinfo{author}{\bibfnamefont{Z.}~\bibnamefont{Fu}},
  \bibinfo{author}{\bibfnamefont{J.}~\bibnamefont{Mao}},
  \bibinfo{author}{\bibfnamefont{T.}~\bibnamefont{Dai}}, \bibnamefont{et~al.},
  \bibinfo{journal}{Nature} \textbf{\bibinfo{volume}{639}},
  \bibinfo{pages}{329} (\bibinfo{year}{2025}).

\bibitem[{\citenamefont{Larsen et~al.}(2025)\citenamefont{Larsen, Bourassa,
  Kocsis, Tasker, Chadwick, González-Arciniegas, Hastrup, Lopetegui-González,
  Miatto, Motamedi et~al.}}]{Larsen2025Integrated}
\bibinfo{author}{\bibfnamefont{M.~V.} \bibnamefont{Larsen}},
  \bibinfo{author}{\bibfnamefont{J.~E.} \bibnamefont{Bourassa}},
  \bibinfo{author}{\bibfnamefont{S.}~\bibnamefont{Kocsis}},
  \bibinfo{author}{\bibfnamefont{J.~F.} \bibnamefont{Tasker}},
  \bibinfo{author}{\bibfnamefont{R.~S.} \bibnamefont{Chadwick}},
  \bibinfo{author}{\bibfnamefont{C.}~\bibnamefont{González-Arciniegas}},
  \bibinfo{author}{\bibfnamefont{J.}~\bibnamefont{Hastrup}},
  \bibinfo{author}{\bibfnamefont{C.~E.} \bibnamefont{Lopetegui-González}},
  \bibinfo{author}{\bibfnamefont{F.~M.} \bibnamefont{Miatto}},
  \bibinfo{author}{\bibfnamefont{A.}~\bibnamefont{Motamedi}},
  \bibnamefont{et~al.}, \bibinfo{journal}{Nature}
  \textbf{\bibinfo{volume}{642}}, \bibinfo{pages}{587} (\bibinfo{year}{2025}).

\bibitem[{\citenamefont{Shen et~al.}(2025{\natexlab{b}})\citenamefont{Shen,
  Hsieh, Srinivasan, Henry, Moille, Sridhar, Restelli, Chang, Srinivasan, Smith
  et~al.}}]{Shen2025HighlySqueezed}
\bibinfo{author}{\bibfnamefont{Y.}~\bibnamefont{Shen}},
  \bibinfo{author}{\bibfnamefont{P.-Y.} \bibnamefont{Hsieh}},
  \bibinfo{author}{\bibfnamefont{D.}~\bibnamefont{Srinivasan}},
  \bibinfo{author}{\bibfnamefont{A.}~\bibnamefont{Henry}},
  \bibinfo{author}{\bibfnamefont{G.}~\bibnamefont{Moille}},
  \bibinfo{author}{\bibfnamefont{S.~K.} \bibnamefont{Sridhar}},
  \bibinfo{author}{\bibfnamefont{A.}~\bibnamefont{Restelli}},
  \bibinfo{author}{\bibfnamefont{Y.-C.} \bibnamefont{Chang}},
  \bibinfo{author}{\bibfnamefont{K.}~\bibnamefont{Srinivasan}},
  \bibinfo{author}{\bibfnamefont{T.~A.} \bibnamefont{Smith}},
  \bibnamefont{et~al.}, \bibinfo{journal}{arXiv preprint arXiv:2505.03734}
  (\bibinfo{year}{2025}{\natexlab{b}}), \eprint{2505.03734},
  \urlprefix\url{https://arxiv.org/abs/2505.03734}.

\bibitem[{\citenamefont{Guidry et~al.}(2022)\citenamefont{Guidry, Lukin, Yang,
  Trivedi, and V{\v{u}}kovi{\'c}}}]{Guidry2022QuantumOptics}
\bibinfo{author}{\bibfnamefont{M.~A.} \bibnamefont{Guidry}},
  \bibinfo{author}{\bibfnamefont{D.~M.} \bibnamefont{Lukin}},
  \bibinfo{author}{\bibfnamefont{K.~Y.} \bibnamefont{Yang}},
  \bibinfo{author}{\bibfnamefont{R.}~\bibnamefont{Trivedi}}, \bibnamefont{and}
  \bibinfo{author}{\bibfnamefont{J.}~\bibnamefont{V{\v{u}}kovi{\'c}}},
  \bibinfo{journal}{Nature Photonics} \textbf{\bibinfo{volume}{16}},
  \bibinfo{pages}{52} (\bibinfo{year}{2022}).

\bibitem[{\citenamefont{Rad et~al.}(2025)\citenamefont{Rad, Ainsworth,
  Alexander, Altieri, Askarani, Baby, Banchi, Baragiola, Bourassa, Chadwick
  et~al.}}]{AghaeeRad2025Scaling}
\bibinfo{author}{\bibfnamefont{H.~A.} \bibnamefont{Rad}},
  \bibinfo{author}{\bibfnamefont{T.}~\bibnamefont{Ainsworth}},
  \bibinfo{author}{\bibfnamefont{R.~N.} \bibnamefont{Alexander}},
  \bibinfo{author}{\bibfnamefont{B.}~\bibnamefont{Altieri}},
  \bibinfo{author}{\bibfnamefont{M.~F.} \bibnamefont{Askarani}},
  \bibinfo{author}{\bibfnamefont{R.}~\bibnamefont{Baby}},
  \bibinfo{author}{\bibfnamefont{L.}~\bibnamefont{Banchi}},
  \bibinfo{author}{\bibfnamefont{B.~Q.} \bibnamefont{Baragiola}},
  \bibinfo{author}{\bibfnamefont{J.~E.} \bibnamefont{Bourassa}},
  \bibinfo{author}{\bibfnamefont{R.~S.} \bibnamefont{Chadwick}},
  \bibnamefont{et~al.}, \bibinfo{journal}{Nature}
  \textbf{\bibinfo{volume}{638}}, \bibinfo{pages}{912} (\bibinfo{year}{2025}).

\bibitem[{\citenamefont{Wu et~al.}(2020)\citenamefont{Wu, Alexander, Liu, and
  Zhang}}]{Wu2020quantum}
\bibinfo{author}{\bibfnamefont{B.-H.} \bibnamefont{Wu}},
  \bibinfo{author}{\bibfnamefont{R.~N.} \bibnamefont{Alexander}},
  \bibinfo{author}{\bibfnamefont{S.}~\bibnamefont{Liu}}, \bibnamefont{and}
  \bibinfo{author}{\bibfnamefont{Z.}~\bibnamefont{Zhang}},
  \bibinfo{journal}{Phys. Rev. Res.} \textbf{\bibinfo{volume}{2}},
  \bibinfo{pages}{023138} (\bibinfo{year}{2020}),
  \urlprefix\url{https://link.aps.org/doi/10.1103/PhysRevResearch.2.023138}.

\bibitem[{\citenamefont{Su et~al.}(2019)\citenamefont{Su, Myers, and
  Sabapathy}}]{Su2019conversion}
\bibinfo{author}{\bibfnamefont{D.}~\bibnamefont{Su}},
  \bibinfo{author}{\bibfnamefont{C.~R.} \bibnamefont{Myers}}, \bibnamefont{and}
  \bibinfo{author}{\bibfnamefont{K.~K.} \bibnamefont{Sabapathy}},
  \bibinfo{journal}{Phys. Rev. A} \textbf{\bibinfo{volume}{100}},
  \bibinfo{pages}{052301} (\bibinfo{year}{2019}),
  \urlprefix\url{https://link.aps.org/doi/10.1103/PhysRevA.100.052301}.

\bibitem[{\citenamefont{Raffaelli et~al.}(2018)\citenamefont{Raffaelli,
  Ferranti, Mahler, Sibson, Kennard, Santamato, Sinclair, Bonneau, Thompson,
  and Matthews}}]{Raffaelli2018AHomodyne}
\bibinfo{author}{\bibfnamefont{F.}~\bibnamefont{Raffaelli}},
  \bibinfo{author}{\bibfnamefont{G.}~\bibnamefont{Ferranti}},
  \bibinfo{author}{\bibfnamefont{D.~H.} \bibnamefont{Mahler}},
  \bibinfo{author}{\bibfnamefont{P.}~\bibnamefont{Sibson}},
  \bibinfo{author}{\bibfnamefont{J.~E.} \bibnamefont{Kennard}},
  \bibinfo{author}{\bibfnamefont{A.}~\bibnamefont{Santamato}},
  \bibinfo{author}{\bibfnamefont{G.}~\bibnamefont{Sinclair}},
  \bibinfo{author}{\bibfnamefont{D.}~\bibnamefont{Bonneau}},
  \bibinfo{author}{\bibfnamefont{M.~G.} \bibnamefont{Thompson}},
  \bibnamefont{and} \bibinfo{author}{\bibfnamefont{J.~C.~F.}
  \bibnamefont{Matthews}}, \bibinfo{journal}{Quantum Science and Technology}
  \textbf{\bibinfo{volume}{3}}, \bibinfo{pages}{025003} (\bibinfo{year}{2018}),
  \urlprefix\url{https://doi.org/10.1088/2058-9565/aaa38f}.

\bibitem[{\citenamefont{Zhang et~al.}(2019)\citenamefont{Zhang, Haw, Cai, Xu,
  Assad, Fitzsimons, Zhou, Zhang, Yu, Wu et~al.}}]{Zhang2019Integrated}
\bibinfo{author}{\bibfnamefont{G.}~\bibnamefont{Zhang}},
  \bibinfo{author}{\bibfnamefont{J.~Y.} \bibnamefont{Haw}},
  \bibinfo{author}{\bibfnamefont{H.}~\bibnamefont{Cai}},
  \bibinfo{author}{\bibfnamefont{F.}~\bibnamefont{Xu}},
  \bibinfo{author}{\bibfnamefont{S.}~\bibnamefont{Assad}},
  \bibinfo{author}{\bibfnamefont{J.~F.} \bibnamefont{Fitzsimons}},
  \bibinfo{author}{\bibfnamefont{X.}~\bibnamefont{Zhou}},
  \bibinfo{author}{\bibfnamefont{Y.}~\bibnamefont{Zhang}},
  \bibinfo{author}{\bibfnamefont{S.}~\bibnamefont{Yu}},
  \bibinfo{author}{\bibfnamefont{J.}~\bibnamefont{Wu}}, \bibnamefont{et~al.},
  \bibinfo{journal}{Nature Photonics} \textbf{\bibinfo{volume}{13}},
  \bibinfo{pages}{839} (\bibinfo{year}{2019}).

\bibitem[{\citenamefont{Tasker et~al.}(2021)\citenamefont{Tasker, Frazer,
  Ferranti, Allen, Brunel, Tanzilli, D'Auria, and
  Matthews}}]{Tasker2021SiliconPhotonics}
\bibinfo{author}{\bibfnamefont{J.~F.} \bibnamefont{Tasker}},
  \bibinfo{author}{\bibfnamefont{J.}~\bibnamefont{Frazer}},
  \bibinfo{author}{\bibfnamefont{G.}~\bibnamefont{Ferranti}},
  \bibinfo{author}{\bibfnamefont{E.~J.} \bibnamefont{Allen}},
  \bibinfo{author}{\bibfnamefont{L.~F.} \bibnamefont{Brunel}},
  \bibinfo{author}{\bibfnamefont{S.}~\bibnamefont{Tanzilli}},
  \bibinfo{author}{\bibfnamefont{V.}~\bibnamefont{D'Auria}}, \bibnamefont{and}
  \bibinfo{author}{\bibfnamefont{J.~C.~F.} \bibnamefont{Matthews}},
  \bibinfo{journal}{Nature Photonics} \textbf{\bibinfo{volume}{15}},
  \bibinfo{pages}{11} (\bibinfo{year}{2021}).

\bibitem[{\citenamefont{Tasker et~al.}(2024)\citenamefont{Tasker, Frazer,
  Ferranti, and Matthews}}]{Tasker2024ABi-CMOS}
\bibinfo{author}{\bibfnamefont{J.~F.} \bibnamefont{Tasker}},
  \bibinfo{author}{\bibfnamefont{J.}~\bibnamefont{Frazer}},
  \bibinfo{author}{\bibfnamefont{G.}~\bibnamefont{Ferranti}}, \bibnamefont{and}
  \bibinfo{author}{\bibfnamefont{J.~C.~F.} \bibnamefont{Matthews}},
  \bibinfo{journal}{Science Advances} \textbf{\bibinfo{volume}{10}},
  \bibinfo{pages}{eadk6890} (\bibinfo{year}{2024}),
  \eprint{https://www.science.org/doi/pdf/10.1126/sciadv.adk6890},
  \urlprefix\url{https://www.science.org/doi/abs/10.1126/sciadv.adk6890}.

\bibitem[{\citenamefont{Alexander et~al.}(2025)\citenamefont{Alexander,
  Benyamini, Black, Bonneau, Burgos, Burridge, Cable, Campbell, Catalano,
  Ceballos et~al.}}]{Alexander2025Manufacturable}
\bibinfo{author}{\bibfnamefont{K.}~\bibnamefont{Alexander}},
  \bibinfo{author}{\bibfnamefont{A.}~\bibnamefont{Benyamini}},
  \bibinfo{author}{\bibfnamefont{D.}~\bibnamefont{Black}},
  \bibinfo{author}{\bibfnamefont{D.}~\bibnamefont{Bonneau}},
  \bibinfo{author}{\bibfnamefont{S.}~\bibnamefont{Burgos}},
  \bibinfo{author}{\bibfnamefont{B.~M.} \bibnamefont{Burridge}},
  \bibinfo{author}{\bibfnamefont{H.}~\bibnamefont{Cable}},
  \bibinfo{author}{\bibfnamefont{G.}~\bibnamefont{Campbell}},
  \bibinfo{author}{\bibfnamefont{G.}~\bibnamefont{Catalano}},
  \bibinfo{author}{\bibfnamefont{A.}~\bibnamefont{Ceballos}},
  \bibnamefont{et~al.}, \bibinfo{journal}{Nature}
  \textbf{\bibinfo{volume}{641}}, \bibinfo{pages}{876} (\bibinfo{year}{2025}).

\bibitem[{\citenamefont{Reddy et~al.}(2020)\citenamefont{Reddy, Nerem, Nam,
  Mirin, and Verma}}]{Reddy2020Superconducting}
\bibinfo{author}{\bibfnamefont{D.~V.} \bibnamefont{Reddy}},
  \bibinfo{author}{\bibfnamefont{R.~R.} \bibnamefont{Nerem}},
  \bibinfo{author}{\bibfnamefont{S.~W.} \bibnamefont{Nam}},
  \bibinfo{author}{\bibfnamefont{R.~P.} \bibnamefont{Mirin}}, \bibnamefont{and}
  \bibinfo{author}{\bibfnamefont{V.~B.} \bibnamefont{Verma}},
  \bibinfo{journal}{Optica} \textbf{\bibinfo{volume}{7}}, \bibinfo{pages}{1649}
  (\bibinfo{year}{2020}).

\bibitem[{\citenamefont{Korzh et~al.}(2020)\citenamefont{Korzh, Zhao, Allmaras,
  Frasca, Autry, Bersin, Beyer, Briggs, Bumble, Colangelo
  et~al.}}]{Korzh2020Sub3psSNSPD}
\bibinfo{author}{\bibfnamefont{B.}~\bibnamefont{Korzh}},
  \bibinfo{author}{\bibfnamefont{Q.-Y.} \bibnamefont{Zhao}},
  \bibinfo{author}{\bibfnamefont{J.~P.} \bibnamefont{Allmaras}},
  \bibinfo{author}{\bibfnamefont{S.}~\bibnamefont{Frasca}},
  \bibinfo{author}{\bibfnamefont{T.~M.} \bibnamefont{Autry}},
  \bibinfo{author}{\bibfnamefont{E.~A.} \bibnamefont{Bersin}},
  \bibinfo{author}{\bibfnamefont{A.~D.} \bibnamefont{Beyer}},
  \bibinfo{author}{\bibfnamefont{R.~M.} \bibnamefont{Briggs}},
  \bibinfo{author}{\bibfnamefont{B.}~\bibnamefont{Bumble}},
  \bibinfo{author}{\bibfnamefont{M.}~\bibnamefont{Colangelo}},
  \bibnamefont{et~al.}, \bibinfo{journal}{Nature Photonics}
  \textbf{\bibinfo{volume}{14}}, \bibinfo{pages}{250} (\bibinfo{year}{2020}).

\bibitem[{\citenamefont{Pernice et~al.}(2012)\citenamefont{Pernice, Schuck,
  Minaeva, Li, Gol'tsman, Sergienko, and Tang}}]{Pernice2012TravellingWave}
\bibinfo{author}{\bibfnamefont{W.~H.~P.} \bibnamefont{Pernice}},
  \bibinfo{author}{\bibfnamefont{C.}~\bibnamefont{Schuck}},
  \bibinfo{author}{\bibfnamefont{O.}~\bibnamefont{Minaeva}},
  \bibinfo{author}{\bibfnamefont{M.}~\bibnamefont{Li}},
  \bibinfo{author}{\bibfnamefont{G.~N.} \bibnamefont{Gol'tsman}},
  \bibinfo{author}{\bibfnamefont{A.~V.} \bibnamefont{Sergienko}},
  \bibnamefont{and} \bibinfo{author}{\bibfnamefont{H.~X.} \bibnamefont{Tang}},
  \bibinfo{journal}{Nature Communications} \textbf{\bibinfo{volume}{3}},
  \bibinfo{pages}{1325} (\bibinfo{year}{2012}).

\bibitem[{\citenamefont{Schuck et~al.}(2013)\citenamefont{Schuck, Pernice, and
  Tang}}]{Schuck2013Waveguide}
\bibinfo{author}{\bibfnamefont{C.}~\bibnamefont{Schuck}},
  \bibinfo{author}{\bibfnamefont{W.~H.~P.} \bibnamefont{Pernice}},
  \bibnamefont{and} \bibinfo{author}{\bibfnamefont{H.~X.} \bibnamefont{Tang}},
  \bibinfo{journal}{Scientific Reports} \textbf{\bibinfo{volume}{3}},
  \bibinfo{pages}{1893} (\bibinfo{year}{2013}).

\bibitem[{\citenamefont{Cheng et~al.}(2023)\citenamefont{Cheng, Zhou, Wang,
  Shen, Taher, and Tang}}]{Cheng2023Hundred}
\bibinfo{author}{\bibfnamefont{R.}~\bibnamefont{Cheng}},
  \bibinfo{author}{\bibfnamefont{Y.}~\bibnamefont{Zhou}},
  \bibinfo{author}{\bibfnamefont{S.}~\bibnamefont{Wang}},
  \bibinfo{author}{\bibfnamefont{M.}~\bibnamefont{Shen}},
  \bibinfo{author}{\bibfnamefont{T.}~\bibnamefont{Taher}}, \bibnamefont{and}
  \bibinfo{author}{\bibfnamefont{H.~X.} \bibnamefont{Tang}},
  \bibinfo{journal}{Nature Photonics} \textbf{\bibinfo{volume}{17}},
  \bibinfo{pages}{112} (\bibinfo{year}{2023}).

\end{thebibliography}

\end{document}